\documentclass[referee,sn-nature]{sn-jnl}

\usepackage{graphicx}%
\usepackage{multirow}%
\usepackage{amsmath,amssymb,amsfonts}%
\usepackage{amsthm}%
\usepackage{mathrsfs}%
\usepackage[title]{appendix}%
\usepackage{xcolor}%
\usepackage{textcomp}%
\usepackage{manyfoot}%
\usepackage{booktabs}%
\usepackage{algorithm}%
\usepackage{algorithmicx}%
\usepackage{algpseudocode}%
\usepackage{listings}%
\usepackage{makecell}
\usepackage{subcaption}

\begin{document}

\title[Article Title]{Enhancing Drug-Target Interaction Prediction through Transfer Learning from Activity Cliff Prediction Tasks}

\author[1]{\fnm{Regina} \sur{Ibragimova}}\email{regina.r.ibragimova@gmail.com}

\author[1]{\fnm{Dimitrios} \sur{Iliadis}}\email{dimitrios.iliadis@ugent.be}

\author*[1]{\fnm{Willem} \sur{Waegeman}}\email{willem.waegeman@ugent.be}

\affil[1]{\orgdiv{Department of Data Analysis and Mathematical Modelling}, \orgname{Ghent University}, \orgaddress{\street{Coupure Links}, \city{Ghent}, \postcode{9000}, \country{Belgium}}}

\abstract{Recently, machine learning (ML) has gained popularity in the early stages of drug discovery. This trend is unsurprising given the increasing volume of relevant experimental data and the continuous improvement of ML algorithms. However, conventional models, which rely on the principle of molecular similarity, often fail to capture the complexities of chemical interactions, particularly those involving activity cliffs (ACs) — compounds that are structurally similar but exhibit evidently different activity behaviors. In this work, we address two distinct yet related tasks: (1) activity cliff (AC) prediction and (2) drug-target interaction (DTI) prediction. 
Leveraging insights gained from the AC prediction task, we aim to improve the performance of DTI prediction through transfer learning. A universal model was developed for AC prediction, capable of identifying activity cliffs across diverse targets. Insights from this model were then incorporated into DTI prediction, enabling better handling of challenging cases involving ACs while maintaining similar overall performance. This approach establishes a strong foundation for integrating AC awareness into predictive models for drug discovery.

\textbf{Scientific Contribution}

This study presents a novel approach that applies transfer learning from AC prediction to enhance DTI prediction, addressing limitations of traditional similarity-based models. By introducing AC-awareness, we improve DTI model performance in structurally complex regions, demonstrating the benefits of integrating compound-specific and protein-contextual information. Unlike previous studies, which treat AC and DTI predictions as separate problems, this work establishes a unified framework to address both data scarcity and prediction challenges in drug discovery.}


\maketitle

\section{Introduction}

Drug development is a challenging, lengthy, and costly process. Although the application of artificial intelligence is creating new opportunities throughout the entire cycle of the drug development process, there is evidence that one of the primary reasons for the high failure rates of drug candidates during clinical development has been a lack of clinical efficacy \cite{Sun2021}. This emphasises the importance of target selection in early drug discovery, drug-target interaction (DTI) prediction, aided by computational methods, offers a more rapid and cost-effective alternative \cite{Vijayan2021}. Moreover, the application of machine learning (ML) in drug-drug interaction and DTI has increased due to the abundance of accessible data, the power of the available tools and services, and the growing demand brought on by high-throughput approaches \cite{Xu2021}. 

Studies on DTI prediction have demonstrated that models with good performance can be constructed by utilizing various optimization or calculation techniques during the phases of dataset acquisition, feature extraction and processing, and task algorithm selection \cite{Yamanishi2008}. Most of the methods are based on the principle of similarity, which states that similar compounds have similar properties and share similar targets \cite{Johnson1990}. One of the most widely used frameworks defines the prediction of DTIs and DDIs as classification problems and formulates various types of similarity functions as input \cite{Yamanishi2008, Ding2013}. However, there is a known weakness referred to as activity cliffs (ACs) \cite{Dimova2016}. Even when an ML model overall has high predictive performance, it may struggle in the case of ACs \cite{Tilborg2022}. 

Predicting ACs is challenging for ML due to three primary problems. Firstly, in ACs, small structural changes in compounds can lead to dramatic changes in activity, resulting in highly discontinuous structure-activity relationships  \cite{Cruz2014}. This phenomenon has two sides. On the one hand, it accelerates drug research by providing important information for drug design and studying drug-drug and drug-target interactions \cite{Dimova2016}. On the other hand, it may lead to the unexpected loss of desired properties \cite{Stumpfe2019}. For ML models, capturing such cases can be difficult, potentially leading to a decrease in performance \cite{Cruz2014, Stumpfe2019}. Secondly, datasets for ACs and non-ACs are highly imbalanced \cite{Chen2022}. Lastly, predictions should be made at the level of compound pairs, in contrast to the usual molecular property predictions, which are made for individual compounds \cite{Chen2022}. 

Despite achieving reasonable overall performance, models often struggle, especially in predicting ACs. While there are examples in the literature of attempts to predict ACs \cite{Chen2022, Tilborg2022, Iqbal2021, Tamura2023, Zhang2023}, to our knowledge, the subsequent application of these predictions to enhance DTI model performance has not yet been extensively explored. A model that lacks AC awareness typically performs worse in the AC area. Integrating such awareness via transfer learning could potentially boost performance in this domain, thereby improving overall model performance. Deep learning models, however, require vast amounts of data for effective training. Traditional ML approaches rely on training and testing datasets with identical input feature spaces. Nevertheless, large volumes of data are often unavailable for certain compounds or targets due to difficulty or high costs, resulting in a lack of models for these cases \cite{Dalkıran2023}.  Transfer learning presents a promising approach to this challenge.  It mitigates this issue by pre-training a model on one task and transferring the knowledge to a related, yet distinct, task \cite{Weiss2016}. By pre-training a model on AC phenomena and subsequently transferring this knowledge to DTI prediction tasks, model performance can be significantly enhanced.

The objective of this research is to improve the performance of DTI prediction models through the application of transfer learning techniques derived from AC prediction tasks. By using the knowledge from AC prediction, we aim to enhance the robustness and accuracy of DTI models, especially in handling complex chemical interactions that traditional models struggle with. Additionally, the implementation of transfer learning will allow the model to better generalize across diverse datasets, even those with limited and scarce data. Ultimately, our goal is to develop a more reliable and efficient technique for DTI prediction that can significantly aid in the early stages of drug discovery and development, potentially reducing the time and cost associated with bringing new drugs to market. 

This paper is organized as follows: we first describe the methodology for the AC prediction and DTI prediction tasks. Subsequently, we discuss the results of our experimental study, organized into subsections where we analyze different aspects of the experiments. We examine the impact of transferring encoders and evaluate model performance in challenging AC-related scenarios using heatmaps. The findings highlight the potential of AC-informed transfer learning to enhance DTI prediction in drug discovery workflows.

\section{Methodology}
\subsection{Datasets and data preprocessing}
\label{datasets_ac}
The KIBA \cite{Tang2014} and BindingDB \cite{Liu2007} datasets were utilized in this study because of their varying numbers of drugs, targets, and measured affinities as well as their popularity in DTI prediction publications. Their brief statistics are shown in Table \ref{tab_datasets}.

\begin{table}[t]
\centering
\caption{Overview of the datasets.}
\begin{tabular}{l c c c c c}
\toprule
 Dataset  & \# Drugs & \# Targets  & \# DTIs  & \# ACs (\%) & \# non-ACs (\%)\\
 \midrule
KIBA & 2068  & 229  & 118254 & 9916 (12\%) & 70601 (88\%)   \\
BindingDB ($K_i$) & 32347  & 1018 & 78628 & 71304 (28\%)  & 181499 (72\%) \\
\bottomrule  
\end{tabular}
\label{tab_datasets}
\end{table} 

We define two compounds as an AC pair if: 1) the compounds are structurally similar; 2) the compounds have different affinities towards the same target. There is no one strict definition of similarity criteria, and several possible methods can be used. The most popular one is the matched molecular pair \cite{Chen2022}, which is a pair of molecules that only differentiate by a single chemical modification at a specific site \cite{Hussain2010}. Another approach is to consider structural similarities and differences pairwise, using several criteria, such as substructure, scaffold, or SMILES similarity \cite{Tilborg2022}. 
Based on previous research \cite{Tilborg2022}, we defined a pair of compounds as an AC if the compounds were at least 90\% similar according to one of the three criteria mentioned above and the affinities of the compounds towards the same target were different at least 10 times.

To identify AC pairs, the dataset was preprocessed through several steps. First, for each target, drugs that interact with the target were paired based on their structural similarity. Pairs with a similarity of over 90\%, according to at least one criterion mentioned above, were then selected. Next, for each selected pair, the differences in drug affinities towards the target were calculated. If the difference in affinities exceeded a predefined affinity threshold, the pair was classified as an AC pair. Otherwise, the pair was classified as a non-AC pair.

The study focuses on two tasks: AC and DTI prediction. While improving DTI prediction is the primary objective, AC problem is leveraged as an auxiliary task to enhance the model's performance. In the preprocessing step for both AC and DTI tasks, Extended Connectivity Fingerprints (ECFPs) \cite{Rogers2010} were generated from the molecules encoded in SMILES \cite{Wigh2022} format using RDKit and used as input features. Proteins were assigned unique labels and encoded using word embeddings \cite{Vinokourov2002}, without incorporating any structural information to simplify the model and reduce computational complexity. This approach allows for easier integration of diverse datasets where structural information might be incomplete or unavailable, making the model more versatile. Furthermore, word embeddings capture relationships between proteins through their labels, providing a meaningful way to represent proteins in a continuous space, while avoiding the challenges associated with structural variability or missing data.

\subsection{Dataset splitting methods}
There are several approaches to splitting the data in ML tasks, each affecting model performance and generalization differently \cite{Stock2016}. The dataset can be split randomly, based on drugs, where each drug appears only in the training or test set, or based on targets, ensuring that each target (protein) is included exclusively in either the training or test set, but not both \cite{Iliadis2022}. Splitting can also involve a combination of these approaches \cite{Iliadis2022}. For the AC task, for example, one could have scenarios with either one novel compound or two novel compounds. While we did not explore all of these possibilities, they are certainly viable options for data splitting. We opted for the setting that we believe best reflects real-world challenges, but other configurations could be equally relevant depending on the certain task.

\begin{figure}[t]%
\centering
\includegraphics[trim={0 100 0 0},clip, scale=0.45]{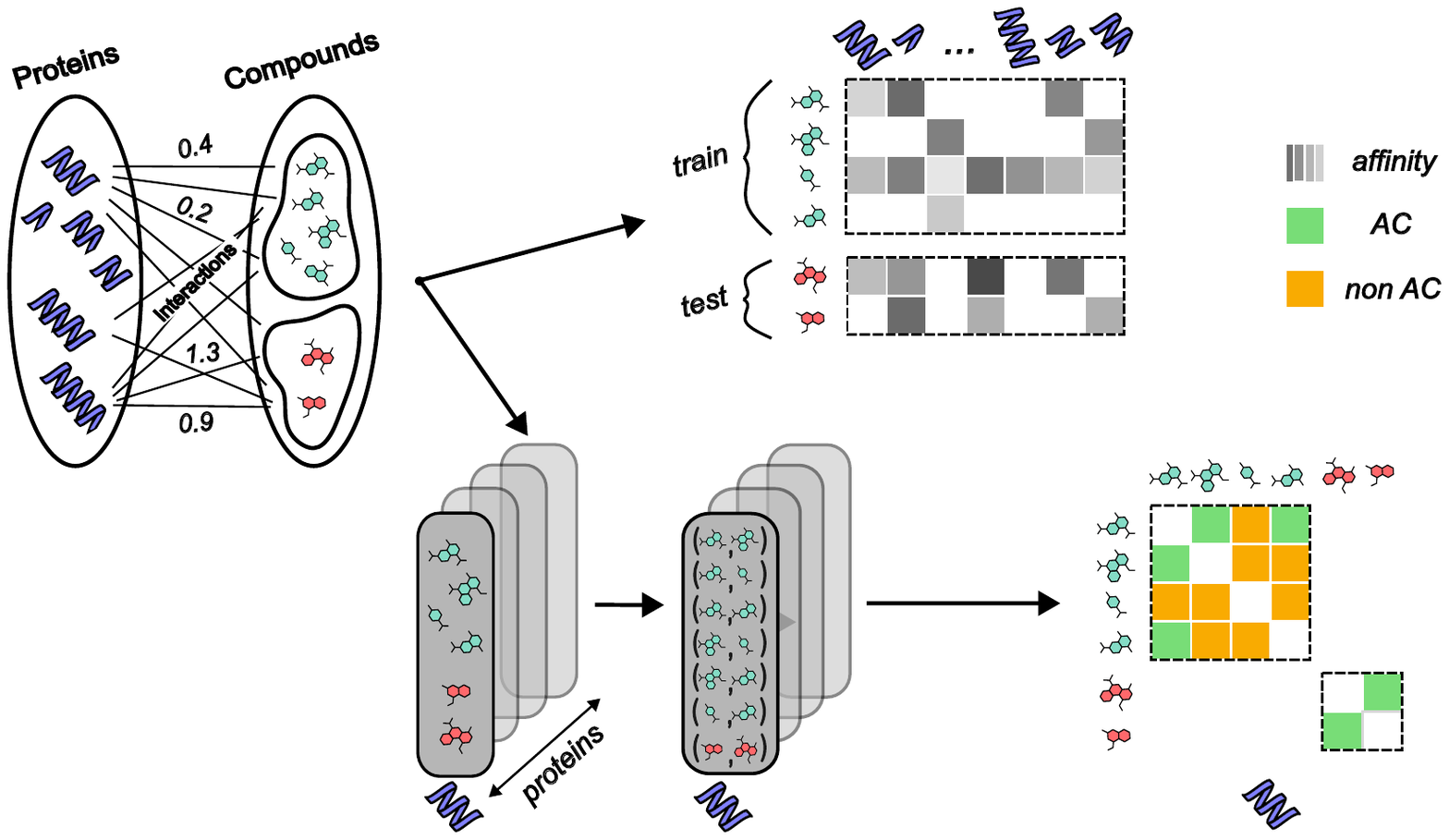}
\caption{Schematic representation of the  compound-based dataset splitting process for the two main tasks considered in this study: DTI and AC prediction. Initially, the dataset is split into training (green) and testing (red) sets. For the DTI task, drug-target pairs are directly generated within these sets. 
In contrast, the AC task involves three additional steps (shown at the bottom): first, identifying all drugs that interact with a specific protein in both the training and testing sets; second, pairing these drugs for each protein based on their structural similarity; and third, categorizing each pair as either an AC or non-AC based on predefined criteria. The labels in the figure indicate AC pairs in green, non-AC pairs in orange, and continuous affinity levels in grayscale.} 
\label{fig:dataset_split}%
\end{figure}

For both the AC and DTI tasks, two dataset splitting methods were utilized: random splitting and compound-based splitting. In the random split, compounds were divided randomly among the train and test sets. However, this approach may result in overoptimistic results, as the same compounds could appear in multiple sets, leading to potential data leakage. In contrast, the compound-based split, as demonstrated in Figure \ref{fig:dataset_split}, ensures no overlap of compounds between the train and test sets. This method maintains distinct sets of compounds for each stage, thereby avoiding data leakage. By ensuring that the same compounds do not appear in different splits, the compound-based approach provides a more reliable and realistic assessment of the model's performance. This approach prevents the model from memorizing specific compound properties, which might appear across different splits, and thus provides a more robust evaluation of its generalization.

As shown in Figure \ref{fig:dataset_split}, the dataset splitting approaches initially follow the same procedure for both the DTI and AC tasks, with drugs being divided into training (shown in green color) and testing sets (shown in red color) in a single step. This initial split of drugs ensures that the same compounds are consistently used across both tasks, thereby maintaining separate training and testing sets and avoiding any data leakage in subsequent steps.

For the DTI task, after the initial drug split, drug-target pairs are formed directly within the training and testing sets. However, for the AC task, additional steps are undertaken within the pre-defined training and testing sets to ensure accurate classification. Specifically, in step a), within each training and testing set, all drugs that interact with a specific protein are selected. Then, in step b), these selected drugs are paired based on their structural similarity, as outlined in Section \ref{datasets_ac}. Pairs are evaluated for similarity and affinity differences to identify potential AC pairs. Finally, in step c), the datasets for all targets are then combined within each training and testing set.
Although this approach may result in fewer available pairs for the AC task, it is crucial for avoiding data leakage and ensuring a robust and reliable evaluation of the models. By consistently maintaining the same training and testing sets for both tasks, this method provides a solid foundation for accurately assessing the models' performance without the risk of overfitting. It is worth noting that by combining data for all targets and using a unified model across these targets, we create a larger dataset. This is beneficial for deep learning models, which typically require vast amounts of data to perform effectively.

\subsection{Model details}
\begin{figure}[!t]
\centering
\includegraphics[trim={0 250 0 0},clip,scale=0.6]{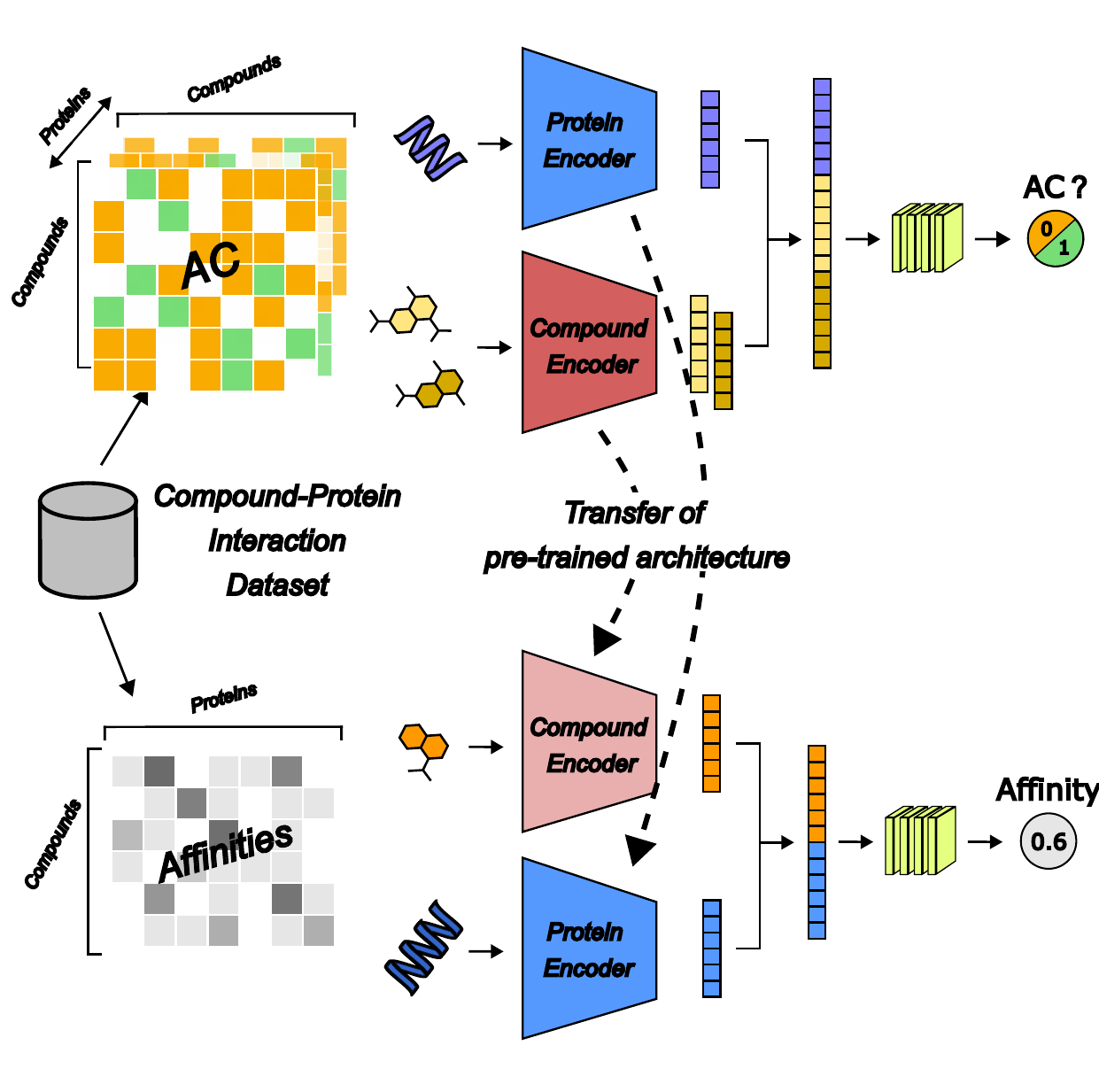}%
\caption{Schematic representation of the model architectures used for AC (top) and DTI (bottom) tasks. Both tasks employ separate encoders for drugs (Compound Encoder) and targets (Protein Encoder). In the AC task, the model processes both drugs through the same Compound Encoder, concatenates their embeddings with the target embedding from the Protein Encoder, and predicts whether the pair represents an AC (labeled as 1) or non-AC (labeled as 0) for a given target. For the DTI task, drug and target embeddings are concatenated to predict drug-target affinity as a continuous value, leveraging transferred features from the AC task to improve prediction accuracy for novel compound-protein interactions.}%
\label{fig:nn_arc}%
\end{figure}

The architecture of the models consists of two branches, as commonly done in DTI prediction models \cite{Huang2020, Lee2019, Ozturk2018, Rifaioglu2020, Iliadis2024}. As shown in Figure \ref{fig:nn_arc}, this design allows for the separate handling drug-related and target-related information, with specific encoders for each of them. The embeddings produced by these encoders are aggregated to create a combined representation, which is then utilized to make a final prediction. 

The ACs task is a binary classification problem, and the aim of the Drug-Drug Cliff (DDC) model is to predict whether a pair of the drugs given is AC (represented as 1) or non-AC (represented as 0) towards the given target. In the DDC model, both drugs are processed through the same compound encoder to generate their respective embeddings, which are then used to assess the AC status. In contrast, the DTI model's architecture, while also employing a two-branch structure, focuses on the interaction between a single drug and a target. Separate encoders are utilized for drugs (Compound Encoder) and targets (Protein Encoder). In this model, the aim is to predict the affinity or activity level of the drug towards the target, which is represented as a continuous value (e.g., 0.6). The model can leverage a pre-trained architecture, transferring learned features from the AC task.

\subsection{Experimental setup and hyper-parameter optimization}
In this study, we tested various combinations of model configurations, datasets, and transfer learning strategies to evaluate the performance of the models under different conditions. Supplementary Table \ref{experimental_setup_table} provides an overview of the experimental setup, summarizing all combinations of datasets, splits, tasks, and training configurations used.

A random search was conducted to find the best configuration of the model parameters in both AC and DTI baseline tasks. For each problem, 100 configurations were evaluated, each with a maximum allowance of 100 epochs and early stopping on the validation loss. Details of the search space are provided in the Supplementary information \ref{hp_ranges}. The Weights and Biases platform was used for logging \citep{wandb}, and the links are provided in Supplementary Table \ref{experimental_setup_table}. The best configuration was trained three times to ensure consistency of the results.

In the transfer learning models, a similar procedure was applied for both the warm starting and freezing weights settings. However, in the case where an additional layer was added to the weight-freezing model, the best parameters obtained from the weight-freezing setup were retained. Additional parameters related to the new layer were then varied using a grid search, resulting in a total of 42 configurations being evaluated. 

\subsection{Performance measures}
\label{averaging}
\subsubsection{AC task}
Given the imbalanced nature of this task, appropriate evaluation metrics are essential for assessing the model’s performance. In this work, the F1-score and Matthews Correlation Coefficient (MCC) were employed as the metrics to evaluate the model. The F1-score provides a balanced assessment of the model’s ability to predict ACs by calculating the harmonic mean of precision and recall. Precision represents how many of the predicted ACs are correct (Equation \ref{precision}), while recall (or true positive rate) measures the proportion of actual ACs that are correctly identified (Equation \ref{recall}). The F1-score combines both metrics, ensuring that the evaluation accounts for both false positives and false negatives, which is crucial in imbalanced datasets like the AC task (Equation \ref{f1}).

\begin{equation}
    Precision = \frac{TP}{TP + FP}  \,,
\label{precision}
\end{equation}

\begin{equation}
    Recall = \frac{TP}{TP + FN}  \,,
\label{recall}
\end{equation}

\begin{equation}
    F1 = 2 \cdot \frac{ Recall \cdot Precision}{Recall + Precision}  \,,
\label{f1}
\end{equation}
where TP, FP, and FN represent the total counts of true positive, false positive, and false negative predictions, respectively. 

The MCC was included to ensure an effective evaluation of model performance in imbalanced datasets \cite{Chicco2020}. MCC considers all prediction categories, including true negatives (TN), and provides a comprehensive measure of model performance across all classes:

\begin{equation}
    MCC = \frac{TP \cdot TN - FP \cdot FN}{\sqrt{(TP + FP)\cdot(TP + FN)\cdot(TN + FP)\cdot(TN + FN)}} \,.
\label{mcc}
\end{equation}

\subsubsection{DTI task}
In order to evaluate the performance of the DTI task, macro- and micro-averaging of well-known metrics was used:
\begin{itemize}
    \item \textit{Micro-averaging} This approach provides an evaluation of the model's performance across the entire dataset by treating all predictions as a single pool, regardless of the target. The micro-averaged Root Mean Square Error (RMSE) is calculated by summing the squared differences between the observed and predicted values for all samples across all targets, dividing by the total number of samples, and then taking the square root:

    \begin{equation}
    RMSE_{micro} = \sqrt{\frac{\sum_{t=1}^{T} \sum_{i=1}^{n_t} \left( y_{it} - \hat{y}_{it} \right)^2}{\sum_{t=1}^{T} n_t}} \,,
    \label{micro_rmse}
\end{equation}
     where $y_{it}$ and $\hat{y}_{it}$ are the true and predicted values for sample $i$ of target $t$, respectively, $n_t$ represents the number of samples for target $t$, and $T$ is the total number of targets.

    \item \textit{Macro-averaging} This approach ensures that every target contributes equally to the performance metric, giving a balanced picture of the model's performance across all targets. Here, RMSE is first calculated for each target individually, and then the average of these RMSE values across all targets is computed:

   \begin{equation}
    RMSE_{macro} = \frac{1}{T} \sum_{t=1}^{T} \sqrt{\frac{1}{n_t} \sum_{i=1}^{n_t} (y_{it} - \hat{y}_{it})^2} \,.
    \label{macro_rmse}
    \end{equation}
\end{itemize}

\section{Results and discussion}

\subsection{AC task}
It is important to note that ACs represent the minority class, making this an imbalanced problem, as outlined in Table \ref{tab_datasets}. Although the AC task is an important aspect of the overall study, our primary focus remains on improving the DTI task, so we present the AC prediction results concisely. For both datasets, the parameters of the best DDC models for random and compound-based splits are provided in Supplementary Table \ref{hps_ddc}. The models demonstrated satisfactory performance, as shown in Supplementary Table \ref{ddc_perf}. In the case of the BindingDB random split, the Uniform Manifold Approximation and Projection visualizations of the input features and hidden states after the hidden layers are provided in Supplementary Figure \ref{fig:umap}.

The two-branch model structure in the AC tasks allowed us to aggregate data across all targets, effectively expanding the dataset size and complexity. This approach is advantageous in deep learning, where substantial data typically enhances model performance. Leveraging data from multiple targets in a unified model enabled better learning of complex patterns and improved generalization, especially in cases where data is limited for individual targets. This structure ultimately enhanced both the robustness and accuracy of our models across diverse patterns.

\subsection{DTI prediction: baseline models}
We evaluated AC-specific DTI prediction performance across varying levels of AC severity using heatmaps that adjust thresholds for compound similarity and affinity differences toward the same target. This approach provides a systematic and detailed visualization of the AC phenomenon compared to traditional ad hoc thresholding methods. It is important to emphasize that the AC phenomenon may not be evident when both AC and non-AC compounds are analyzed together due to the significantly higher number of non-ACs. By progressively filtering out structurally dissimilar compounds and concentrating on those with higher structural similarity, the AC effect becomes more pronounced, enabling for a clearer distinction between AC and non-AC compounds. This method is particularly advantageous in limited data scenarios, although extreme AC cases may lack representativeness due to the small number of compound pairs.

Supplementary Figures \ref{fig:n_pairs_kiba} and \ref{fig:n_pairs_bdb} present heatmaps showing the summed number of pairs for all targets at different thresholds in the test set of both datasets. Subgroups with fewer than 100 pairs were masked in gray to indicate insufficient data.  

\begin{figure}[t]
\centering
\includegraphics[scale=0.3]{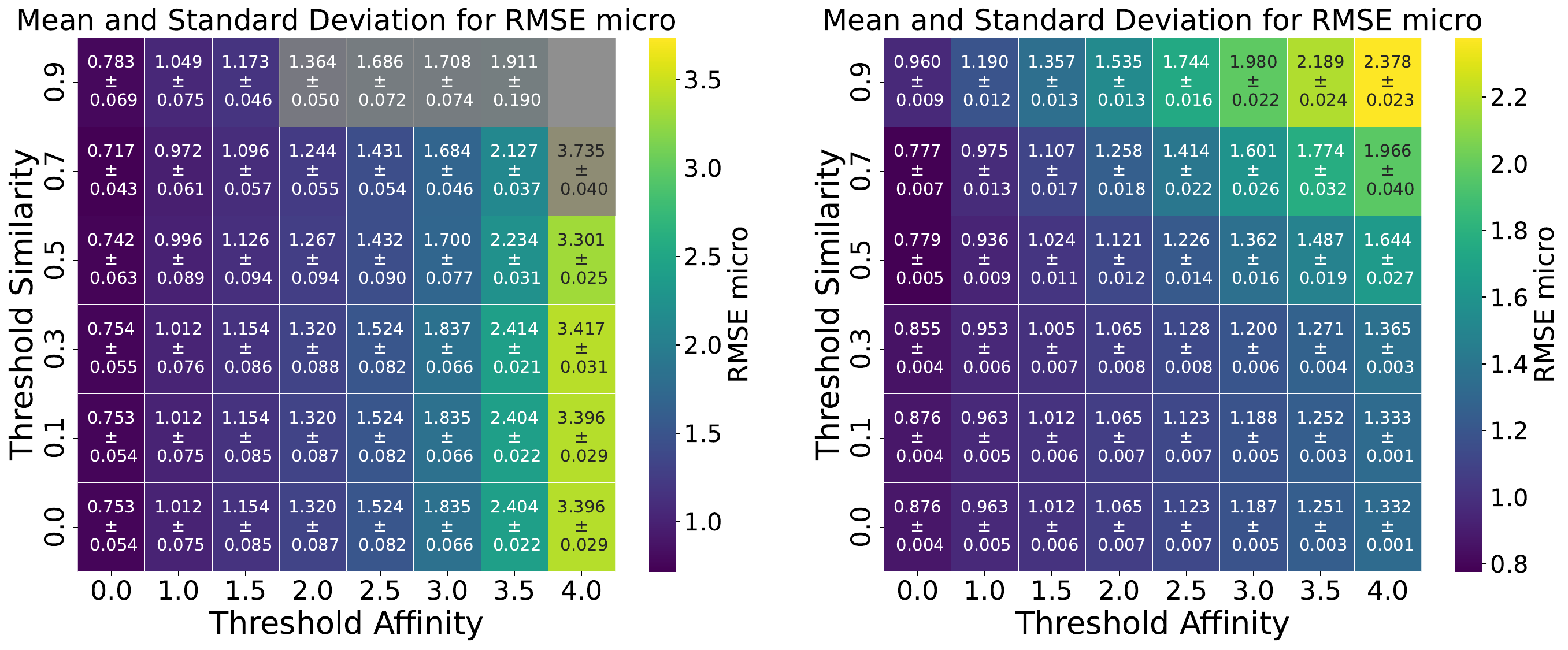}%
\caption{The heatmap of the $RMSE_{micro}$ for the best DTI model trained from scratch for the KIBA (left) and BindingDB (right) datasets in the case of a compound-based splits, showing groups of compounds split by similarity and affinity thresholds. The values represent the mean ± standard deviation based on 3 experiments. The groups with fewer than 100 pairs are masked in gray. The model performs well in groups containing both ACs and non-ACs, particularly in the bottom left regions of the heatmaps. However, as the non-ACs are gradually filtered out and the model is faced with more difficult AC predictions, performance significantly declines, particularly in the upper right areas where thresholds for affinity and similarity are high.}%
\label{fig:DTI_KIBA_BDB_cb_bl_hm}%
\end{figure}

The hyperparameters of the best DTI models for both datasets under random and compound-based splits are shown in Supplementary Tables \ref{hps_dti_comb_rs} and \ref{hps_dti_comb_cb}, respectively. The model's performance on the filtered groups of the KIBA and BindingDB datasets for compound-based split is illustrated in Figure \ref{fig:DTI_KIBA_BDB_cb_bl_hm}. In both datasets, the model demonstrates decent performance in groups containing both ACs and non-ACs (bottom left of the heatmaps). However, performance declines as non-ACs are progressively filtered out, with the model struggling to accurately predict interactions in groups with fewer non-ACs and more challenging AC cases (upper right areas of the heatmaps, where both affinity and similarity thresholds are high).

For the KIBA dataset, some upper-right groups lacked sufficient data, making these findings less robust. In contrast, the BindingDB dataset has a higher number of compound pairs, leading to a more distinct pattern for the compound-based split (Figure \ref{fig:DTI_KIBA_BDB_cb_bl_hm}, right). However, this improvement is not consistent across all targets, as shown in Supplementary Figure \ref{fig:DTI_KIBA_BDB_cb_bl_macro_hm}.

A similar pattern can be observed for the random split, as shown in Supplementary Figures \ref{fig:DTI_KIBA_rs_bl_hm} and \ref{fig:DTI_BDB_rs_bl_hm}. In both datasets, the random split demonstrates higher performance compared to the compound-based split, reflecting a setting where the training and test sets share a higher degree of similarity. While random splits are useful for evaluating model performance in scenarios with closely related data, the compound-based split represents a more challenging setting where the model must generalize to entirely novel compounds, making it more reflective in applications where unseen compounds are encountered.

\subsection{Transfer learning settings}
To enhance the prediction of affinities in novel compound-protein interactions, we employed transfer learning to leverage features learned from the AC task. This approach allows the DTI model to better handle the complexities of DTIs, improving its ability to generalize and deliver more accurate predictions, especially in challenging cases. For the KIBA dataset, we explored three different configurations for transfer learning:
\begin{itemize}
    \item Warm starting: This method involves fine-tuning, meaning that all the layers of the pre-trained model are updated during training.
    \item Frozen weights: In this technique, the weights of the pre-trained model are kept frozen and do not update during the training. Only the layers on top of the pre-trained model are updated. 
    \item Frozen weights with an additional layer: This approach is similar to the previous method of frozen weights, but with the addition of an extra layer in the drug branch before concatenation. This allows the frozen weights to be fine-tuned in the extra layer before the concatenation.
\end{itemize}

Among these configurations, warm starting achieved the best performance on the KIBA dataset, so this approach was also applied for transfer learning on the BindingDB dataset. The experiments involved transferring either only the drug encoder or both the drug and target encoders as shown in Supplementary Table  \ref{experimental_setup_table}. The idea is that The AC problem is largely compound-specific but is influenced by the context of the target protein. To account for this relationship, we explored the impact of incorporating a target-specific branch in the transfer learning framework.
The parameters of the best DTI models for both datasets, across random and compound-based splits, are presented in Supplementary Tables \ref{hps_dti_comb_rs} and \ref{hps_dti_comb_cb} for the case of transferring only the drug encoder. For the case of transferring both drug and target encoders, the parameters are shown in Supplementary Tables \ref{hps_dti_comb_rs_tenc} and \ref{hps_dti_comb_cb_tenc}, respectively. The heatmaps with models' performance are shown in Supplementary information  \ref{ad_file:TL_d_enc} and \ref{ad_file:TL_both_enc}, resembling a pattern similar to that of the baseline DTI models. The effect of transfer learning is evaluated using differential heatmaps, which are discussed in the next section.

\subsection{Evaluation of the transfer learning effect}
Differential heatmaps were created to evaluate the impact of transfer learning by comparing the best baseline model, trained from scratch, with the best transfer learning model. The evaluation process involves subtracting the RMSE values of the transfer learning model from those of the baseline model within each subgroup. In the context of RMSE, where lower values indicate better performance, a positive difference in the heatmap suggests that the transfer learning model outperforms the baseline by achieving lower RMSE values, thus providing more accurate predictions.

\begin{figure}[t]
\centering
\includegraphics[scale=0.3]{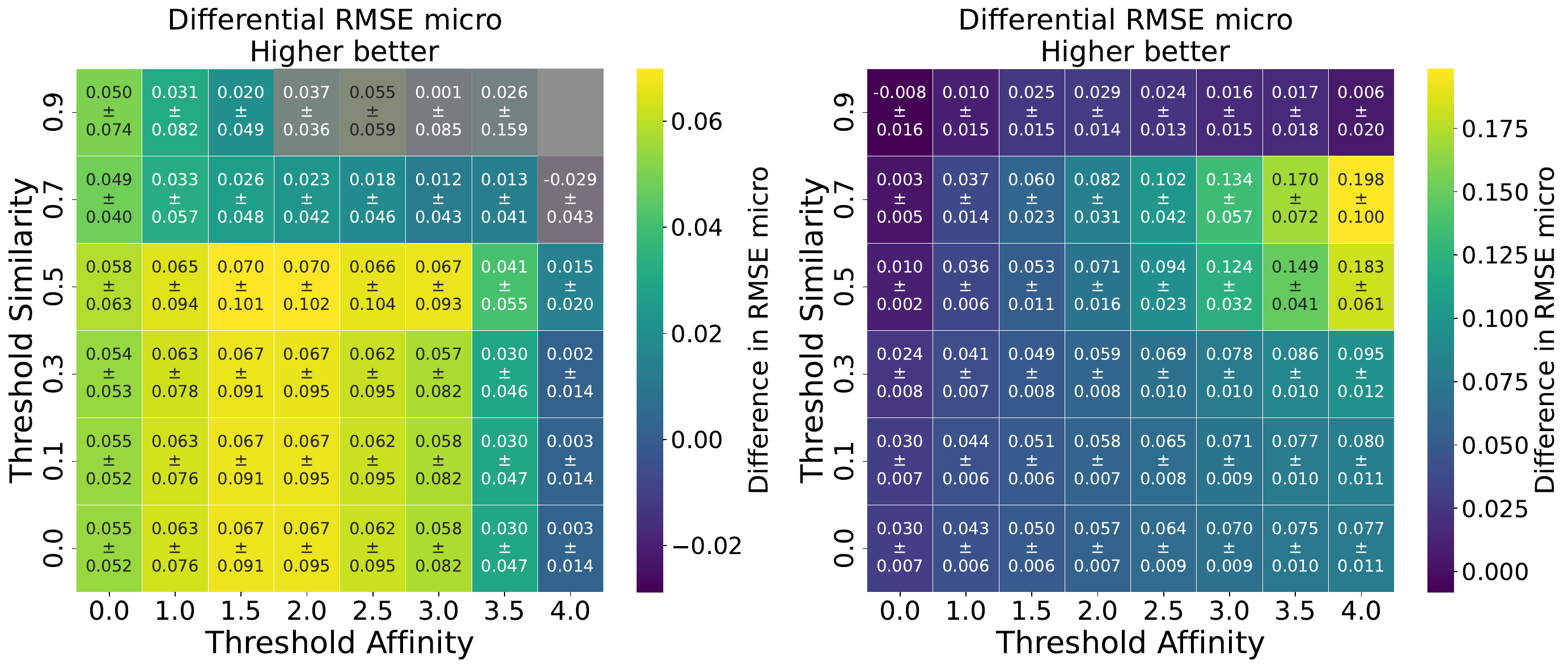}%
\caption{The differential heatmap of the $RMSE_{micro}$ for the best DTI model(transfer learning involving both drug and target encoders, \textbf{warm start}) for the KIBA (left) and BindingDB (right) datasets in the case of a compound-based splits, showing groups of compounds split by similarity and affinity thresholds. Transfer learning models outperform baseline models at lower affinity thresholds (left column), as indicated by positive RMSE differences. However, negative RMSE values at higher thresholds (top right) suggest poorer performance in cases with very similar compounds and large affinity variations. This indicates that transfer learning works well for simpler cases but struggles with more challenging ACs.}%
\label{fig:DTI_KIBA_BDB_cb_tl_ws_t_enc_micro_hm_diff}%
\end{figure}

For example, to evaluate the effect of transferring both the drug and target encoders on the KIBA dataset in case of the compound-based split and warm starting setting, the RMSE values from the transfer learning model (Supplementary Figure \ref{fig:DTI_KIBA_cb_tl_t_enc_ws_hm}) were subtracted from those of the baseline model (Figure \ref{fig:DTI_KIBA_BDB_cb_bl_hm}). This calculation produced the differential heatmap presented in Figure \ref{fig:DTI_KIBA_BDB_cb_tl_ws_t_enc_micro_hm_diff}, left. The same approach was used to generate all differential heatmaps. 

In the case of the KIBA dataset, for both random and compound-based splits, transfer learning with the warm starting setting consistently outperformed the baseline model, whereas models using frozen weights (both with and without an additional layer) did not show improved performance. This highlights the importance of fine-tuning all layers of the pre-trained model to effectively adapt the learned features for the DTI task.

Notably, transferring only the drug encoder resulted in worse performance across all settings and splits. This outcome is unsurprising, as the AC task, while primarily compound-specific, inherently relies on the contextual information provided by the target protein. Without the target encoder, the transfer learning model lacks the critical context needed to effectively generalize to the DTI task. These findings underscore the anticipated importance of incorporating both drug and target encoders in transfer learning approaches for DTI tasks, aligning with the understanding that drug-target interactions are inherently dependent on the interplay between compound structure and protein context.

In the compound-based split of the KIBA dataset (Figure \ref{fig:DTI_KIBA_BDB_cb_tl_ws_t_enc_micro_hm_diff}, left), transfer learning with warm starting setting and both the drug and target encoders transferred shows improvements at lower affinity thresholds, as indicated by positive $RMSE_{micro}$ differences in the leftmost columns, demonstrating effectiveness in simpler cases. However, performance declines at higher affinity and similarity thresholds (upper right), reflecting challenges in handling extreme AC cases. Furthermore, the standard deviations are often comparable to or larger than the mean, highlighting variability in the model's predictions and reduced stability. 

In the case of random split but with the same warm starting setting and transfer of both encoders, the results are slightly better for the KIBA dataset (Supplementary Figure \ref{fig:DTI_KIBA_rs_tl_t_enc_ws_hm_diff}) compared to the compound-based split. This indicates that the random split provides a less stringent evaluation, as the overlap of compounds between training and testing sets allows the model to encounter familiar compounds, potentially resulting in data leakage and inflated performance metrics.
In contrast, with the compound-based split, where the same compounds are consistently used across both AC and DTI tasks, we observed a performance drop. While this drop is expected, it provides a more realistic evaluation of the model’s generalization abilities, offering a more accurate measure of how the model will perform on new, unseen data.

However, one major challenge when generalizing to new drugs and proteins is that the dataset sizes significantly diminish. As we apply various rules and constraints to ensure proper generalization, the dataset becomes much smaller, which can impact the robustness of the models and their ability to make accurate predictions. In some cases, particularly for extreme AC instances, the limited number of samples makes it difficult to draw meaningful conclusions, as the small sample size is not representative. On the other hand, having a large dataset is advantageous, as it provides more robust data for analysis compared to a smaller dataset focused on a single protein \cite{Zhang2023}.

The compound-based split of the BindingDB dataset (Figure \ref{fig:DTI_KIBA_BDB_cb_tl_ws_t_enc_micro_hm_diff}, right) presents a more favorable trend, with transfer learning showing consistent improvements across most regions. Positive $RMSE_{micro}$ differences are observed even at moderate and higher affinity thresholds, suggesting that the larger dataset size of BindingDB provides better support for generalization in complex scenarios. Nonetheless, similar to KIBA, performance gains are reduced at the highest thresholds of similarity, indicating that extreme AC cases remain challenging. While the standard deviations in the BindingDB dataset are generally smaller and more controlled compared to those in the KIBA dataset, particularly at higher thresholds, they are still relatively large in some regions. This is especially noticeable in groups with high similarity and affinity thresholds, where variability remains significant. These observations highlight that, although the larger dataset size of BindingDB reduces variability and enhances stability in most regions, challenging subgroups with extreme thresholds still present inconsistencies in the model's predictive performance.

However, for the BindingDB dataset (Supplementary Figure \ref{fig:DTI_BDB_rs_tl_t_enc_ws_hm_diff}), the results are slightly worse under the random split compared to the compound-based split. This suggests that the larger and more diverse nature of the BindingDB dataset may expose limitations in the model's generalization when evaluated in a random split scenario. 

The $RMSE_{macro}$ values for both datasets are in the case of both encoders transfer learning and warm starting setting is illustrated in Supplementary Figure \ref{fig:DTI_KIBA_BDB_cb_tl_ws_t_enc_macro_hm_diff}. Positive values of differences in $RMSE_{macro}$ are dispersed across the heatmap but are generally lower in magnitude compared to the micro-averaging heatmap. This indicates that while the transfer learning model shows improvement in certain areas, these improvements are less consistent across different groups. This inconsistency may be attributed to the nature of the proteins or variations in the amount of affinity data available for different targets.

\subsection{Further implications}

Prior work in DTI prediction has often overlooked the impact of ACs or treated AC prediction as a separate challenge, focusing primarily on molecular similarity-based approaches. These methods, while effective in many cases, struggle to account for the complex, discontinuous structure-activity relationships introduced by AC phenomena. This work stands apart by directly integrating AC awareness into DTI prediction through transfer learning. By leveraging knowledge from AC prediction tasks, our approach not only addresses these challenges but also improves the generalizability and accuracy of predictive models, particularly in scenarios involving structurally similar yet functionally distinct compounds.

Moreover, the incorporation of both drug and target encoders in transfer learning highlights the interdependent nature of these two elements in accurately modeling DTIs. By illustrating the limitations of transferring only the drug encoder, our study provides theoretical support for the development of multi-faceted models that leverage contextual target information. This insight aligns with the fact that molecular similarity alone is insufficient for predicting biological interactions, emphasizing the importance of incorporating protein-specific context in predictive models.

Furthermore, the development of a universal model for AC prediction across all targets simplifies the modeling process. This approach reduces the need for target-specific models, which often require substantial computational and data resources. This study underscores the importance of further exploring the role of transfer learning in handling data scarcity and imbalance, which are common challenges in biomedical datasets. Future research could build on our findings by investigating additional transfer learning configurations, such as domain-specific pre-training or incorporating structural protein information into target encoders. Additionally, the variability observed in the performance of transfer learning models across datasets and splits suggests that further work is needed to tailor these approaches to the specific characteristics of datasets.

\section{Conclusion}

In this work, we investigated whether transfer learning techniques from AC prediction tasks can improve the overall performance of the DTI prediction model. First, we developed DDC models to predict ACs, with the BindingDB dataset yielding better performance than KIBA, likely due to its larger size. The DDC models provided a universal framework for identifying ACs without the need for separate models per target. Next, we created DTI baseline models and confirmed that predicting DTI in the context of ACs remains a significant challenge, as models struggled with AC pairs. To address this, we applied a transfer learning approach by incorporating knowledge from the AC prediction task into the DTI model. This involved three strategies: warm starting, freezing weights, and freezing weights with an additional layer. Differential heatmaps were used to assess the performance of the transfer learning models against the baseline models.

For both the KIBA and BindingDB datasets, transfer learning models with warm starting showed consistent improvements over the baseline, particularly in scenarios involving high similarity thresholds and varying affinity differences. In contrast, the models using frozen weights, either with or without an additional layer, generally performed the same or worse than the baseline models, failing to provide notable improvements. Moreover, when comparing the transfer of only the drug encoder versus both the drug and target encoders, transferring both encoders yielded superior results. This was especially evident in the case of high similarity thresholds, where the full transfer led to better generalization. However, some regions still exhibited less consistent gains, particularly in the KIBA dataset, where certain subgroups showed comparable or slightly worse performance. This variability underscores the dependency of transfer learning effectiveness on the specific characteristics of the data, such as the nature of the targets and the availability of affinity data.

Overall, our findings demonstrate that transfer learning from AC prediction tasks can significantly improve DTI model performance, particularly in predicting ACs. This approach contributes to more accurate and robust drug-target interaction predictions, with promising implications for real-world drug discovery and development.

\backmatter

\section{Supplementary Information} 
The online version contains supplementary material available at [link].

\section{Availability of data and materials} 
\label{sec5}
The code of this study was developed using PyTorch Lightning module and is accessible via the GitHub repository \url{https://github.com/reginaib/AC-DTI}. This repository also includes the scripts for data preprocessing and postprocessing, as well as the datasets and pre-trained models, ensuring the reproducibility of the results.

\section{Abbreviations}

\begin{description}
    \item[\textbf{AC:}] Activity cliff
    \item[\textbf{DDC:}] Drug-drug cliff
    \item[\textbf{DTI:}] Drug-target interaction
    \item[\textbf{ECFP:}] Extended Connectivity Fingerprint
    \item[\textbf{ML:}] Machine learning
    \item[\textbf{RMSE:}] Root mean square error
    \item[\textbf{UMAP:}] Uniform Manifold Approximation and Projection 
\end{description}

\section{Competing interests}
No competing interests.

\section{Funding}
D.I. and W.W. were supported by the Flemish Government under the Flanders AI Research Program. 

\section{Authors' contributions}
R.I. implemented the neural networks and training scripts, conducted the experiments, analyzed the results, and drafted the manuscript. D.I. designed and supervised the research, prepared Figures 1 and 2, and contributed to manuscript revisions. W.W. provided guidance on the experiments and reviewed the manuscript. All authors read and approved the final version of the manuscript. 

\section{Acknowledgements}
Not applicable.


\bibliography{sn-bibliography}

\begin{thebibliography}{10}
\expandafter\ifx\csname url\endcsname\relax
  \def\url#1{\burl{#1}}\fi
\expandafter\ifx\csname urlprefix\endcsname\relax\def\urlprefix{URL }\fi
\providecommand{\bibinfo}[2]{#2}
\providecommand{\eprint}[2][]{\url{#2}}
\providecommand{\doi}[1]{\url{https://doi.org/#1}}
\bibcommenthead

\bibitem{Sun2021}
\bibinfo{author}{Sun, D.}, \bibinfo{author}{Gao, W.}, \bibinfo{author}{Hu, H.} \& \bibinfo{author}{Zhou, S.}
\newblock \bibinfo{title}{Why 90\% of clinical drug development fails and how to improve it?}
\newblock \emph{\bibinfo{journal}{Acta Pharmaceutica Sinica B}} \textbf{\bibinfo{volume}{12}}, \bibinfo{pages}{3049--3062} (\bibinfo{year}{2022}).

\bibitem{Vijayan2021}
\bibinfo{author}{Vijayan, R. S.~K.}, \bibinfo{author}{Kihlberg, J.}, \bibinfo{author}{Cross, J.~B.} \& \bibinfo{author}{Poongavanam, V.}
\newblock \bibinfo{title}{Enhancing preclinical drug discovery with artificial intelligence}.
\newblock \emph{\bibinfo{journal}{Drug Discovery Today}} \textbf{\bibinfo{volume}{27}}, \bibinfo{pages}{967--984} (\bibinfo{year}{2022}).

\bibitem{Xu2021}
\bibinfo{author}{Xu, L.}, \bibinfo{author}{Ru, X.} \& \bibinfo{author}{Song, R.}
\newblock \bibinfo{title}{Application of machine learning for drug–target interaction prediction}.
\newblock \emph{\bibinfo{journal}{Front. Genet.}} \textbf{\bibinfo{volume}{12}}, \bibinfo{pages}{6801176} (\bibinfo{year}{2021}).

\bibitem{Yamanishi2008}
\bibinfo{author}{Yamanishi, Y.}, \bibinfo{author}{Araki, M.}, \bibinfo{author}{Gutteridge, A.}, \bibinfo{author}{Honda, W.} \& \bibinfo{author}{Kanehisa, M.}
\newblock \bibinfo{title}{Prediction of drug–target interaction networks from the integration of chemical and genomic spaces}.
\newblock \emph{\bibinfo{journal}{Bioinformatics}} \textbf{\bibinfo{volume}{24}}, \bibinfo{pages}{i232--i240} (\bibinfo{year}{2008}).

\bibitem{Johnson1990}
\bibinfo{editor}{Johnson, M.~A.} \& \bibinfo{editor}{Maggiora, G.~M.} (eds) \emph{\bibinfo{title}{Concepts and {A}pplications of {M}olecular {S}imilarity}}  (\bibinfo{publisher}{Wiley}, \bibinfo{address}{New {Y}ork}, \bibinfo{year}{1990}).

\bibitem{Ding2013}
\bibinfo{author}{Ding, H.}, \bibinfo{author}{Takigawa, I.}, \bibinfo{author}{Mamitsuka, H.} \& \bibinfo{author}{Zhu, S.}
\newblock \bibinfo{title}{Similarity-based machine learning methods for predicting drug-target interactions: a brief review}.
\newblock \emph{\bibinfo{journal}{Brief Bioinform.}} \textbf{\bibinfo{volume}{15}}, \bibinfo{pages}{734--747} (\bibinfo{year}{2014}).

\bibitem{Dimova2016}
\bibinfo{author}{Dimova, D.} \& \bibinfo{author}{Bajorath, J.}
\newblock \bibinfo{title}{Advances in {A}ctivity {C}liff {R}esearch}.
\newblock \emph{\bibinfo{journal}{Mol. Inf.}} \textbf{\bibinfo{volume}{35}}, \bibinfo{pages}{181--191} (\bibinfo{year}{2016}).

\bibitem{Tilborg2022}
\bibinfo{author}{van Tilborg, D.}, \bibinfo{author}{Alenicheva, A.} \& \bibinfo{author}{Grisoni, F.}
\newblock \bibinfo{title}{Exposing the {L}imitations of {M}olecular {M}achine {L}earning with {A}ctivity {C}liffs}.
\newblock \emph{\bibinfo{journal}{J. Chem. Inf. Model.}} \textbf{\bibinfo{volume}{62}}, \bibinfo{pages}{5938--5951} (\bibinfo{year}{2022}).

\bibitem{Cruz2014}
\bibinfo{author}{Cruz-Monteagudo, M.} \emph{et~al.}
\newblock \bibinfo{title}{Activity cliffs in drug discovery: Dr {J}ekyll or {M}r {H}yde?}
\newblock \emph{\bibinfo{journal}{Drug Discovery Today}} \textbf{\bibinfo{volume}{19}}, \bibinfo{pages}{1069--1080} (\bibinfo{year}{2014}).

\bibitem{Stumpfe2019}
\bibinfo{author}{Stumpfe, D.}, \bibinfo{author}{Hu, H.} \& \bibinfo{author}{Bajorath, J.}
\newblock \bibinfo{title}{Evolving {C}oncept of {A}ctivity {C}liffs}.
\newblock \emph{\bibinfo{journal}{ACS Omega}} \textbf{\bibinfo{volume}{4}}, \bibinfo{pages}{14360--14368} (\bibinfo{year}{2019}).

\bibitem{Chen2022}
\bibinfo{author}{Chen, H.}, \bibinfo{author}{Vogt, M.} \& \bibinfo{author}{Bajorath, J.}
\newblock \bibinfo{title}{Deep{AC} – conditional transformer-based chemical language model for the prediction of activity cliffs formed by bioactive compounds}.
\newblock \emph{\bibinfo{journal}{Digital Discovery}} \textbf{\bibinfo{volume}{1}}, \bibinfo{pages}{898--909} (\bibinfo{year}{2022}).

\bibitem{Iqbal2021}
\bibinfo{author}{Iqbal, J.}, \bibinfo{author}{Vogt, M.} \& \bibinfo{author}{Bajorath, J.}
\newblock \bibinfo{title}{Prediction of activity cliffs on the basis of images using convolutional neural networks}.
\newblock \emph{\bibinfo{journal}{J Comput Aided Mol Des}} \textbf{\bibinfo{volume}{35}}, \bibinfo{pages}{1157--1164} (\bibinfo{year}{2021}).

\bibitem{Tamura2023}
\bibinfo{author}{Tamura, S.}, \bibinfo{author}{Miyao, T.} \& \bibinfo{author}{Bajorath, J.}
\newblock \bibinfo{title}{Large-scale prediction of activity cliffs using machine and deep learning methods of increasing complexity}.
\newblock \emph{\bibinfo{journal}{J. Cheminform}} \textbf{\bibinfo{volume}{15}} (\bibinfo{year}{2023}).

\bibitem{Zhang2023}
\bibinfo{author}{Zhang, Z.}, \bibinfo{author}{Zhao, B.}, \bibinfo{author}{Xie, A.}, \bibinfo{author}{Bian, Y.} \& \bibinfo{author}{Zhou, S.}
\newblock \bibinfo{title}{Activity cliff prediction: Dataset and benchmark}.
\newblock \emph{\bibinfo{journal}{arXiv preprint arXiv:2302.07541}}  (\bibinfo{year}{2023}).

\bibitem{Dalkıran2023}
\bibinfo{author}{Dalkıran, A.} \emph{et~al.}
\newblock \bibinfo{title}{Transfer learning for drug–target interaction prediction}.
\newblock \emph{\bibinfo{journal}{Bioinformatics}} \textbf{\bibinfo{volume}{39}}, \bibinfo{pages}{i103–i110} (\bibinfo{year}{2023}).

\bibitem{Weiss2016}
\bibinfo{author}{Weiss, K.}, \bibinfo{author}{Khoshgoftaar, T.~M.} \& \bibinfo{author}{Wang, D.}
\newblock \bibinfo{title}{A survey of transfer learning}.
\newblock \emph{\bibinfo{journal}{J Big Data}} \textbf{\bibinfo{volume}{3}}, \bibinfo{pages}{i103–i110} (\bibinfo{year}{2016}).

\bibitem{Tang2014}
\bibinfo{author}{Tang, J.} \emph{et~al.}
\newblock \bibinfo{title}{Making {S}ense of {L}arge-{S}cale {K}inase {I}nhibitor {B}ioactivity {D}ata {S}ets: A {C}omparative and {I}ntegrative {A}nalysis}.
\newblock \emph{\bibinfo{journal}{J. Chem. Inf. Model.}} \textbf{\bibinfo{volume}{54}}, \bibinfo{pages}{735--743} (\bibinfo{year}{2014}).

\bibitem{Liu2007}
\bibinfo{author}{Liu, T.}, \bibinfo{author}{Lin, Y.}, \bibinfo{author}{Wen, X.}, \bibinfo{author}{Jorissen, R.~N.} \& \bibinfo{author}{Gilson, M.~K.}
\newblock \bibinfo{title}{Binding{DB}: a web-accessible database of experimentally determined protein–ligand binding affinities}.
\newblock \emph{\bibinfo{journal}{Nucleic Acids Research}} \textbf{\bibinfo{volume}{35}}, \bibinfo{pages}{D198–D201} (\bibinfo{year}{2007}).

\bibitem{Hussain2010}
\bibinfo{author}{Hussain, J.} \& \bibinfo{author}{Rea, C.}
\newblock \bibinfo{title}{Computationally {E}fficient {A}lgorithm to {I}dentify {M}atched {M}olecular {P}airs ({MMP}s) in {L}arge {D}ata {S}ets}.
\newblock \emph{\bibinfo{journal}{J. Chem. Inf. Model.}} \textbf{\bibinfo{volume}{50}}, \bibinfo{pages}{339--348} (\bibinfo{year}{2010}).

\bibitem{Rogers2010}
\bibinfo{author}{Rogers, D.} \& \bibinfo{author}{Hahn, M.}
\newblock \bibinfo{title}{Extended-{C}onnectivity {F}ingerprints}.
\newblock \emph{\bibinfo{journal}{J. Chem. Inf. Model.}} \textbf{\bibinfo{volume}{50}}, \bibinfo{pages}{742–754} (\bibinfo{year}{2010}).

\bibitem{Wigh2022}
\bibinfo{author}{Wigh, D.}, \bibinfo{author}{Goodman, J.~M.} \& \bibinfo{author}{Lapkin, A.~A.}
\newblock \bibinfo{title}{A review of molecular representation in the age of machine learning}.
\newblock \emph{\bibinfo{journal}{WIREs Comput Mol Sci.}} \textbf{\bibinfo{volume}{12}}, \bibinfo{pages}{e1603} (\bibinfo{year}{2022}).

\bibitem{Vinokourov2002}
\bibinfo{author}{Vinokourov, A.}, \bibinfo{author}{Cristianini, N.} \& \bibinfo{author}{Shawe-Taylor, J.}
\newblock \bibinfo{editor}{Becker, S.}, \bibinfo{editor}{Thrun, S.} \& \bibinfo{editor}{Obermayer, K.} (eds) \emph{\bibinfo{title}{Inferring a {S}emantic {R}epresentation of {T}ext via {C}ross-{L}anguage {C}orrelation {A}nalysis}}.
\newblock (eds \bibinfo{editor}{Becker, S.}, \bibinfo{editor}{Thrun, S.} \& \bibinfo{editor}{Obermayer, K.}) \emph{\bibinfo{booktitle}{Advances in Neural Information Processing Systems}}, Vol.~\bibinfo{volume}{15} (\bibinfo{publisher}{MIT Press}, \bibinfo{year}{2002}).

\bibitem{Stock2016}
\bibinfo{author}{Van~Peer, G.} \emph{et~al.}
\newblock \bibinfo{title}{mi{STAR}: mi{RNA} target prediction through modeling quantitative and qualitative mirna binding site information in a stacked model structure}.
\newblock \emph{\bibinfo{journal}{Nucleic Acids Research}} \textbf{\bibinfo{volume}{45}}, \bibinfo{pages}{e51--e51} (\bibinfo{year}{2016}).

\bibitem{Iliadis2022}
\bibinfo{author}{Iliadis, D.}, \bibinfo{author}{De~Baets, B.} \& \bibinfo{author}{Waegeman, W.}
\newblock \bibinfo{title}{Multi‑target prediction for dummies using two‑branch neural networks}.
\newblock \emph{\bibinfo{journal}{Bioinformatics}} \textbf{\bibinfo{volume}{111}}, \bibinfo{pages}{651--684} (\bibinfo{year}{2022}).

\bibitem{Huang2020}
\bibinfo{author}{Huang, K.} \emph{et~al.}
\newblock \bibinfo{title}{Deep{P}urpose: a deep learning library for drug–target interaction prediction}.
\newblock \emph{\bibinfo{journal}{Bioinformatics}} \textbf{\bibinfo{volume}{36}}, \bibinfo{pages}{5545–5547} (\bibinfo{year}{2014}).

\bibitem{Lee2019}
\bibinfo{author}{Lee, I.}, \bibinfo{author}{Keum, J.} \& \bibinfo{author}{Nam, H.}
\newblock \bibinfo{title}{Deep{C}onv-{DTI}: Prediction of drug-target interactions via deep learning with convolution on protein sequences}.
\newblock \emph{\bibinfo{journal}{PLoS Comput Biol}} \textbf{\bibinfo{volume}{15}}, \bibinfo{pages}{e1007129} (\bibinfo{year}{2019}).

\bibitem{Ozturk2018}
\bibinfo{author}{Öztürk, H.}, \bibinfo{author}{Özgür, A.} \& \bibinfo{author}{Ozkirimli, E.}
\newblock \bibinfo{title}{Deep{DTA}: deep drug–target binding affinity prediction}.
\newblock \emph{\bibinfo{journal}{Bioinformatics}} \textbf{\bibinfo{volume}{34}}, \bibinfo{pages}{i821--i829} (\bibinfo{year}{2018}).

\bibitem{Rifaioglu2020}
\bibinfo{author}{Rifaioglu, A.~S.} \emph{et~al.}
\newblock \bibinfo{title}{{MD}ee{P}red: novel multi-channel protein featurization for deep learning-based binding affinity prediction in drug discovery}.
\newblock \emph{\bibinfo{journal}{Bioinformatics}} \textbf{\bibinfo{volume}{37}}, \bibinfo{pages}{693--704} (\bibinfo{year}{2020}).

\bibitem{Iliadis2024}
\bibinfo{author}{Iliadis, D.}, \bibinfo{author}{De~Baets, B.}, \bibinfo{author}{Pahikkala, T.} \& \bibinfo{author}{Waegeman, W.}
\newblock \bibinfo{title}{A comparison of embedding aggregation strategies in drug–target interaction prediction}.
\newblock \emph{\bibinfo{journal}{BMC Bioinformatics}} \textbf{\bibinfo{volume}{25}} (\bibinfo{year}{2024}).

\bibitem{wandb}
\bibinfo{author}{Biewald, L.}
\newblock \bibinfo{title}{Experiment tracking with weights and biases} (\bibinfo{year}{2020}).
\newblock \urlprefix\url{https://www.wandb.com/}.
\newblock \bibinfo{note}{Software available from wandb.com}.

\bibitem{Chicco2020}
\bibinfo{author}{Chicco, D.} \& \bibinfo{author}{Jurman, G.}
\newblock \bibinfo{title}{The advantages of the {M}atthews correlation coefficient ({MCC}) over {F}1 score and accuracy in binary classification evaluation}.
\newblock \emph{\bibinfo{journal}{BMC Genomics}} \textbf{\bibinfo{volume}{21}}, \bibinfo{pages}{1--13} (\bibinfo{year}{2020}).

\bibitem{McInnes2018}
\bibinfo{author}{McInnes, L.}, \bibinfo{author}{Healy, J.}, \bibinfo{author}{Saul, N.} \& \bibinfo{author}{Großberger, L.}
\newblock \bibinfo{title}{{UMAP}: {U}niform {M}anifold {A}pproximation and {P}rojection}.
\newblock \emph{\bibinfo{journal}{Journal of Open Source Software}} \textbf{\bibinfo{volume}{3}}, \bibinfo{pages}{861} (\bibinfo{year}{2018}).

\end{thebibliography}


\newpage
\section{Supplementary information -- Enhancing Drug-Target Interaction Prediction through Transfer Learning from Activity Cliff Prediction Tasks}
\label{sup_info}

\setcounter{figure}{0}
\setcounter{table}{0}

\begin{center}
Regina Ibragimova\textsuperscript{[1]}, Dimitrios Iliadis\textsuperscript{[1]}, Willem Waegeman\textsuperscript{[1]}  

\textsuperscript{[1]} Department of Data Analysis and Mathematical Modelling, Ghent  
University, Coupure Links, Ghent, 9000, Belgium.

Email: regina.r.ibragimova@gmail.com, dimitrios.iliadis@ugent.be, willem.waegeman@ugent.be
\end{center}

\subsection{Overview of Experimental Setup}
This section provides an overview of the experimental setup utilized in this study, as summarized in Table \ref{experimental_setup_table}. The table outlines the tasks performed, datasets used (including their splitting strategies), transfer learning settings, transferred components (drug and/or target encoders), and links to the corresponding experiments conducted on the Weights and Biases platform.

For the AC task, models were trained separately for two datasets (KIBA and BindingDB) using random and compound-based splits. For the DTI task, both baseline models (trained from scratch) and transfer learning models were evaluated on the same datasets. Transfer learning involved three strategies—warm starting, freezing weights, and freezing weights with an additional layer—applied to either the drug encoder alone or both drug and target encoders. 

This setup ensured a comprehensive evaluation of the models under various conditions, providing insights into the performance of baseline and transfer learning approaches across different datasets and experimental settings.

\begin{sidewaystable}
\centering
\caption{Overview of experimental setup with Weights and Biases projects' links.}
\label{experimental_setup_table}
\begin{tabular}{@{}lllll@{}}
\toprule
\textbf{Task} & \textbf{Dataset (Split)} & \textbf{TL\footnotemark[1] Setting} & \textbf{Transferred Encoder(s)} & \textbf{Links} \\ 
\midrule

\multirow{4}{*}{AC} 
    & KIBA (Random) & - & - & \href{https://wandb.ai/reginaib/DDC_KIBA_rs_sweep}{HPO\footnotemark[2]}, \href{https://wandb.ai/reginaib/DDC_KIBA_rs_best_train}{Best\footnotemark[3]} \\
    & KIBA (Compound-based) & - & - & \href{https://wandb.ai/reginaib/DDC_KIBA_cb_sweep}{HPO},
    \href{https://wandb.ai/reginaib/DDC_KIBA_cb_best_train}{Best} \\
    & BindingDB (Random) & - & - & \href{https://wandb.ai/reginaib/DDC_BDB_rs_sweep}{HPO}, 
    \href{https://wandb.ai/reginaib/DDC_BDB_rs_best_train}{Best}  \\
    & BindingDB (Compound-based) & - & - &  \href{https://wandb.ai/reginaib/DDC_BDB_cb_sweep}{HPO}, 
    \href{https://wandb.ai/reginaib/DDC_BDB_cb_best_train}{Best} \\
\midrule
\multirow{4}{*}{DTI (Baseline)} 
    & KIBA (Random) & - & - & \href{https://wandb.ai/reginaib/DTI_KIBA_rs_bl_sweep}{HPO}, \href{https://wandb.ai/reginaib/DTI_KIBA_rs_bl_best_train}{Best} \\
    & KIBA (Compound-based) & - & - & \href{https://wandb.ai/reginaib/DTI_KIBA_cb_bl_sweep}{HPO}, \href{https://wandb.ai/reginaib/DTI_KIBA_cb_bl_best_train}{Best} \\
    & BindingDB (Random) & - & - & \href{https://wandb.ai/reginaib/DTI_BDB_rs_bl_sweep}{HPO}, 
    \href{https://wandb.ai/reginaib/DTI_BDB_rs_bl_best_train}{Best} \\
    & BindingDB (Compound-based) & - & - & \href{https://wandb.ai/reginaib/DTI_BDB_cb_bl_sweep}{HPO}, 
    \href{https://wandb.ai/reginaib/DTI_BDB_cb_bl_best_train}{Best} \\
\midrule
\multirow{16}{*}{DTI (TL)} 
    & \multirow{3}{*}{KIBA (Random)} 
    & Warm Starting & \multirow{3}{*}{Drug} & \href{https://wandb.ai/reginaib/DTI_KIBA_rs_tl_ws_sweep}{HPO}, \href{https://wandb.ai/reginaib/DTI_KIBA_rs_tl_ws_best_train}{Best} \\
    & & Freezing Weights & & \href{https://wandb.ai/reginaib/DTI_KIBA_rs_tl_f_sweep}{HPO}, 
    \href{https://wandb.ai/reginaib/DTI_KIBA_rs_tl_f_best_train}{Best} \\
    & & Freezing Weights + Layer & & \href{https://wandb.ai/reginaib/DTI_KIBA_rs_tl_f_el_sweep}{HPO},  
    \href{https://wandb.ai/reginaib/DTI_KIBA_rs_tl_f_el_best_train}{Best} \\
    \cmidrule(lr){3-5}
    & & Warm Starting & \multirow{3}{*}{Drug + Target} & \href{https://wandb.ai/reginaib/DTI_KIBA_rs_tl_t_enc_ws_sweep}{HPO},
    \href{https://wandb.ai/reginaib/DTI_KIBA_rs_tl_t_enc_ws_best_train}{Best} \\
    & & Freezing Weights & & \href{https://wandb.ai/reginaib/DTI_KIBA_rs_tl_t_enc_f_sweep}{HPO},
    \href{https://wandb.ai/reginaib/DTI_KIBA_rs_tl_t_enc_f_best_train}{Best}  \\
    & & Freezing Weights + Layer & & \href{https://wandb.ai/reginaib/DTI_KIBA_rs_tl_t_enc_f_el_sweep}{HPO}, 
    \href{https://wandb.ai/reginaib/DTI_KIBA_rs_tl_t_enc_f_el_best_train}{Best} \\
\cmidrule(lr){2-5}
    & \multirow{3}{*}{KIBA (Compound-based)} 
    & Warm Starting & \multirow{3}{*}{Drug} &  \href{https://wandb.ai/reginaib/DTI_KIBA_cb_tl_ws_sweep}{HPO}, \href{https://wandb.ai/reginaib/DTI_KIBA_cb_tl_ws_best_train}{Best} \\
    & & Freezing Weights & &  \href{https://wandb.ai/reginaib/DTI_KIBA_cb_tl_f_sweep}{HPO}, 
    \href{https://wandb.ai/reginaib/DTI_KIBA_cb_tl_f_best_train}{Best} \\
    & & Freezing Weights + Layer & & \href{https://wandb.ai/reginaib/DTI_KIBA_cb_tl_f_el_sweep}{HPO},  
    \href{https://wandb.ai/reginaib/DTI_KIBA_cb_tl_f_el_best_train}{Best} \\
    \cmidrule(lr){3-5}
    & & Warm Starting & \multirow{3}{*}{Drug + Target} & \href{https://wandb.ai/reginaib/DTI_KIBA_cb_tl_t_enc_ws_sweep}{HPO},
    \href{https://wandb.ai/reginaib/DTI_KIBA_cb_tl_t_enc_ws_best_train}{Best} \\
    & & Freezing Weights & & \href{https://wandb.ai/reginaib/DTI_KIBA_cb_tl_t_enc_f_sweep}{HPO},
    \href{https://wandb.ai/reginaib/DTI_KIBA_cb_tl_t_enc_f_best_train}{Best}  \\
    & & Freezing Weights + Layer & &  \href{https://wandb.ai/reginaib/DTI_KIBA_cb_tl_t_enc_f_el_sweep}{HPO}, 
    \href{https://wandb.ai/reginaib/DTI_KIBA_cb_tl_t_enc_f_el_best_train}{Best} \\
\cmidrule(lr){2-5}
    & \multirow{2}{*}{BindingDB (Random)} 
    & Warm Starting & Drug  & \href{https://wandb.ai/reginaib/DTI_BDB_rs_tl_ws_sweep}{HPO},
    \href{https://wandb.ai/reginaib/DTI_BDB_rs_tl_ws_best_train}{Best} \\
    & & Warm Starting & Drug + Target &  \href{https://wandb.ai/reginaib/DTI_BDB_rs_tl_t_enc_ws_sweep}{HPO},
    \href{https://wandb.ai/reginaib/DTI_BDB_rs_tl_t_enc_ws_best_train}{Best} \\
\cmidrule(lr){2-5}
    & \multirow{2}{*}{BindingDB (Compound-based)} 
    & Warm Starting & Drug & \href{https://wandb.ai/reginaib/DTI_BDB_cb_tl_ws_sweep}{HPO}, 
\href{https://wandb.ai/reginaib/DTI_BDB_cb_tl_ws_best_train}{Best}  \\
    & & Warm Starting & Drug + Target  &  \href{https://wandb.ai/reginaib/DTI_BDB_cb_tl_t_enc_ws_sweep}{HPO},
\href{https://wandb.ai/reginaib/DTI_BDB_cb_tl_t_enc_ws_best_train}{Best} \\
\bottomrule
\end{tabular}
\footnotetext[1]{TL: Transfer Learning.}
\footnotetext[2]{HPO: Hyper-parameter optimization.}
\footnotetext[3]{Best: Training of the best model.}
\end{sidewaystable}

\newpage
\subsection{Hyper-parameter ranges}
\label{hp_ranges}

In the following section, we provide the hyper-parameter space we explore in both AC and DTI tasks:
\begin{itemize}
    \item number of hidden layers in drug encoder: [1, 2, 3, 4]
    \item drug hidden layer size: [32, 64, 128, 256, 512, 768, 1024]
    \item target embedding size: [32, 64, 128, 256, 512, 768, 1024]
    \item head hidden layer size: [32, 64, 128, 256, 512, 768, 1024]
    \item learning rate: [0.00001, 0.00003, 0.0001, 0.0003, 0.001]
    \item dropout rate: [0.01, 0.1, 0.2, 0.3, 0.4, 0.5]
\end{itemize}

In the transfer learning setting involving freezing weights and adding an extra layer, the following hyper-parameters were explored:
\begin{itemize}
    \item drug additional hidden layer size: [32, 64, 128, 256, 512, 768, 1024]
    \item dropout rate: [0.01, 0.1, 0.2, 0.3, 0.4, 0.5]
\end{itemize}

\subsection{Hyper-parameters and performance of best DDC models}

\begin{table}[h]
\centering
\begin{tabular}{l c c c c}
\toprule
 & \multicolumn{2}{c}{Datasets (random split)} & \multicolumn{2}{c}{Datasets (compound-based split)}\\
 \cmidrule(r){2-3} \cmidrule(l){4-5}
  Hyper-parameters  & KIBA & BindingDB & KIBA & BindingDB \\
\midrule
Number of hidden layers in drug encoder & 1 & 4 & 2 & 1 \\
Drug hidden layer size & 1024 & 1024 & 128 & 1024 \\
Target embedding size & 768 & 512 & 256 & 512 \\
Head hidden layer size & 768 & 1024 & 768 & 1024  \\
Learning rate & 0.0003 & 0.0003 & 0.001 & 0.001 \\
Dropout rate & 0.01 & 0.2 & 0.2 & 0.4 \\
\bottomrule  
\end{tabular}
\caption{Combinations of hyper-parameters of best DDC models on KIBA and BindindDB datasets}
\label{hps_ddc}
\end{table} 


\begin{table}[h]
\centering
\begin{tabular}{l c c c c}
\toprule
 & \multicolumn{2}{c}{Random split} & \multicolumn{2}{c}{Compound-based split} \\
 \cmidrule(r){2-3} \cmidrule(l){4-5} 
 Metric  & KIBA & BindingDB & KIBA & BindingDB \\
\midrule
Precision & 0.627 & 0.664 & 0.338 & 0.385 \\
Recall & 0.644 & 0.856 & 0.386 & 0.622 \\
F1-score & 0.636 & 0.748 & 0.360 & 0.476 \\
MCC & 0.583 & 0.641 & 0.293 & 0.235 \\
\bottomrule  
\end{tabular}
\caption{Performance of best DDC models on KIBA and BindindDB datasets}
\label{ddc_perf}
\end{table}

\subsection{Hidden state visualization}

\begin{figure}[H]%
\centering
\includegraphics[scale=0.6]{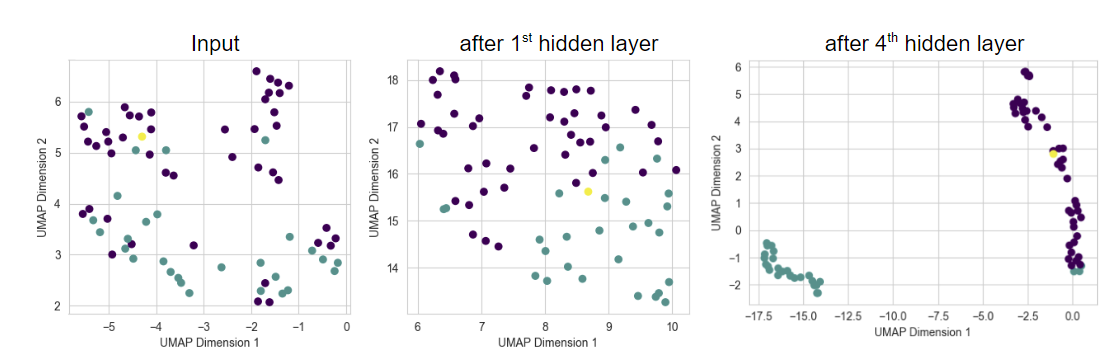}%
\caption{UMAP of input features and hidden states of the DDC model's drug encoder (on the BindingDB dataset, random split).}%
\label{fig:umap}%
\end{figure}

\begin{figure}[H]%
    \centering
    \begin{subfigure}[b]{0.4\textwidth}  
        \centering
        \includegraphics[scale=0.3]{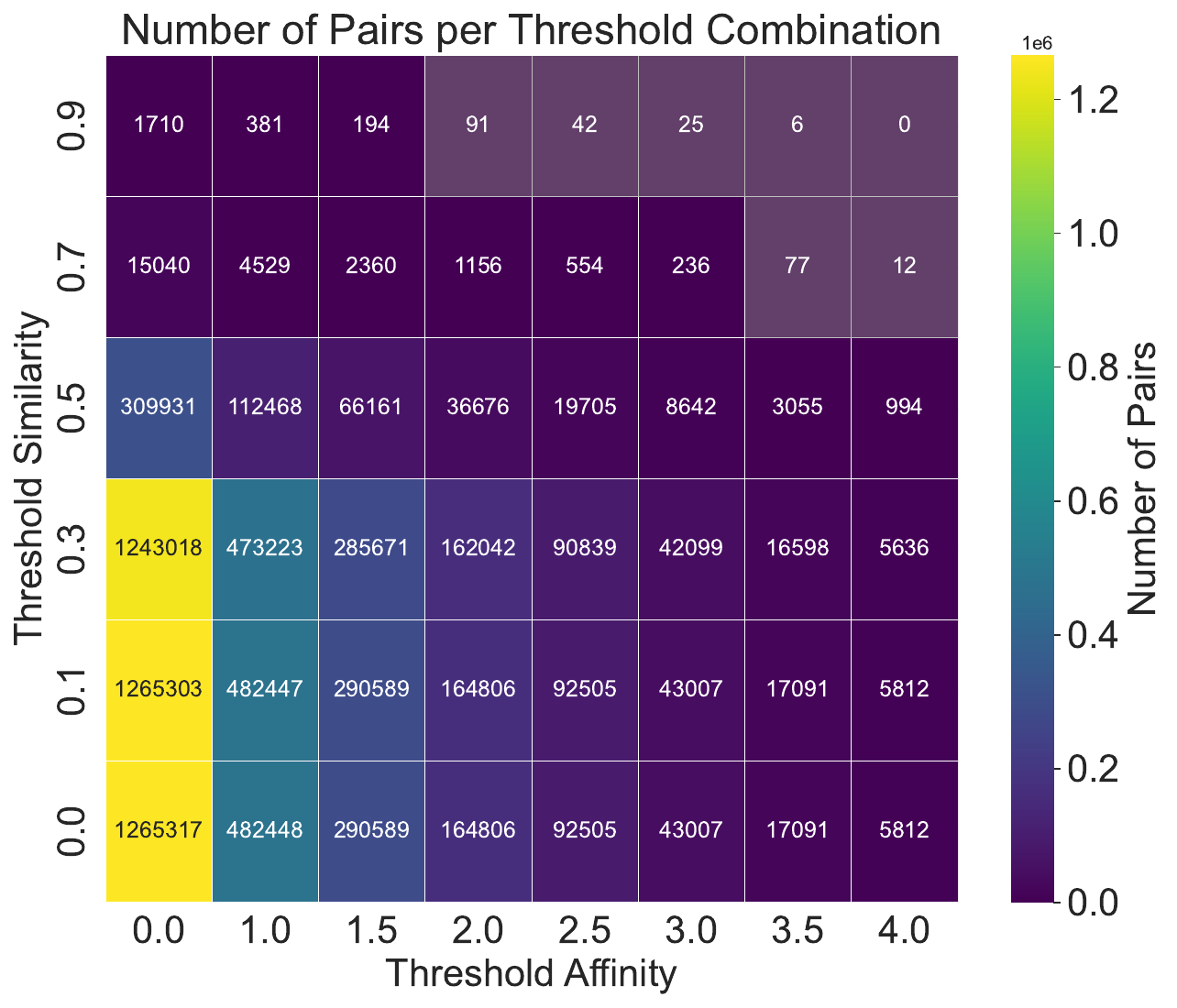}%
        \caption{} 
    \end{subfigure}
    \hfill  
    \begin{subfigure}[b]{0.4\textwidth}  
        \centering
        \includegraphics[scale=0.3]{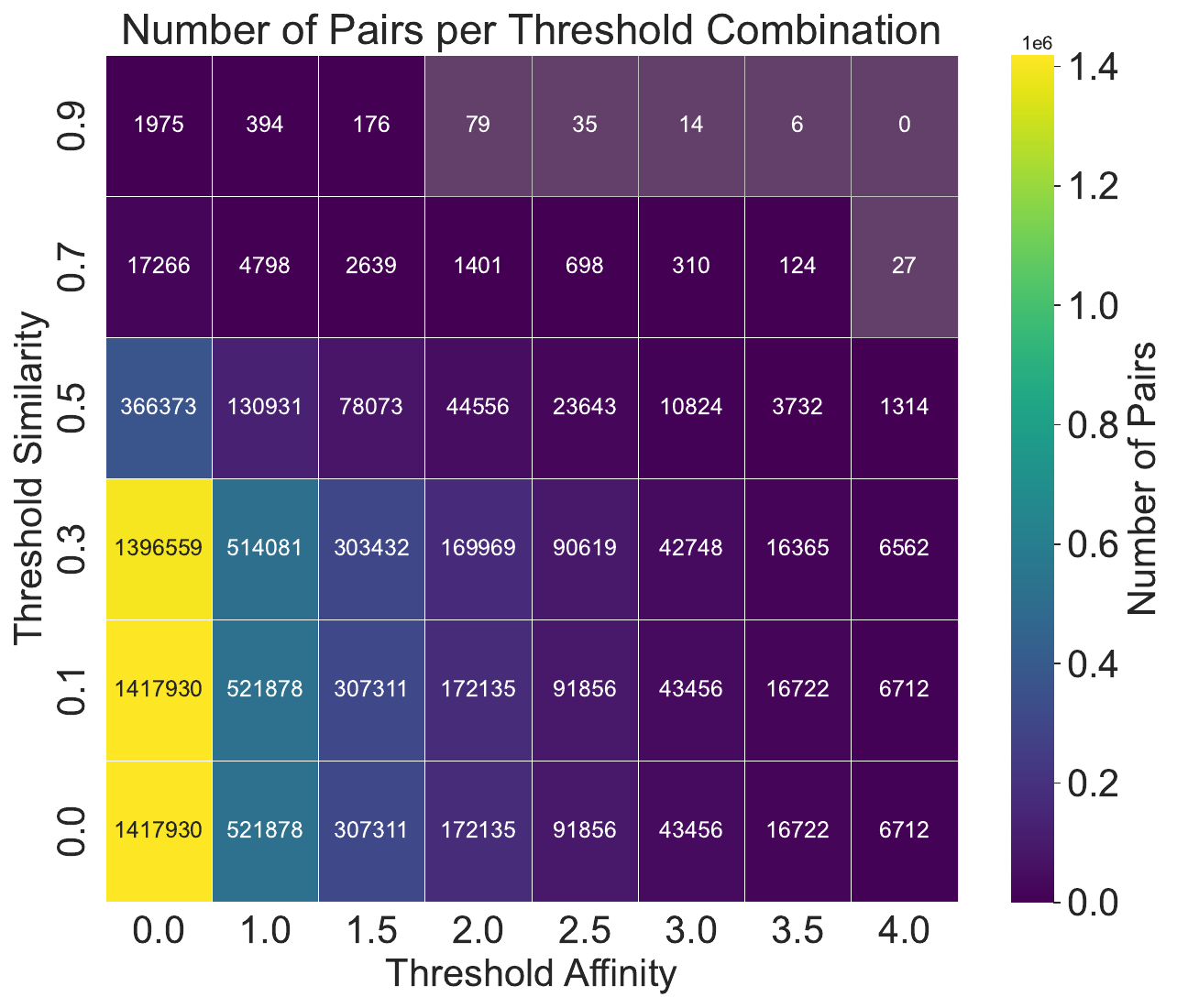}%
        \caption{}       
    \end{subfigure}
    \caption{Number of pairs per groups in test sets in KIBA random (a) and compound-based (b) splits. It was found that not all the subgroups had a sufficient number of pairs, so those below 100 pairs were masked by grey color. The numbers are shown for pairs of compounds targeting the same protein, meaning that targets with an affinity for only one drug were excluded.}
    \label{fig:n_pairs_kiba}
\end{figure}

\begin{figure}[H]%
    \centering
    \begin{subfigure}[b]{0.4\textwidth}  
        \centering
        \includegraphics[scale=0.3]{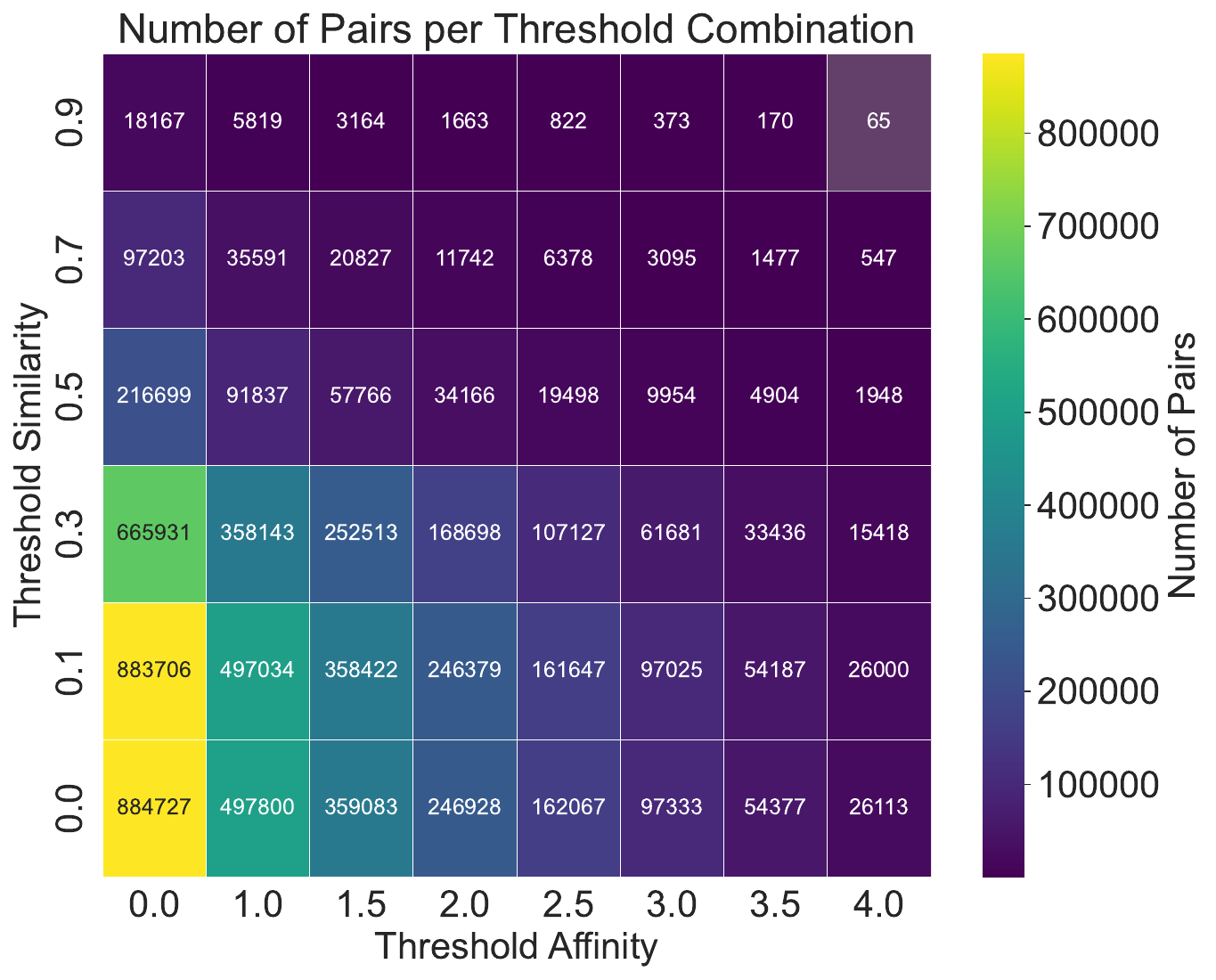}%
        \caption{} 
    \end{subfigure}
    \hfill  
    \begin{subfigure}[b]{0.4\textwidth}  
        \centering
        \includegraphics[scale=0.3]{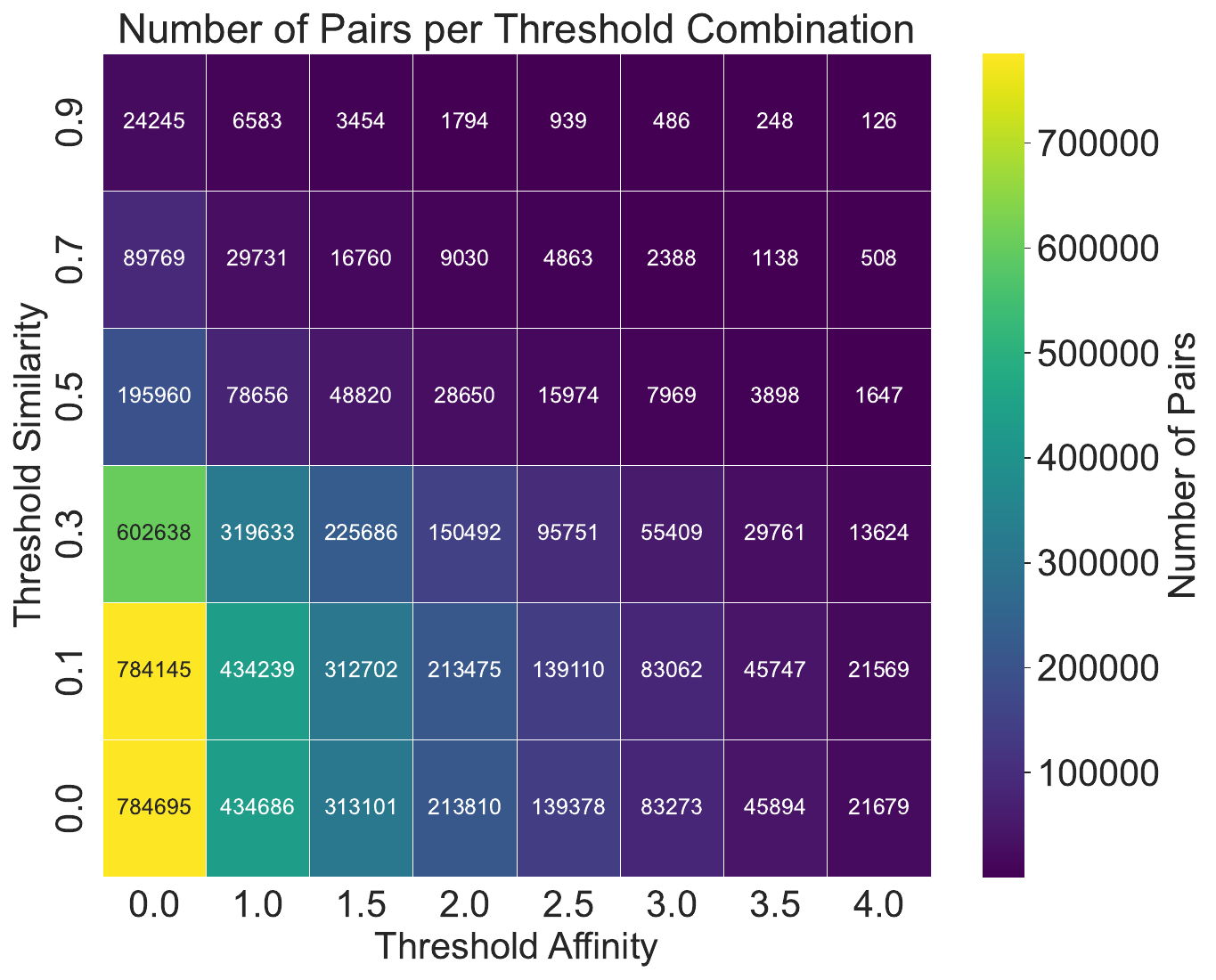}%
        \caption{} 
    \end{subfigure}
    \caption{Number of pairs per groups in test sets in BindingDB random (a) and compound-based (b) splits. Groups with less than  100 pairs were masked by grey color.} 
    \label{fig:n_pairs_bdb}
\end{figure}

To better understand the model's ability to discriminate between ACs and non-ACs, the input and the hidden states after the first and fourth layers of the drug encoder in the DDC model trained on the BindingDB dataset (random split) were visualized using Uniform Manifold Approximation and Projection (UMAP) \citep{McInnes2018}. For a specific protein, all pairs involving the same compound (referred to as the 'Reference Compound' and colored yellow in the plot) were extracted and plotted. The other compounds are colored based on their pairing with the reference compound: those in AC cliff pairs are colored teal, while non-AC pairs are colored purple. As seen in Figure \ref{fig:umap}, the distinction is not immediately present in the feature space. However, as the representations pass through layers, the model increasingly succeeds in discriminating the points, resulting in clearer clusters. Additionally, it can be observed that the reference compound is closer to non-ACs, while ACs are further away, indicating that the model is able to effectively distinguish ACs. 

\subsection{Hyper-parameters and performance of best DTI models}
\begin{sidewaystable}
\caption{The parameters of the best models obtained using transfer learning (random split datasets, transferring only drug encoder)}
\label{hps_dti_comb_rs}
\begin{tabular*}{\textheight}{@{\extracolsep\fill}lcccccc}
\toprule%
& \multicolumn{4}{@{}c@{}}{KIBA}& \multicolumn{2}{@{}c@{}}{BindingDB} 
\\\cmidrule{2-5}\cmidrule{6-7}%
Projectile & Baseline & TL warm starting & TL FW & TL FW + Layer & Baseline & TL warm starting \\
\midrule
\# Hidden Layers in Drug Encoder & 1 & 1 & 1 & 1  & 1 & 4 \\
Drug Hidden Layer Size  & 256 & 1024  & 1024 & 1024 & 1024 & 1024 \\
Target Embedding Size  & 256 & 128 & 256 & 256 & 32 & 256 \\
Head Hidden Layer Size  & 1024 & 768 & 1024 & 1024 & 768 & 1024 \\
Learning Rate  & 0.0001 & 0.001 & 0.0003 & 0.0003 & 0.0003 & 0.0001 \\
Dropout Rate   & 0.01 & 0.2 & 0.1 & 0.1 & 0.1 & 0.3 \\
Additional layer size  & - & -  & - & 1024  & - & -  \\
Additional layer dropout  & - & -  & - & 0.5  & - & - \\
\botrule
\end{tabular*}
\footnotetext{TL: Transfer Learning, FW: Freezing Weights.}
\vspace{1cm}
\caption{The parameters of the best models obtained using transfer learning (compound-based split datasets, transferring only drug encoder)}
\label{hps_dti_comb_cb}
\begin{tabular*}{\textheight}{@{\extracolsep\fill}lcccccc}
\toprule%
& \multicolumn{4}{@{}c@{}}{KIBA}& \multicolumn{2}{@{}c@{}}{BindingDB} 
\\\cmidrule{2-5}\cmidrule{6-7}%
Projectile & Baseline & TL\footnotemark[1] warm starting  & TL FW\footnotemark[1]  & TL FW + Layer & Baseline & TL warm starting \\
\midrule
\# Hidden Layers in Drug Encoder & 1 & 2 & 2 & 2  & 1 & 1 \\
Drug Hidden Layer Size  & 768 & 128  & 128 & 128 & 768 & 1024 \\
Target Embedding Size  & 768 & 128 & 64 & 64 & 512 & 512 \\
Head Hidden Layer Size  & 512 & 1024 & 1024 & 1024 & 1024 & 256 \\
Learning Rate  & 0.001 & 0.0001 & 0.0001 & 0.0001 & 0.00003 & 0.00003 \\
Dropout Rate   & 0.1 & 0.2 & 0.1 & 0.1 & 0.5 & 0.2 \\
Additional layer size  & - & -  & - & 1024  & - & -  \\
Additional layer dropout  & - & -  & - & 0.4  & - & - \\
\botrule
\end{tabular*}
\footnotetext{TL: Transfer Learning, FW: Freezing Weights.}
\end{sidewaystable}

\begin{sidewaystable}
\caption{The parameters of the best models obtained using transfer learning (random split datasets, transferring both drug and target encoders)}
\label{hps_dti_comb_rs_tenc}
\begin{tabular*}{\textheight}{@{\extracolsep\fill}lcccccc}
\toprule%
& \multicolumn{4}{@{}c@{}}{KIBA}& \multicolumn{2}{@{}c@{}}{BindingDB} 
\\\cmidrule{2-5}\cmidrule{6-7}%
Projectile & Baseline & TL warm starting & TL FW & TL FW + Layer & Baseline & TL warm starting \\
\midrule
\# Hidden Layers in Drug Encoder & 1 & 1 & 1 & 1  & 1 & 1 \\
Drug Hidden Layer Size  & 256 & 1024  & 1024 & 1024 & 1024 & 1024 \\
Target Embedding Size  & 256 & 768 & 768 & 768 & 32 & 512 \\
Head Hidden Layer Size  & 1024 & 1024 & 1024 & 1024 & 768 & 1024 \\
Learning Rate  & 0.0001 & 0.0003 & 0.0001 & 0.0001 & 0.0003 & 0.00003 \\
Dropout Rate   & 0.01 & 0.2 & 0.1 & 0.1 & 0.1 & 0.3 \\
Additional layer size  & - & -  & - & 1024  & - & -  \\
Additional layer dropout  & - & -  & - & 0.01  & - & - \\
\botrule
\end{tabular*}
\footnotetext{TL: Transfer Learning, FW: Freezing Weights.}
\vspace{1cm}
\caption{The parameters of the best models obtained using transfer learning (compound-based split datasets, transferring both drug and target encoders)}
\label{hps_dti_comb_cb_tenc}
\begin{tabular*}{\textheight}{@{\extracolsep\fill}lcccccc}
\toprule%
& \multicolumn{4}{@{}c@{}}{KIBA}& \multicolumn{2}{@{}c@{}}{BindingDB} 
\\\cmidrule{2-5}\cmidrule{6-7}%
Projectile & Baseline & TL warm starting  & TL FW & TL FW + Layer & Baseline & TL warm starting \\
\midrule
\# Hidden Layers in Drug Encoder & 1 & 2 & 2 & 2 & 1 & 1 \\
Drug Hidden Layer Size  & 768 & 128  & 128 & 128 & 768 & 1024 \\
Target Embedding Size  & 768 & 256 & 256 & 256 & 512 & 512 \\
Head Hidden Layer Size  & 512 & 1024 & 768 & 768 & 1024 & 768 \\
Learning Rate  & 0.001 & 0.0003 & 0.0003 & 0.0003 & 0.00003 & 0.0001 \\
Dropout Rate   & 0.1 & 0.1 & 0.2 & 0.2 & 0.5 & 0.2 \\
Additional layer size  & - & -  & - & 1024  & - & -  \\
Additional layer dropout  & - & -  & - & 0.2  & - & - \\
\botrule
\end{tabular*}
\footnotetext{TL: Transfer Learning, FW: Freezing Weights.}
\end{sidewaystable}

\newpage
\subsection{Number of pairs heatmaps}
\label{heatmaps}

\subsection{Performance of the baseline models with compound-based split}

\begin{figure}[H]
\centering
\includegraphics[scale=0.3]{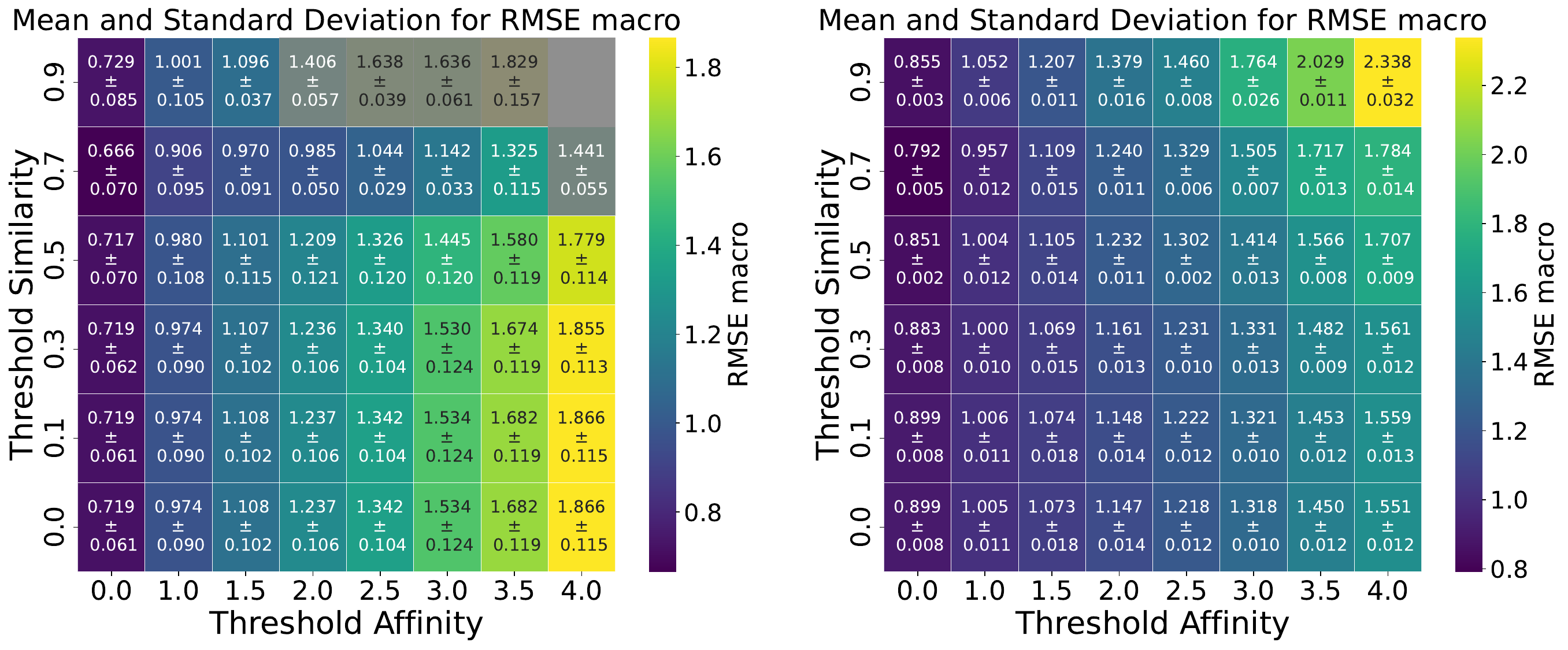}%
\caption{The heatmap of the $RMSE_{macro}$ for the best DTI model trained from scratch for the KIBA (left) and BindingDB (right) datasets in the case of a compound-based splits, showing groups of compounds split by similarity and affinity thresholds. The values represent the mean ± standard deviation based on 3 experiments. The groups with fewer than 100 pairs are masked in gray.}%
\label{fig:DTI_KIBA_BDB_cb_bl_macro_hm}%
\end{figure}

\subsection{Performance of the baseline models with random split}

\begin{figure}[H]
\centering
\includegraphics[scale=0.3]{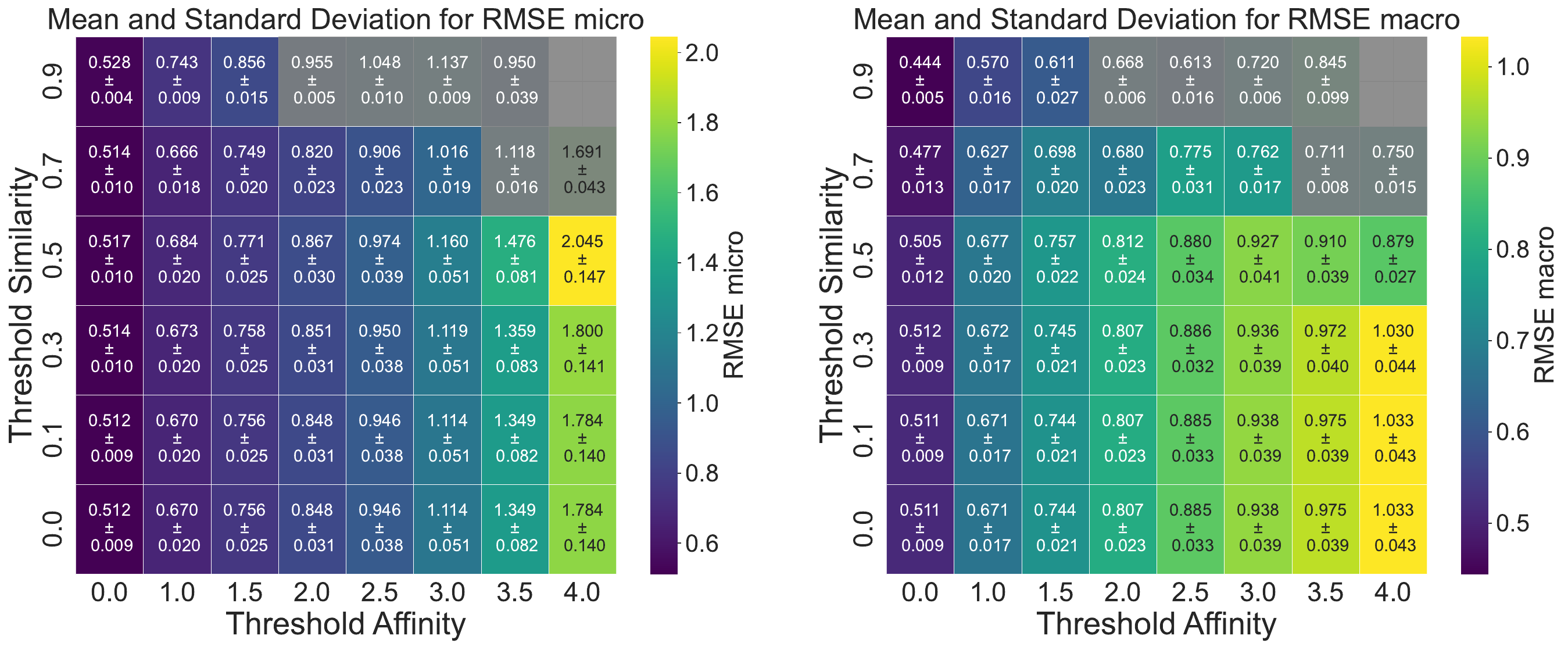}%
\caption{The heatmap of the $RMSE_{micro}$ (left) and $RMSE_{macro}$ (right) for the best DTI model trained from scratch, showing groups of compounds split by similarity and affinity thresholds for the KIBA dataset in the case of a random split. The values represent the mean ± standard deviation based on 3 experiments. The groups with fewer than 100 pairs are masked in gray.}%
\label{fig:DTI_KIBA_rs_bl_hm}%
\end{figure}

\begin{figure}[H]
\centering
\includegraphics[scale=0.3]{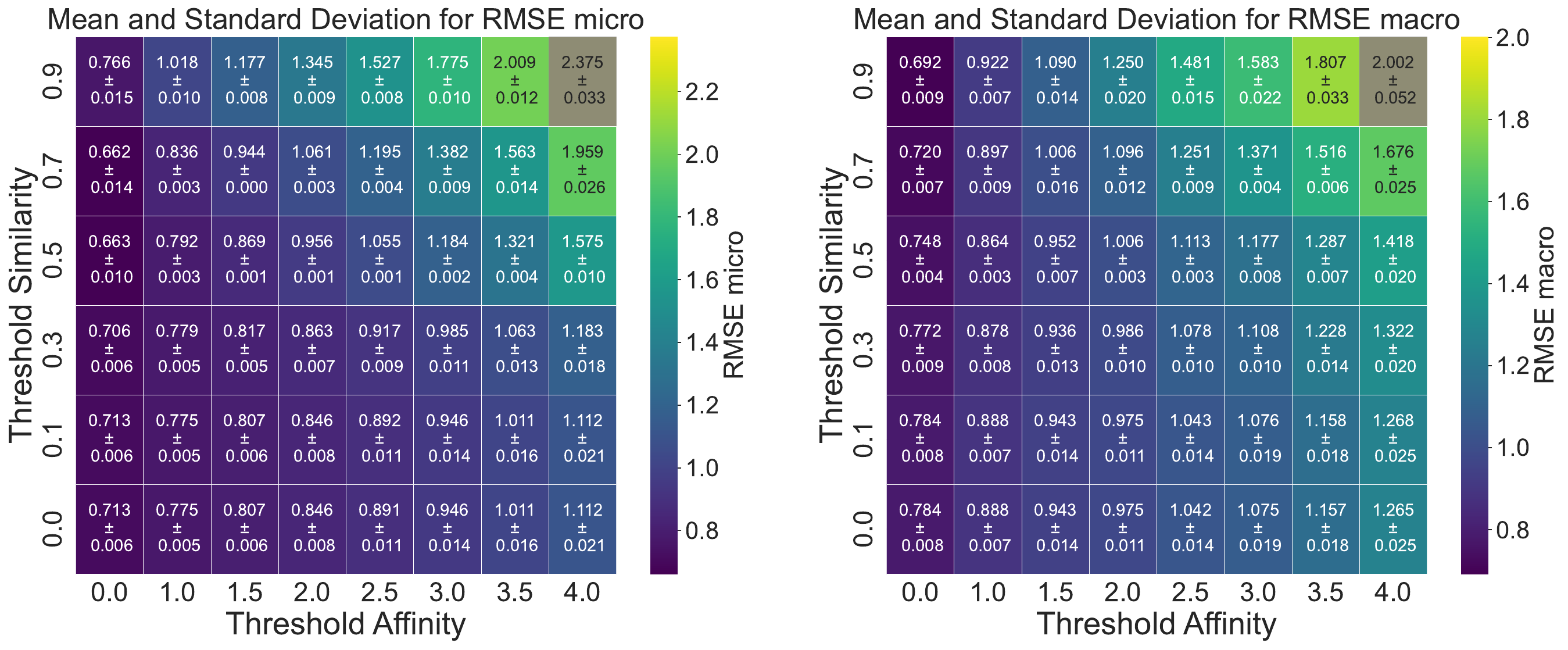}%
\caption{The heatmap of the $RMSE_{micro}$ (left) and $RMSE_{macro}$ (right) for the best DTI model trained from scratch, showing groups of compounds split by similarity and affinity thresholds for the BindingDB dataset in the case of a random split. The values represent the mean ± standard deviation based on 3 experiments. The groups with fewer than 100 pairs are masked in gray.}%
\label{fig:DTI_BDB_rs_bl_hm}%
\end{figure}

\subsection{Performance of the models when transferring only the drug encoder}
\label{ad_file:TL_d_enc}

\subsubsection{KIBA dataset}

\begin{figure}[H]
\centering
\includegraphics[scale=0.3]{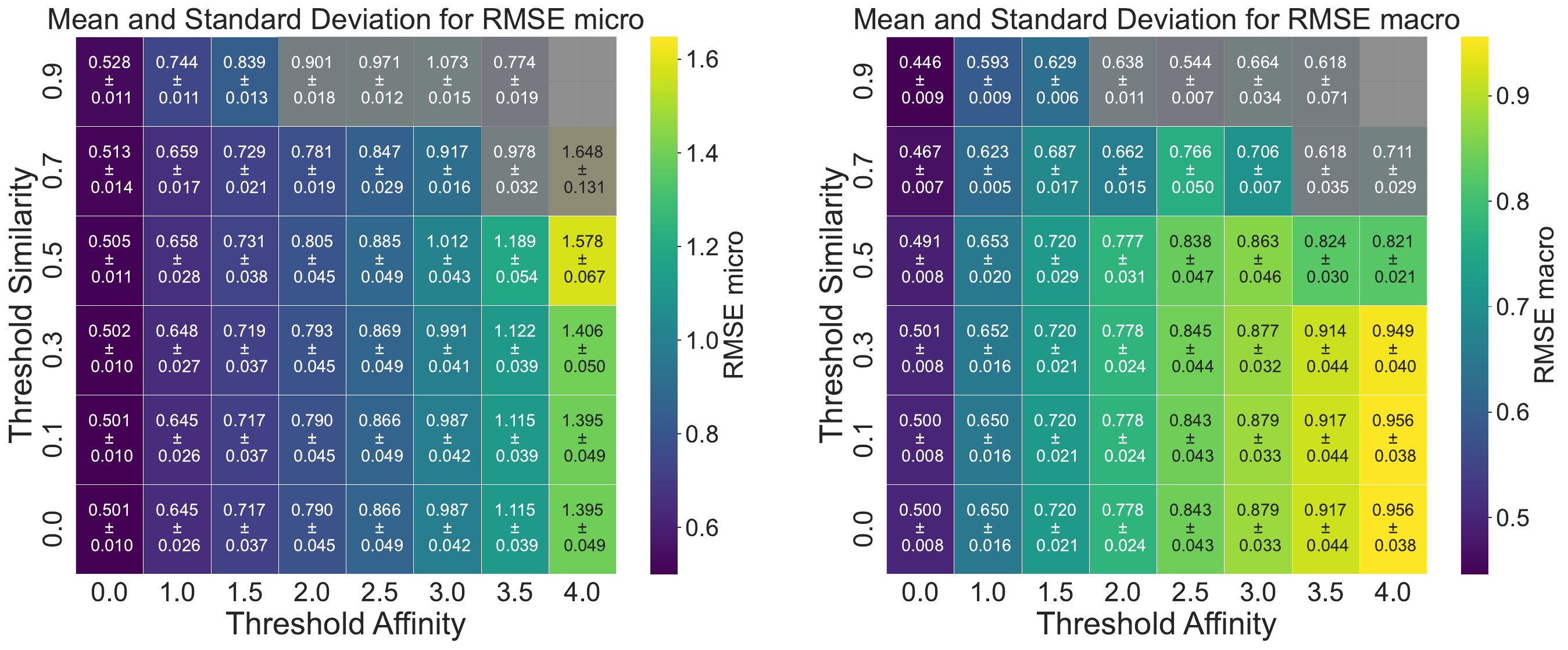}%
\caption{The heatmap of the $RMSE_{micro}$ (left) and $RMSE_{macro}$ (right) of the best DTI model (transfer learning involving only the drug encoder, \textbf{warm start}) for groups of compounds split by similarity and affinity thresholds for KIBA (random split).}%
\label{fig:DTI_KIBA_rs_tl_ws_hm}%
\end{figure}

\begin{figure}[H]
\centering
\includegraphics[scale=0.3]{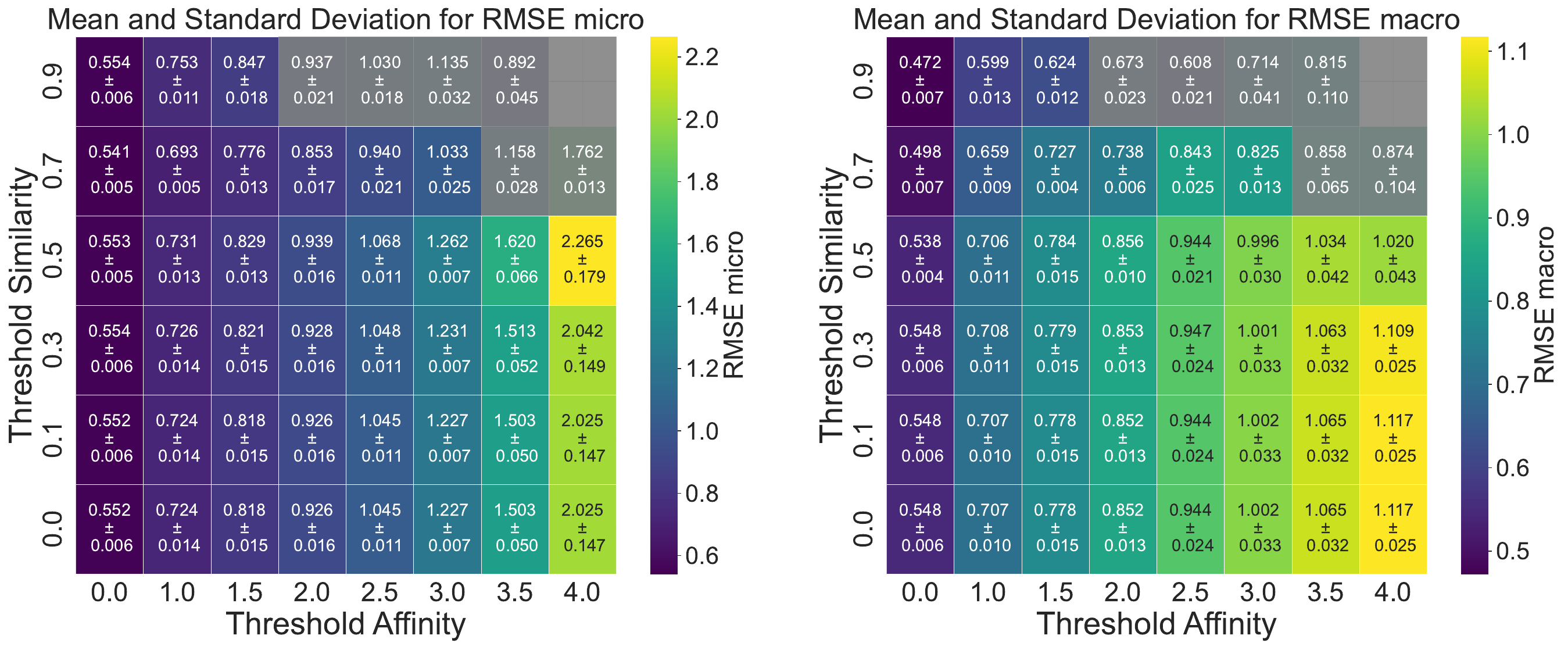}%
\caption{The heatmap of the $RMSE_{micro}$ (left) and $RMSE_{macro}$ (right) of the best DTI model (transfer learning involving only the drug encoder, \textbf{with freezing weights}) for groups of compounds split by similarity and affinity thresholds for KIBA (random split).}%
\label{fig:DTI_KIBA_rs_tl_f_hm}%
\end{figure}

\begin{figure}[H]
\centering
\includegraphics[scale=0.3]{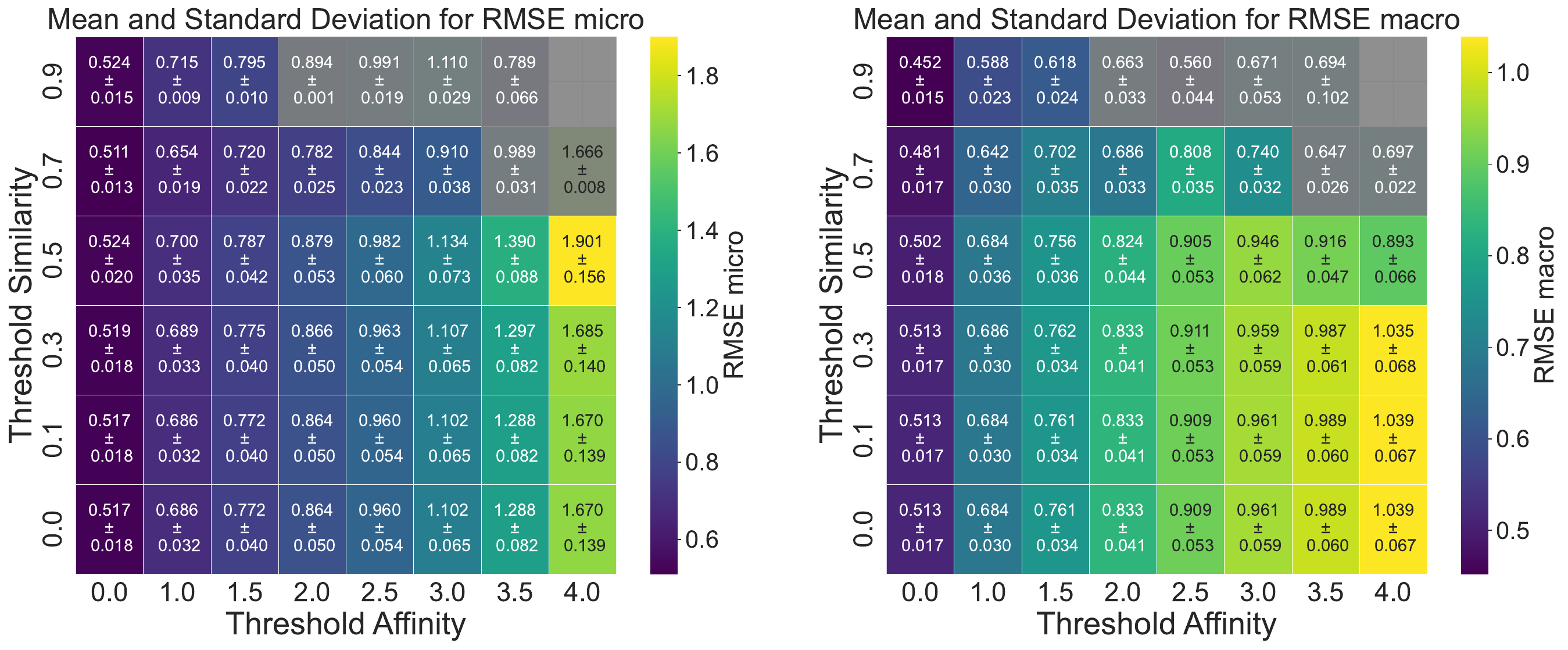}%
\caption{The heatmap of the $RMSE_{micro}$ (left) and $RMSE_{macro}$ (right) of the best DTI model (transfer learning involving only the drug encoder, \textbf{with freezing weights and adding an extra layer}) for groups of compounds split by similarity and affinity thresholds for KIBA (random split split).}%
\label{fig:DTI_KIBA_rs_tl_f_el_hm}%
\end{figure}

\begin{figure}[H]
\centering
\includegraphics[scale=0.3]{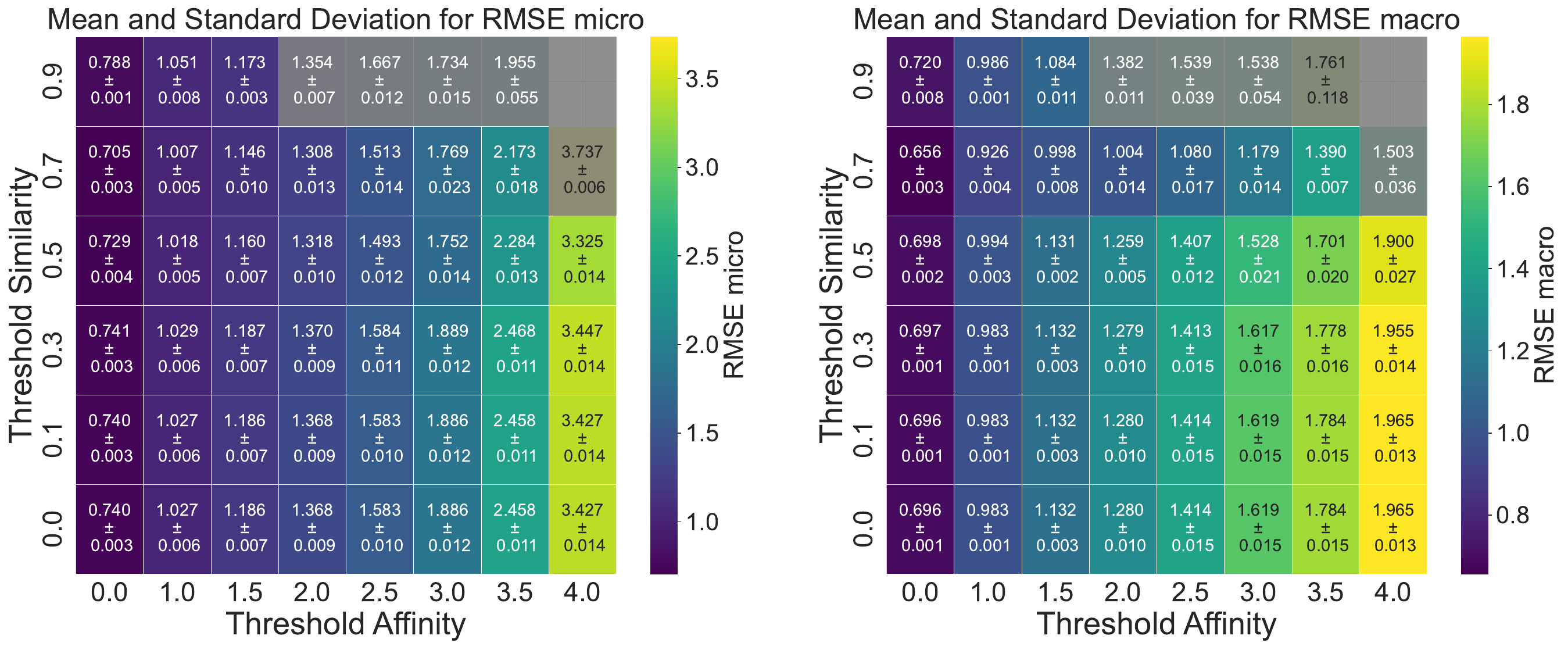}%
\caption{The heatmap of the $RMSE_{micro}$ (left) and $RMSE_{macro}$ (right) of the best DTI model (transfer learning involving only the drug encoder, \textbf{warm start}) for groups of compounds split by similarity and affinity thresholds for KIBA (compound-based split).}%
\label{fig:DTI_KIBA_cb_tl_ws_hm}%
\end{figure}

\begin{figure}[H]
\centering
\includegraphics[scale=0.3]{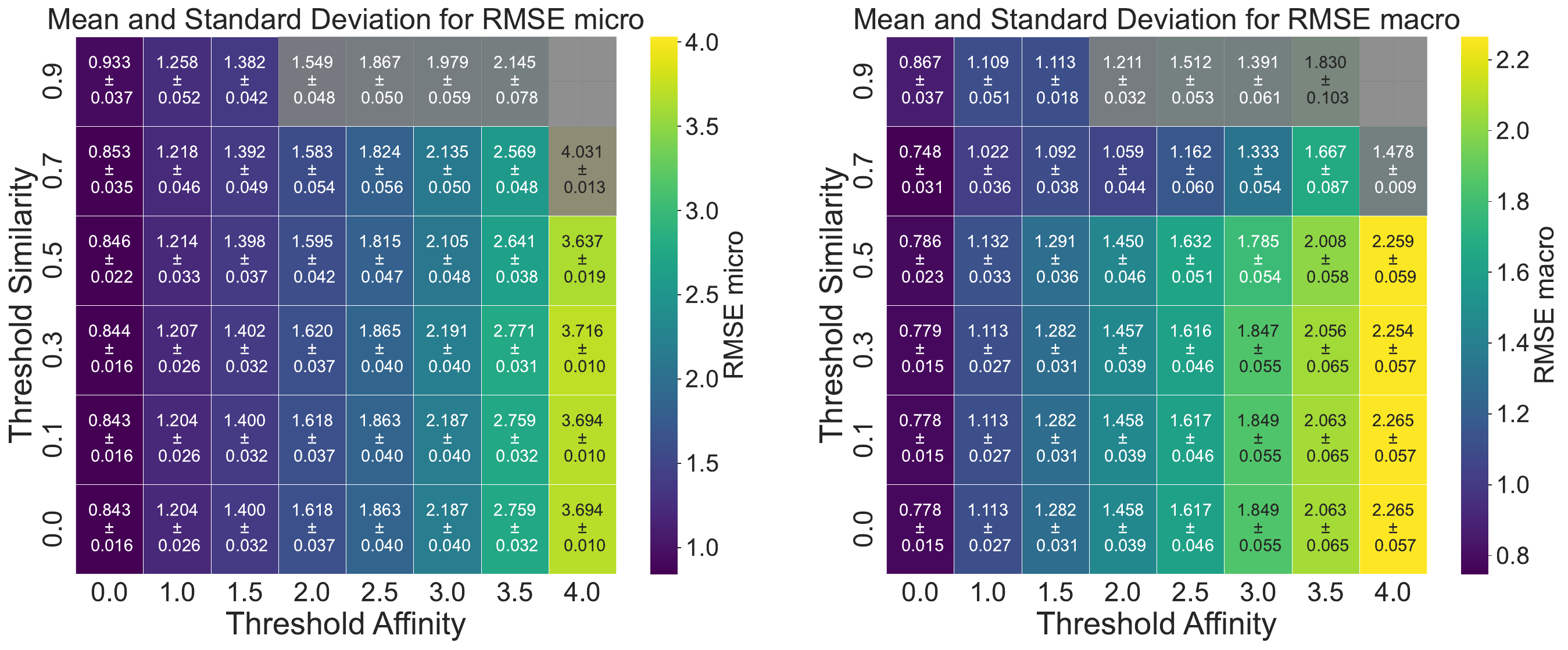}%
\caption{The heatmap of the $RMSE_{micro}$ (left) and $RMSE_{macro}$ (right) of the best DTI model (transfer learning involving only the drug encoder, \textbf{with freezing weights}) for groups of compounds split by similarity and affinity thresholds for KIBA (compound-based split).}%
\label{fig:DTI_KIBA_cb_tl_f_hm}%
\end{figure}

\begin{figure}[H]
\centering
\includegraphics[scale=0.3]{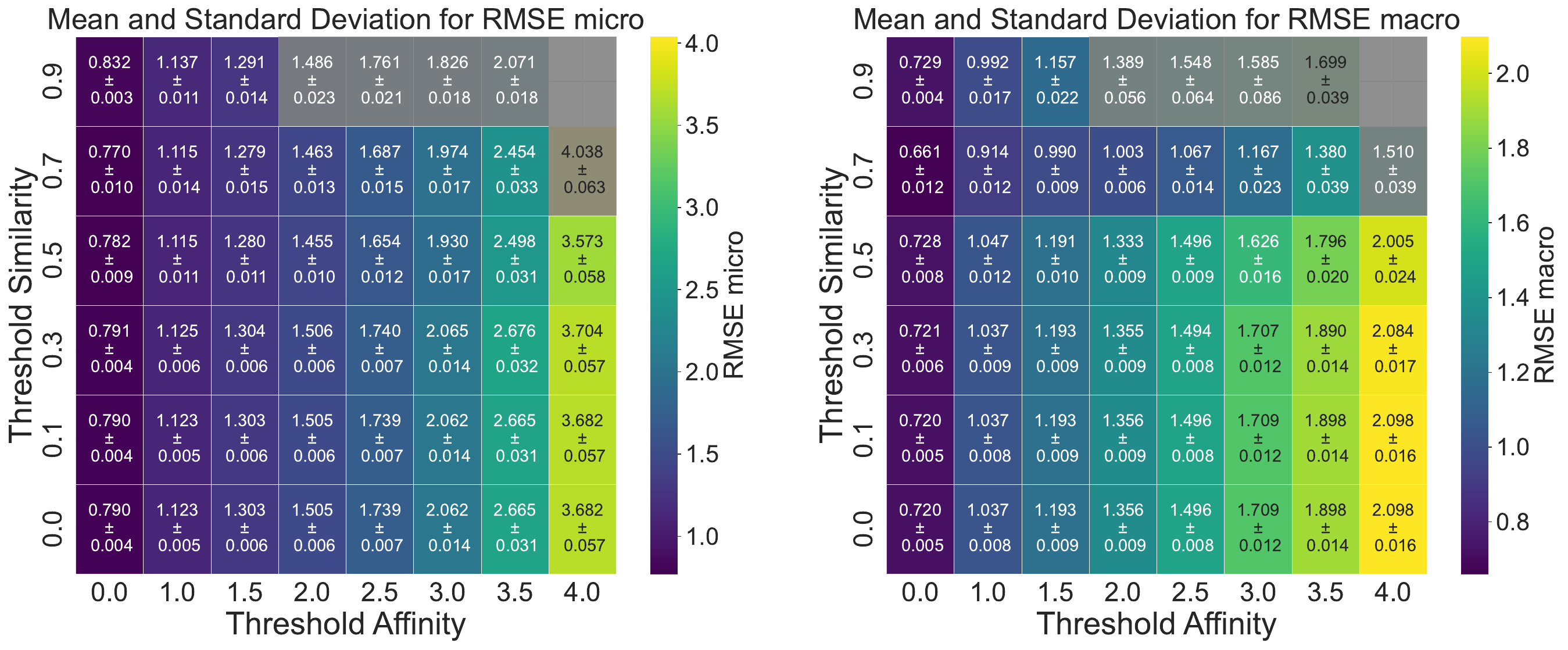}%
\caption{The heatmap of the $RMSE_{micro}$ (left) and $RMSE_{macro}$ (right) of the best DTI model (transfer learning involving only the drug encoder, \textbf{with freezing weights and adding an extra layer}) for groups of compounds split by similarity and affinity thresholds for KIBA (compound-based split).}%
\label{fig:DTI_KIBA_cb_tl_f_el_hm}%
\end{figure}

\subsubsection{BindingDB dataset}
\label{ad_file:BDB_TL_d_enc}

\begin{figure}[H]
\centering
\includegraphics[scale=0.3]{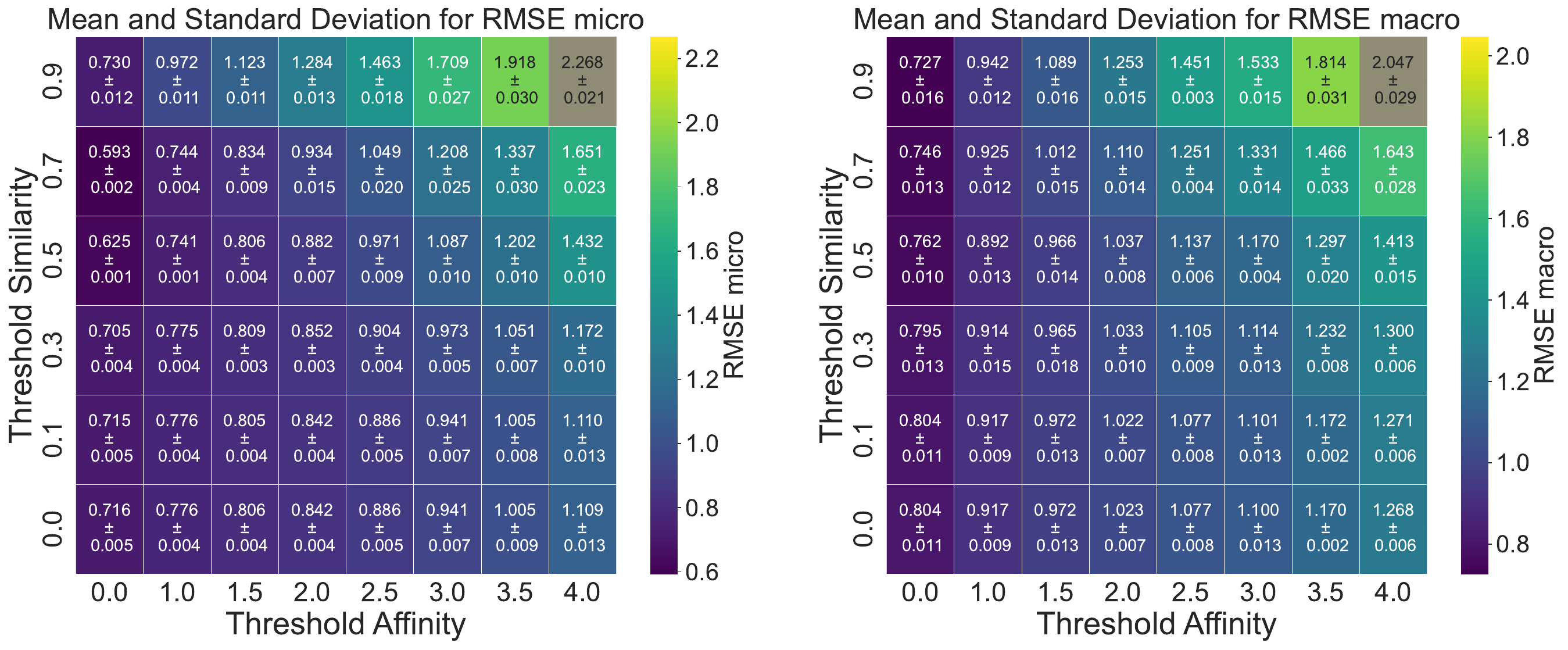}%
\caption{The heatmap of the $RMSE_{micro}$ (left) and $RMSE_{macro}$ (right) of the best DTI model (transfer learning involving only the drug encoder, \textbf{warm start}) for groups of compounds split by similarity and affinity thresholds for BindingDB (random split).}%
\label{fig:DTI_BDB_rs_tl_ws_hm}%
\end{figure}

\begin{figure}[H]
\centering
\includegraphics[scale=0.3]{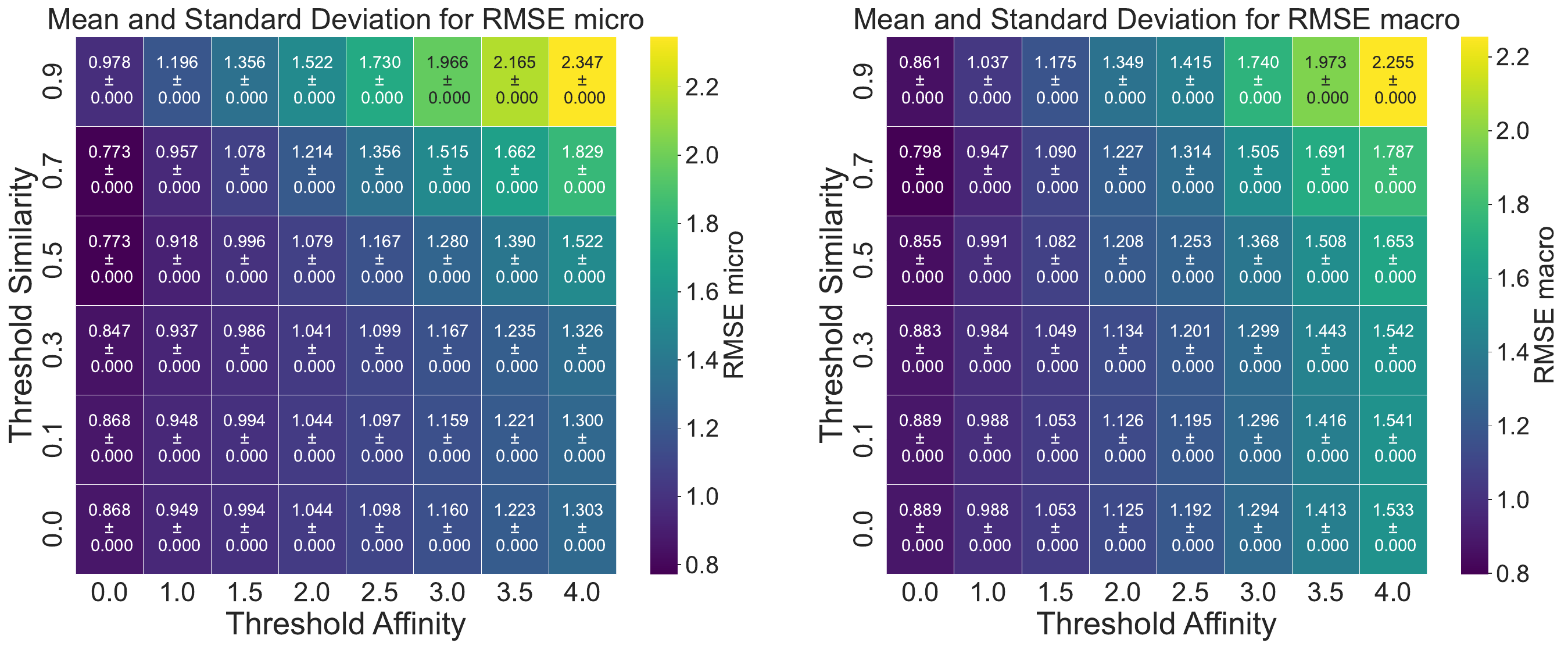}%
\caption{The heatmap of the $RMSE_{micro}$ (left) and $RMSE_{macro}$ (right) of the best DTI model (transfer learning involving only the drug encoder, \textbf{warm start}) for groups of compounds split by similarity and affinity thresholds for BindingDB (compound-based split).}%
\label{fig:DTI_BDB_cb_tl_ws_hm}%
\end{figure}

\subsection{Performance of the models when transferring both drug and target encoders}
\label{ad_file:TL_both_enc}

\subsubsection{KIBA dataset}

\begin{figure}[H]
\centering
\includegraphics[scale=0.3]{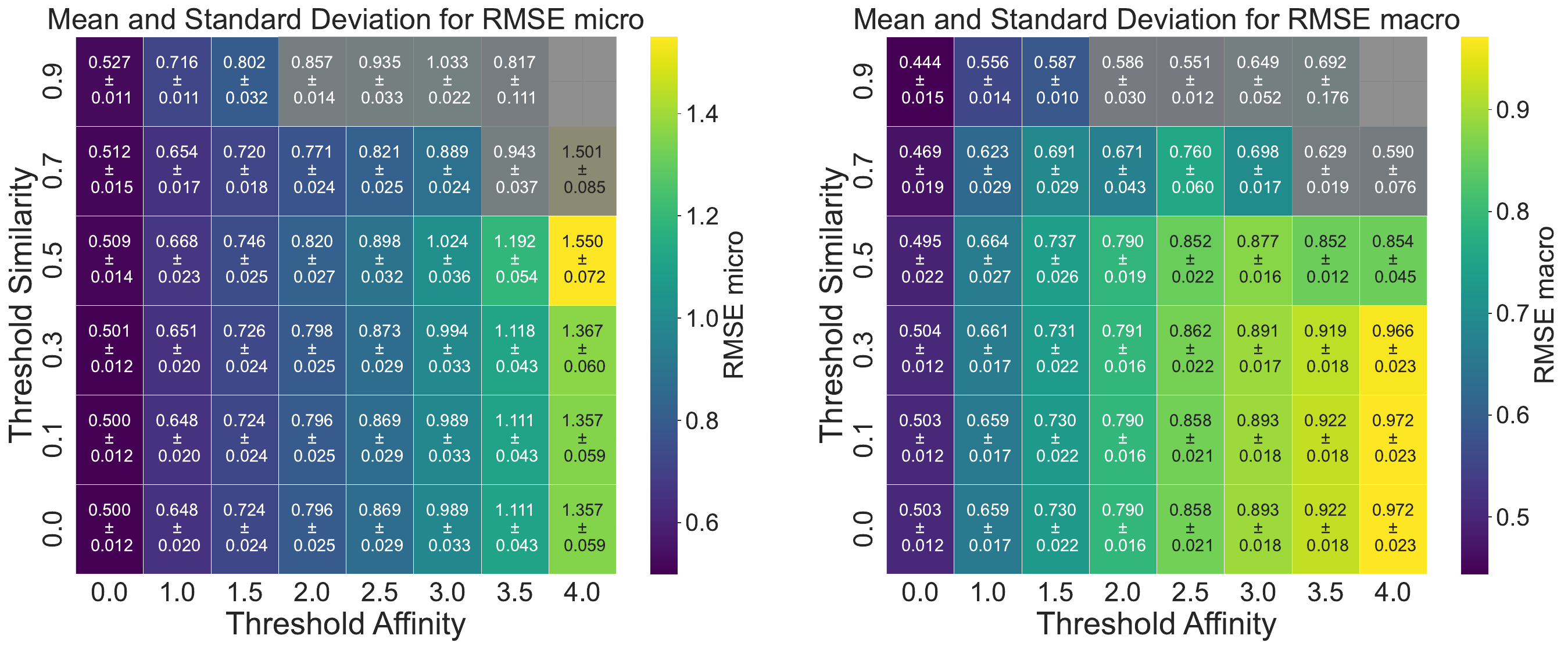}%
\caption{The heatmap of the $RMSE_{micro}$ (left) and $RMSE_{macro}$ (right) of the best DTI model (transfer learning involving both drug and target encoders, \textbf{warm start}) for groups of compounds split by similarity and affinity thresholds for KIBA (random split).}%
\label{fig:DTI_KIBA_rs_tl_t_enc_ws_hm}%
\end{figure}

\begin{figure}[H]
\centering
\includegraphics[scale=0.3]{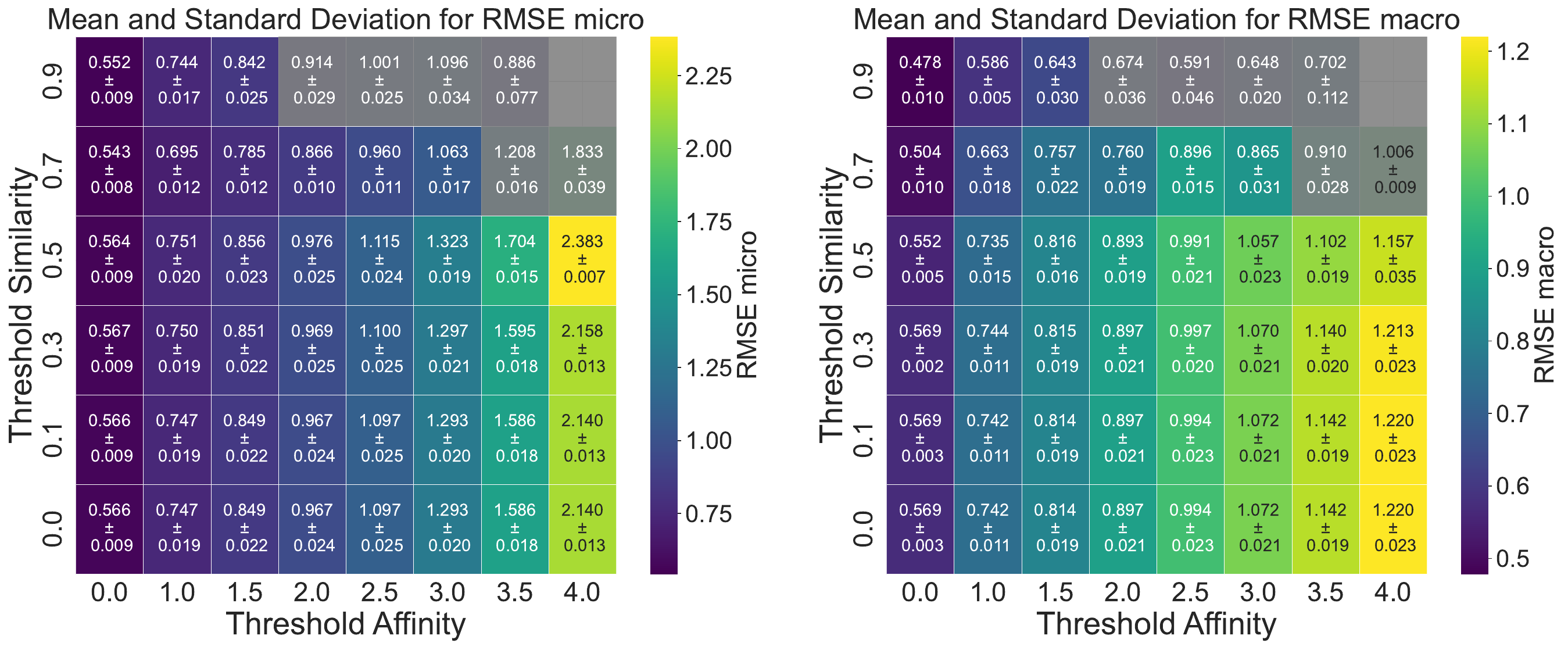}%
\caption{The heatmap of the $RMSE_{micro}$ (left) and $RMSE_{macro}$ (right) of the best DTI model (transfer learning involving both drug and target encoders, \textbf{with freezing weights}) for groups of compounds split by similarity and affinity thresholds for KIBA (random split).}%
\label{fig:DTI_KIBA_rs_tl_t_enc_f_hm}%
\end{figure}

\begin{figure}[H]
\centering
\includegraphics[scale=0.3]{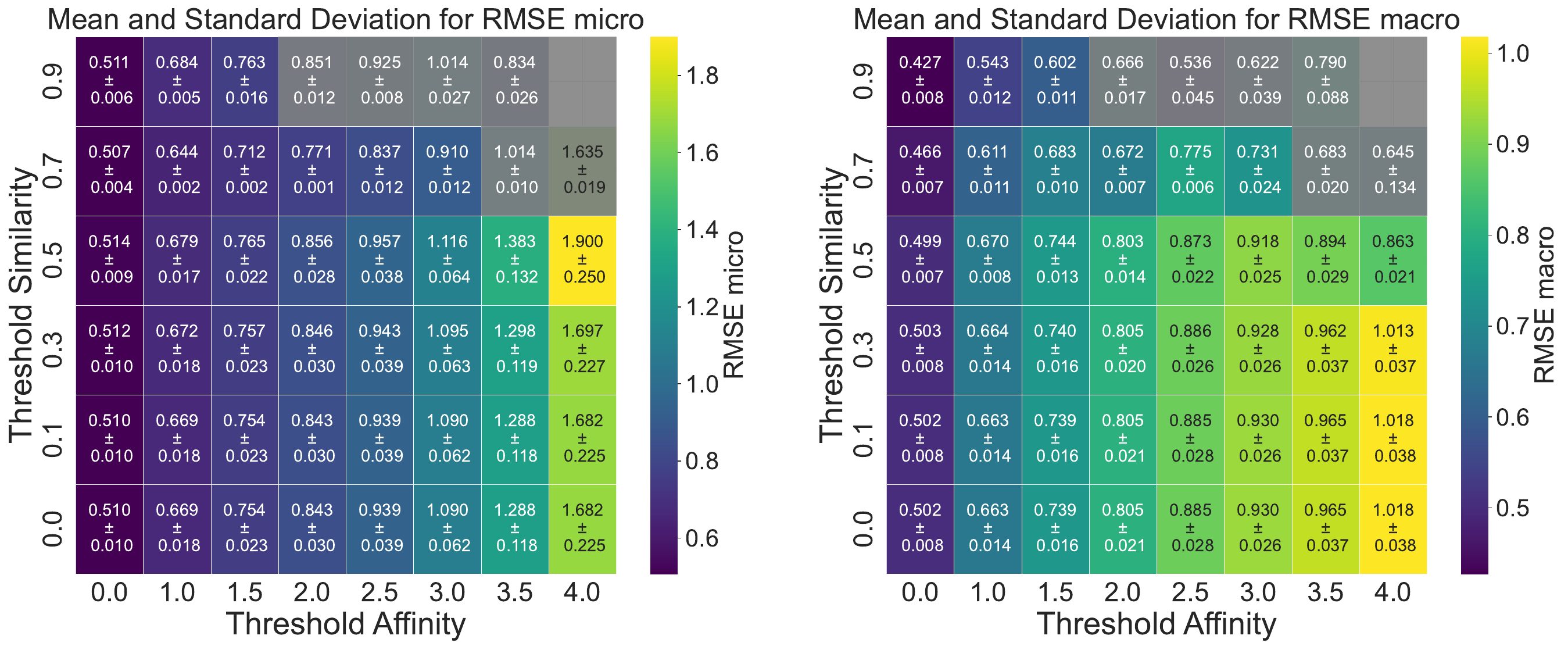}%
\caption{The heatmap of the $RMSE_{micro}$ (left) and $RMSE_{macro}$ (right) of the best DTI model (transfer learning involving both drug and target encoders, \textbf{with freezing weights and adding an extra layer}) for groups of compounds split by similarity and affinity thresholds for KIBA (random split split).}%
\label{fig:DTI_KIBA_rs_tl_t_enc_f_el_hm}%
\end{figure}

\begin{figure}[H]
\centering
\includegraphics[scale=0.3]{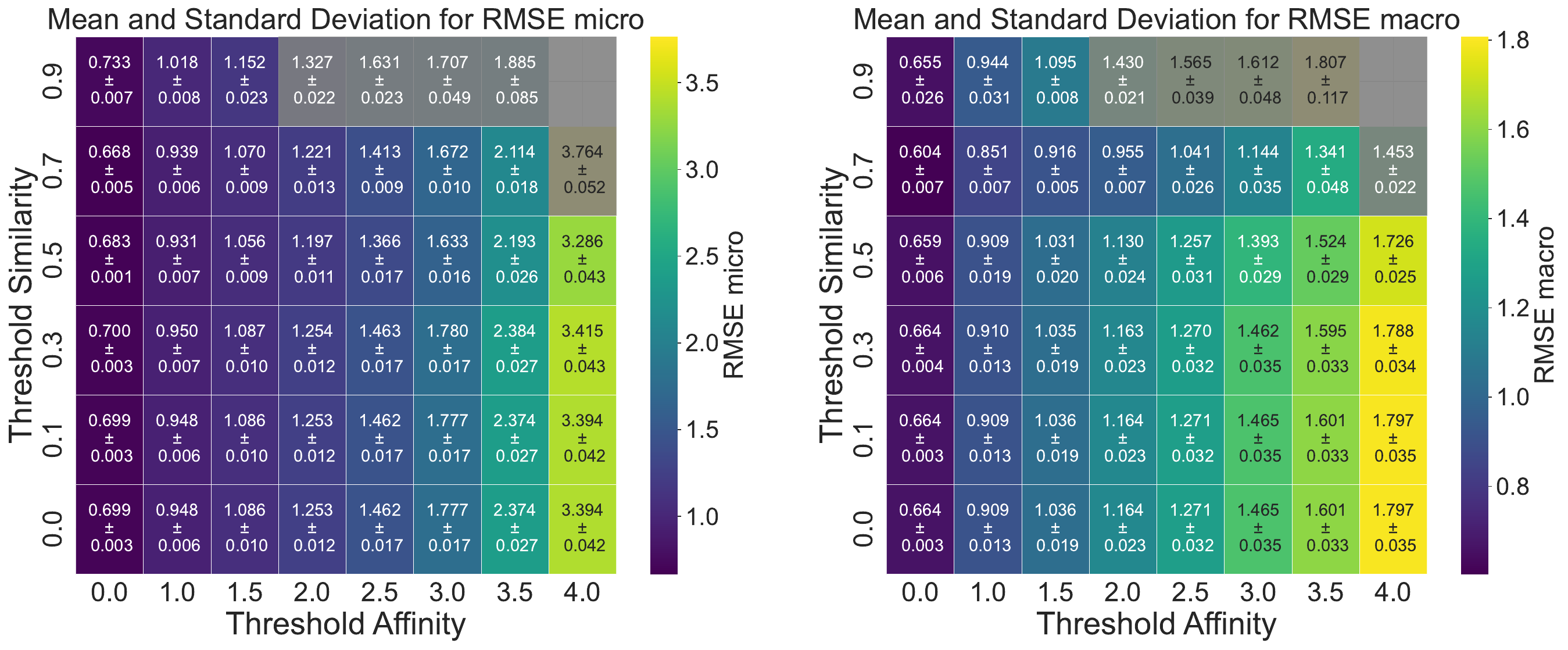}%
\caption{The heatmap of the $RMSE_{micro}$ (left) and $RMSE_{macro}$ (right) of the best DTI model (transfer learning involving both drug and target encoders, \textbf{warm start}) for groups of compounds split by similarity and affinity thresholds for KIBA (compound-based split).}%
\label{fig:DTI_KIBA_cb_tl_t_enc_ws_hm}%
\end{figure}

\begin{figure}[H]
\centering
\includegraphics[scale=0.3]{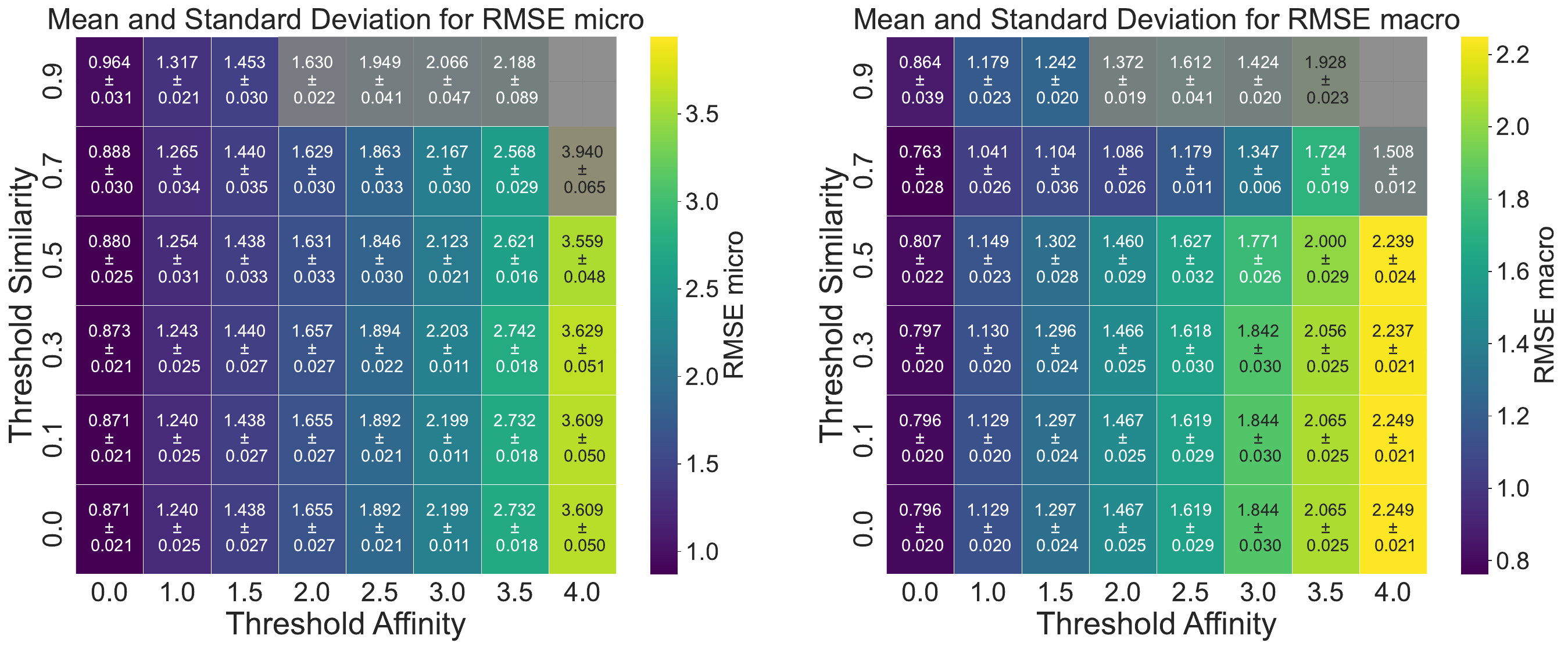}%
\caption{The heatmap of the $RMSE_{micro}$ (left) and $RMSE_{macro}$ (right) of the best DTI model (transfer learning involving both drug and target encoders, \textbf{with freezing weights}) for groups of compounds split by similarity and affinity thresholds for KIBA (compound-based split).}%
\label{fig:DTI_KIBA_cb_tl_t_enc_f_hm}%
\end{figure}

\begin{figure}[H]
\centering
\includegraphics[scale=0.3]{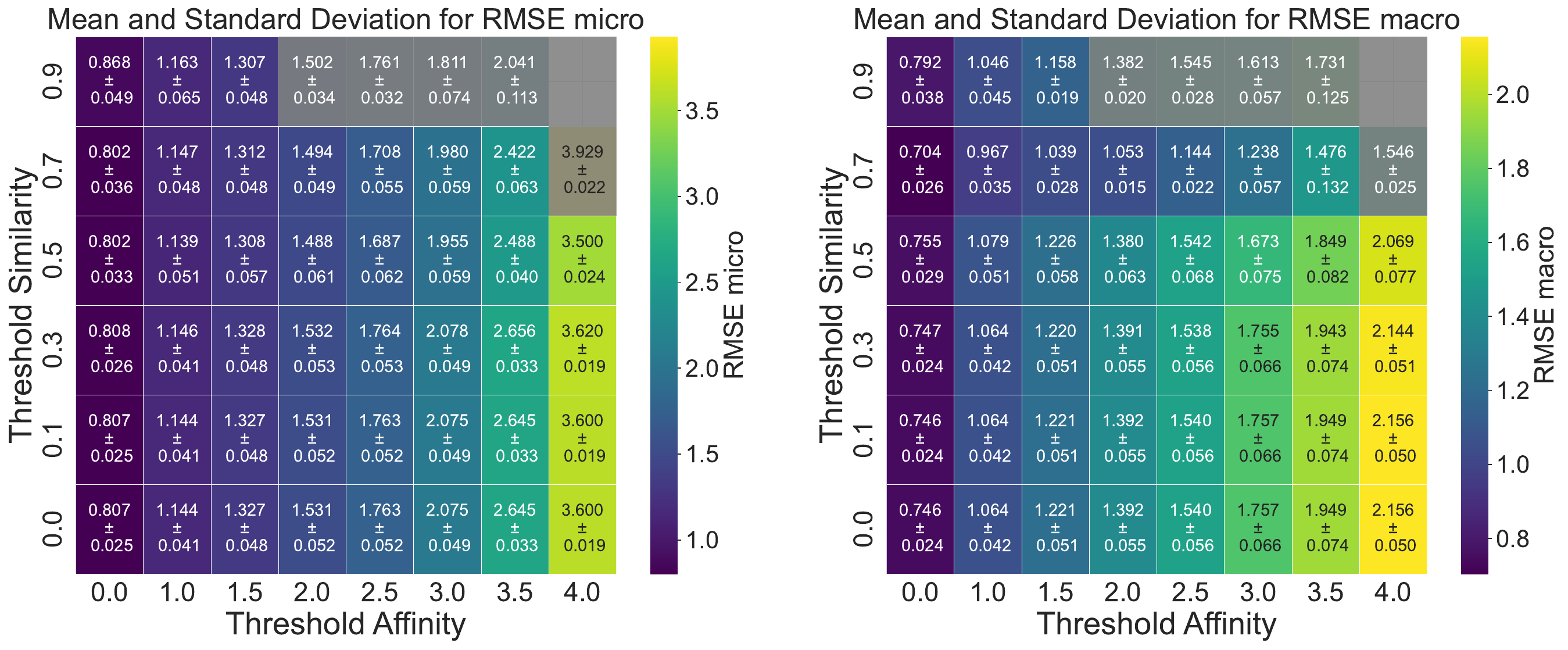}%
\caption{The heatmap of the $RMSE_{micro}$ (left) and $RMSE_{macro}$ (right) of the best DTI model (transfer learning involving both drug and target encoders, \textbf{with freezing weights and adding an extra layer}) for groups of compounds split by similarity and affinity thresholds for KIBA (compound-based split).}%
\label{fig:DTI_KIBA_cb_tl_t_enc_f_el_hm}%
\end{figure}

\subsubsection{BindingDB dataset}
\begin{figure}[H]
\centering
\includegraphics[scale=0.3]{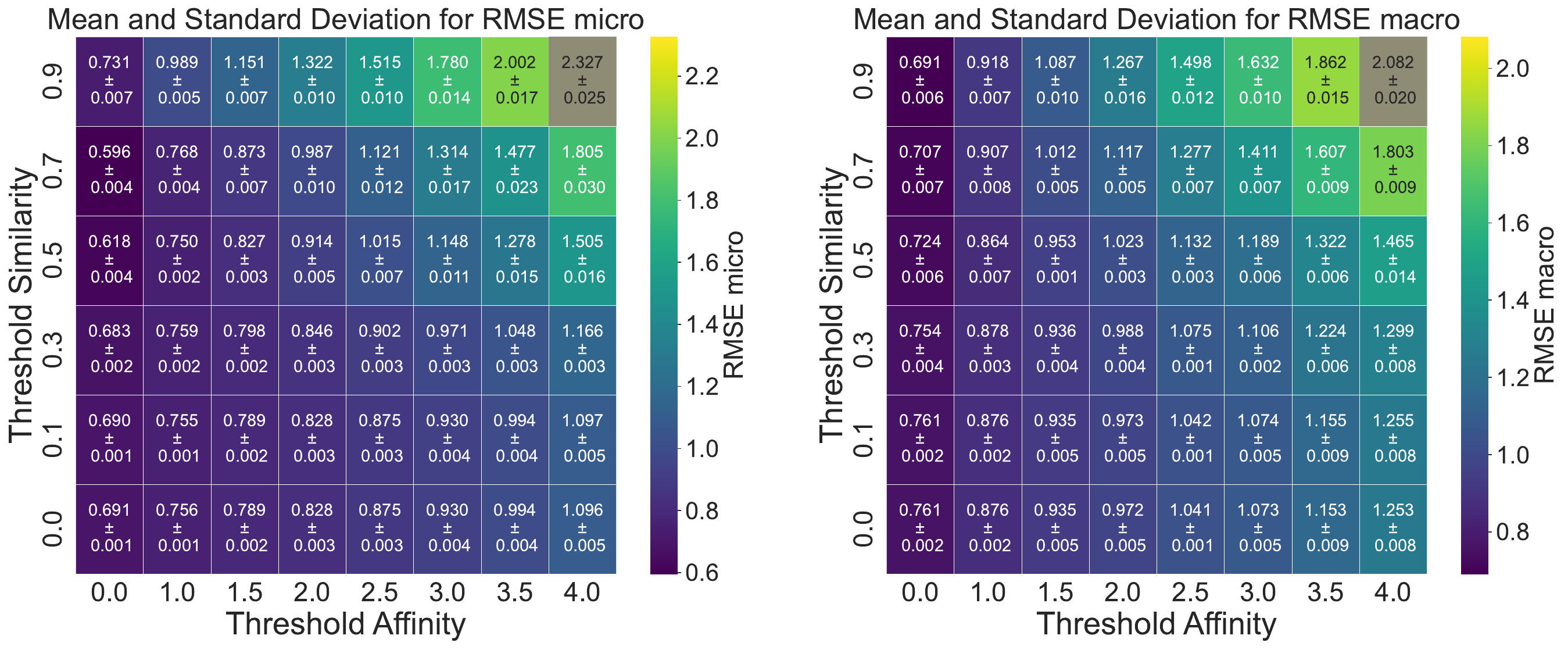}%
\caption{The heatmap of the $RMSE_{micro}$ (left) and $RMSE_{macro}$ (right) of the best DTI model (transfer learning involving both drug and target encoders, \textbf{warm start}) for groups of compounds split by similarity and affinity thresholds for BindingDB (random split).}%
\label{fig:DTI_BDB_rs_tl_t_enc_ws_hm}%
\end{figure}

\begin{figure}[H]
\centering
\includegraphics[scale=0.3]{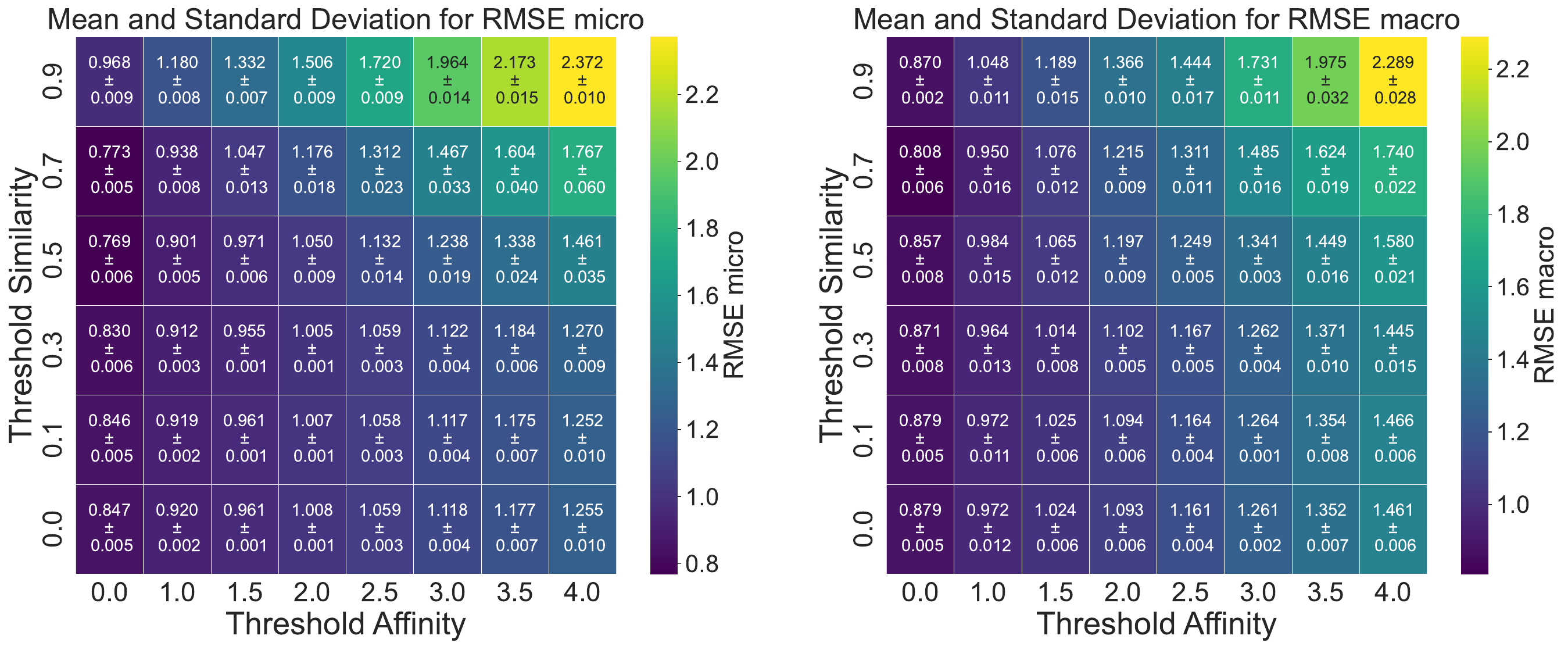}%
\caption{The heatmap of the $RMSE_{micro}$ (left) and $RMSE_{macro}$ (right) of the best DTI model (transfer learning involving both drug and target encoders, \textbf{warm start}) for groups of compounds split by similarity and affinity thresholds for BindingDB (compound-based split).}%
\label{fig:DTI_BDB_cb_tl_t_enc_ws_hm}%
\end{figure}

\subsection{Differential heatmaps when transferring only the drug encoder}

\begin{figure}[H]
\centering
\includegraphics[scale=0.3]{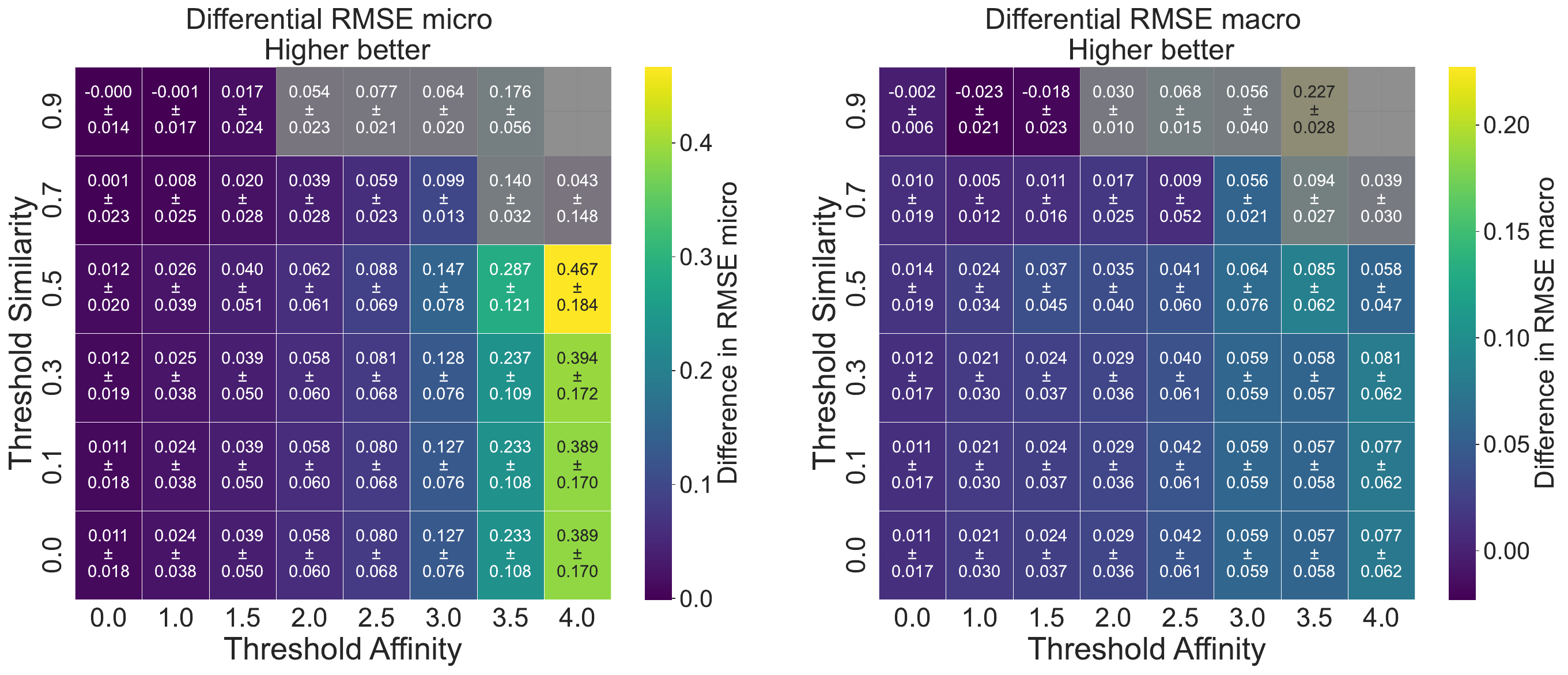}%
\caption{The differential heatmap of the $RMSE_{micro}$ (left) and $RMSE_{macro}$ (right) of the best DTI model (transfer learning involving only the drug encoder, \textbf{warm start}) for groups of compounds split by similarity and affinity thresholds for KIBA (random split).}%
\label{fig:DTI_KIBA_rs_tl_ws_hm_diff}%
\end{figure}

\begin{figure}[H]
\centering
\includegraphics[scale=0.3]{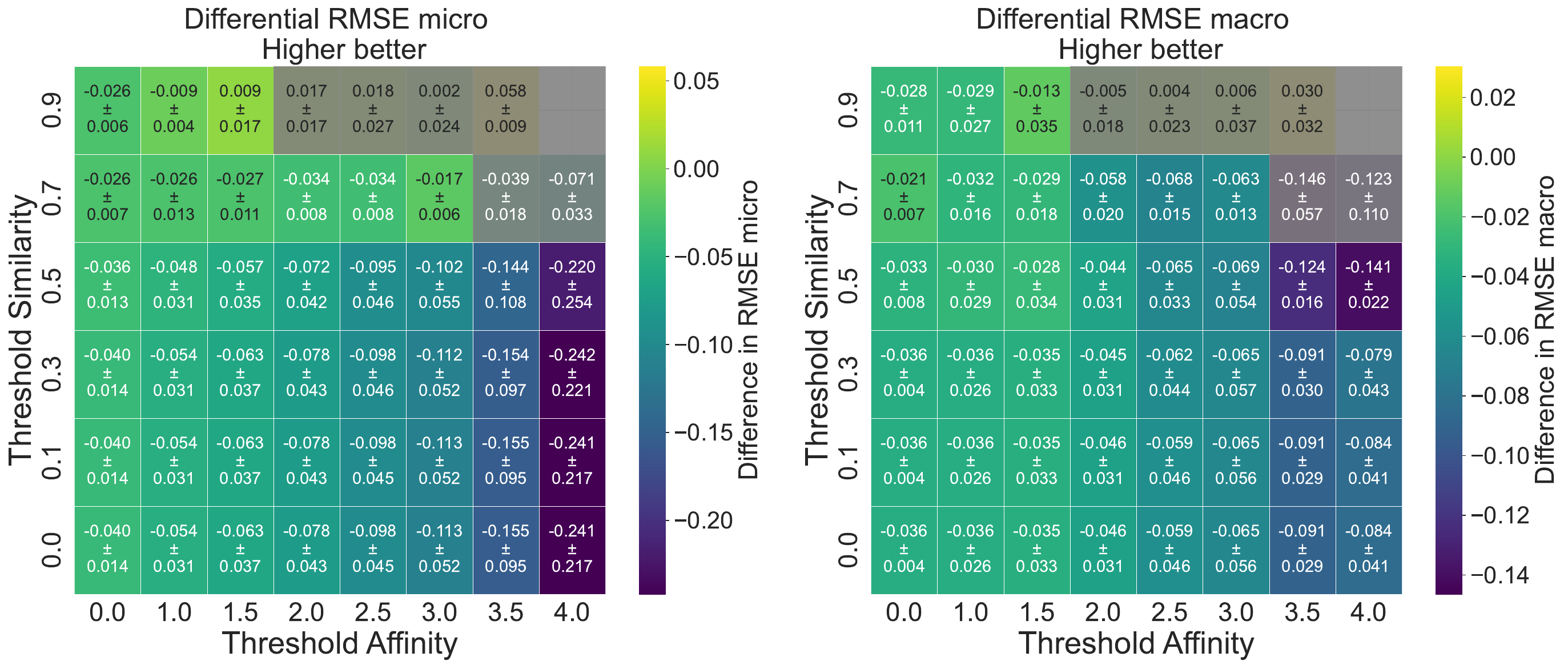}%
\caption{The differential heatmap of the $RMSE_{micro}$ (left) and $RMSE_{macro}$ (right) of the best DTI model (transfer learning involving only the drug encoder, \textbf{with freezing weights}) for groups of compounds split by similarity and affinity thresholds for KIBA (random split).}%
\label{fig:DTI_KIBA_rs_tl_f_hm_diff}%
\end{figure}

\begin{figure}[H]
\centering
\includegraphics[scale=0.3]{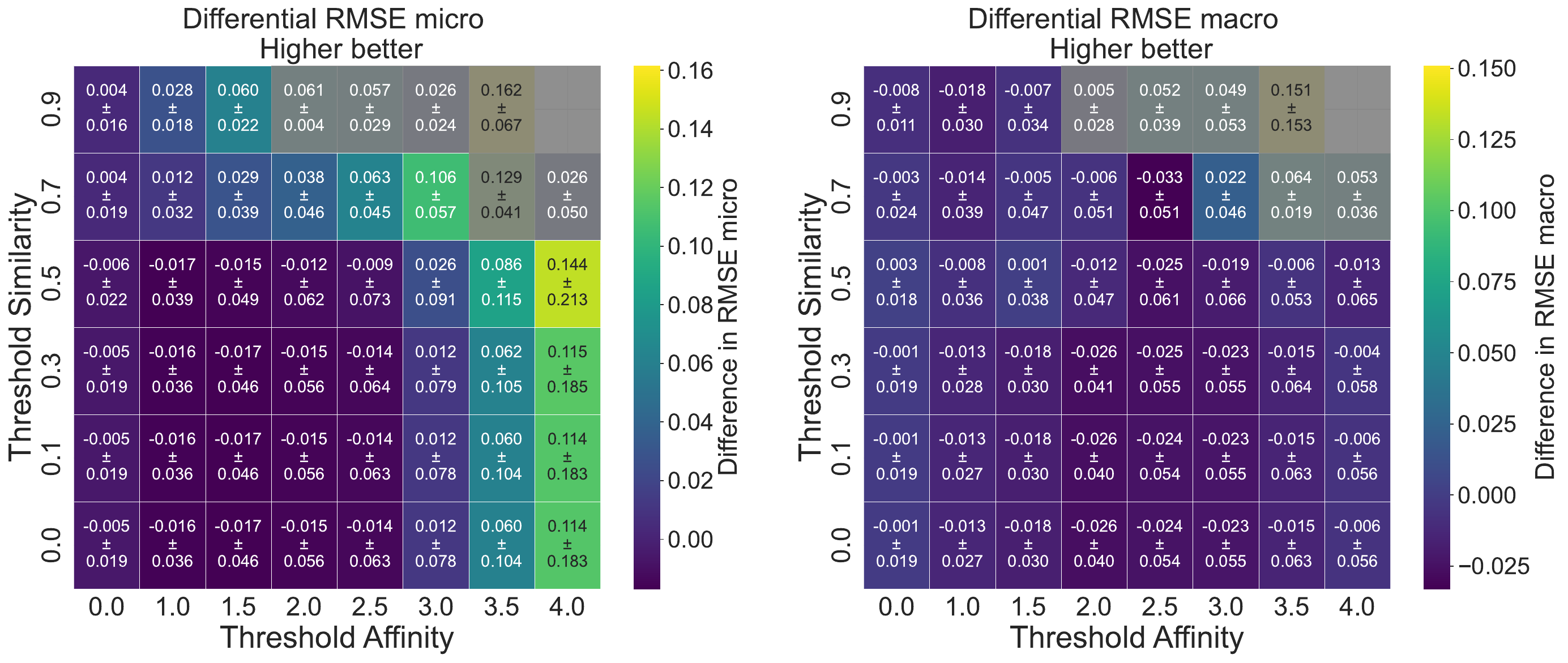}%
\caption{The differential heatmap of the $RMSE_{micro}$ (left) and $RMSE_{macro}$ (right) of the best DTI model (transfer learning involving only the drug encoder, \textbf{with freezing weights and adding an extra layer}) for groups of compounds split by similarity and affinity thresholds for KIBA (random split).}%
\label{fig:DTI_KIBA_rs_tl_f_el_hm_diff}%
\end{figure}

\begin{figure}[H]
\centering
\includegraphics[scale=0.3]{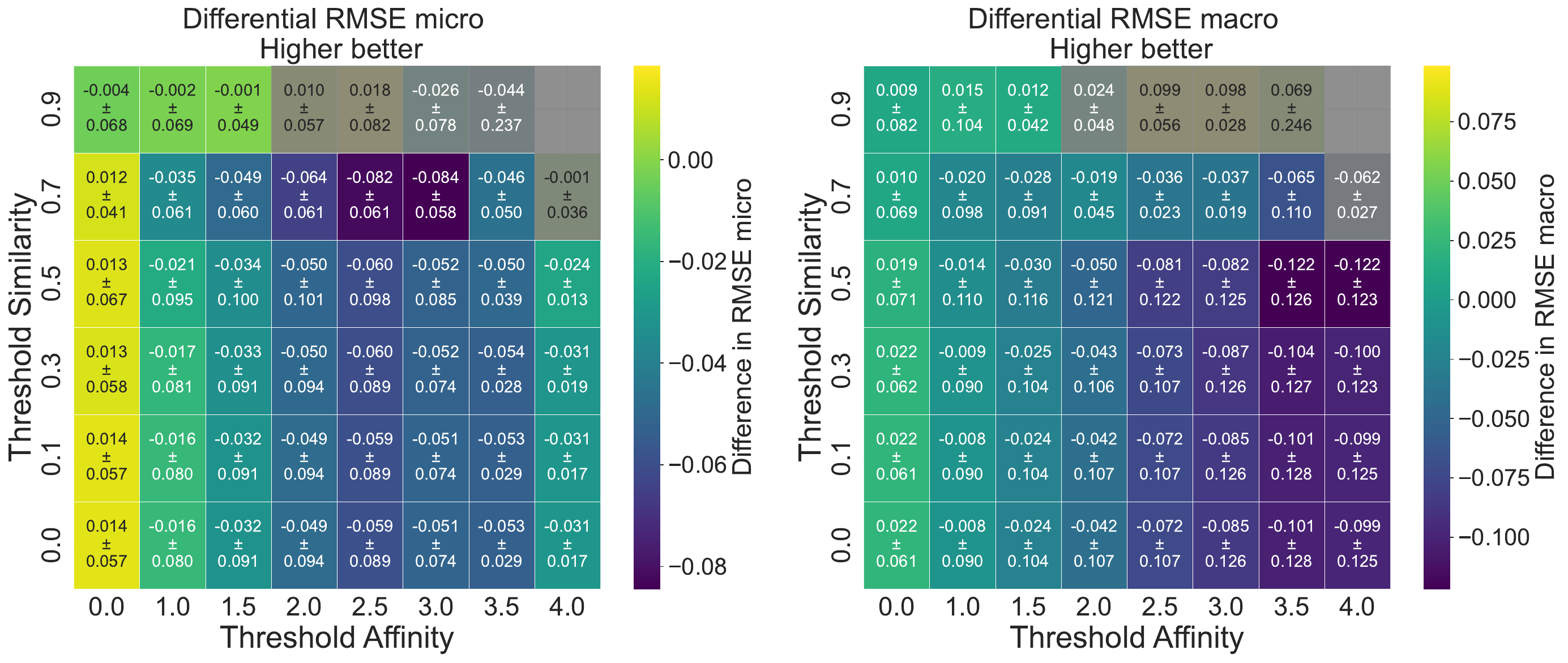}%
\caption{The differential heatmap of the $RMSE_{micro}$ (left) and $RMSE_{macro}$ (right) of the best DTI model (transfer learning involving only the drug encoder, \textbf{warm start}) for groups of compounds split by similarity and affinity thresholds for KIBA (compound-based split).}%
\label{fig:DTI_KIBA_cb_tl_ws_hm_diff}%
\end{figure}

\begin{figure}[H]
\centering
\includegraphics[scale=0.3]{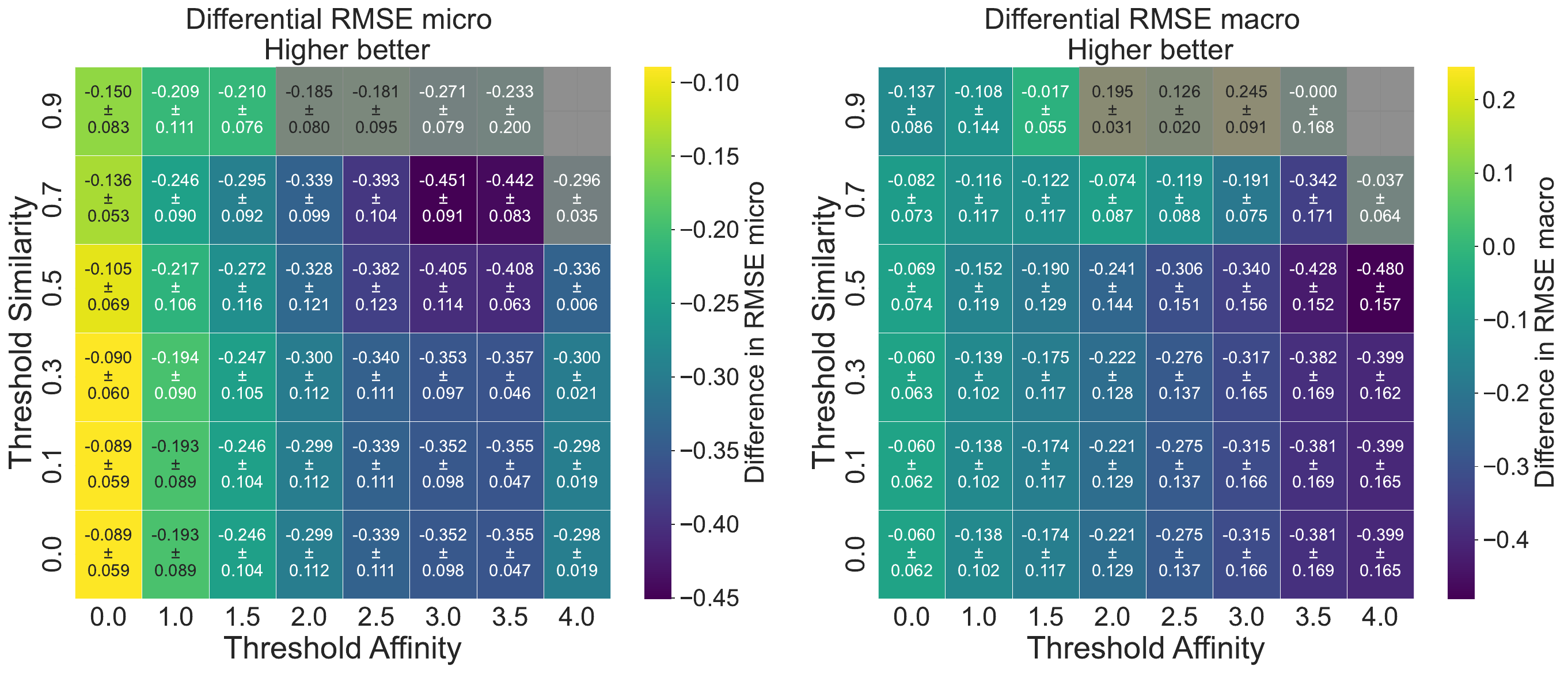}%
\caption{The differential heatmap of the $RMSE_{micro}$ (left) and $RMSE_{macro}$ (right) of the best DTI model (transfer learning involving only the drug encoder, \textbf{with freezing weights}) for groups of compounds split by similarity and affinity thresholds for KIBA (compound-based split).}%
\label{fig:DTI_KIBA_cb_tl_f_hm_diff}%
\end{figure}

\begin{figure}[H]
\centering
\includegraphics[scale=0.3]{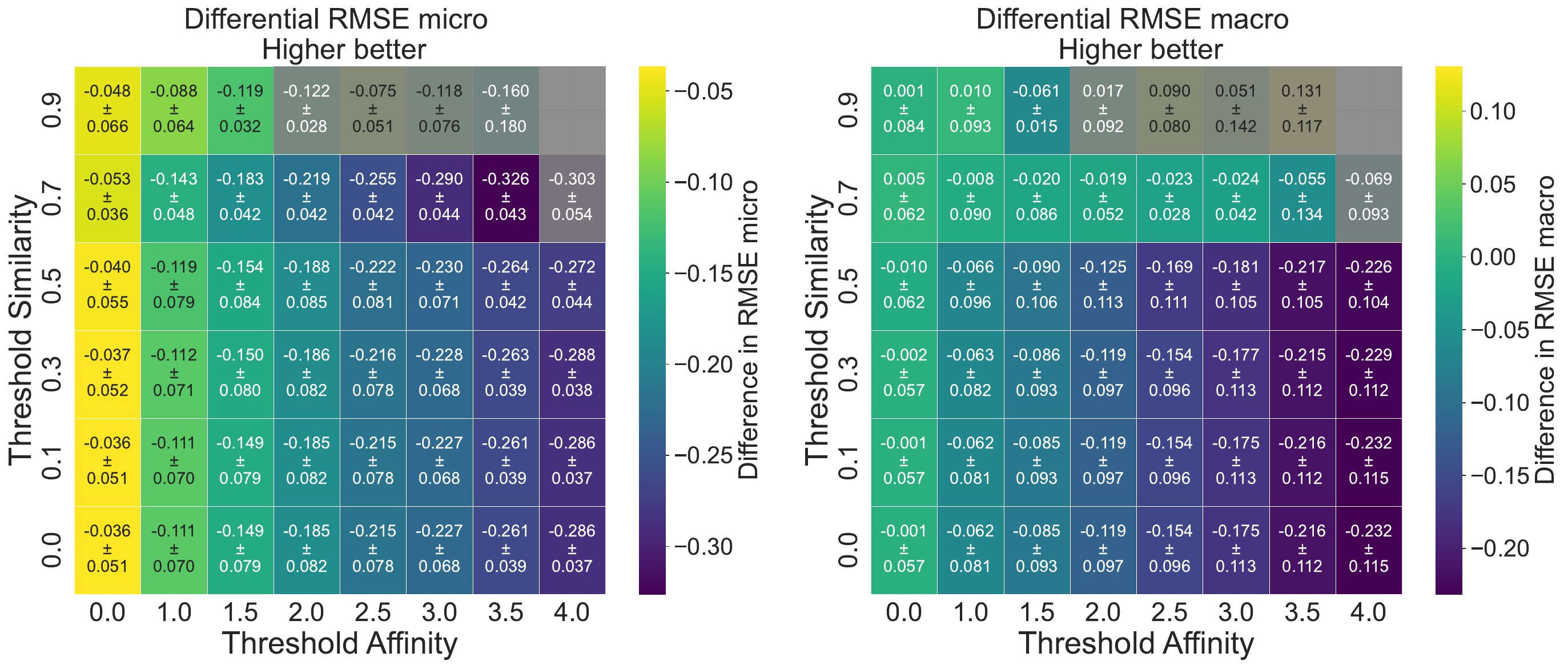}%
\caption{The differential heatmap of the $RMSE_{micro}$ (left) and $RMSE_{macro}$ (right) of the best DTI model (transfer learning involving only the drug encoder, \textbf{with freezing weights and adding an extra layer}) for groups of compounds split by similarity and affinity thresholds for KIBA (compound-based split).}%
\label{fig:DTI_KIBA_cb_tl_f_el_hm_diff}%
\end{figure}

\begin{figure}[H]
\centering
\includegraphics[scale=0.3]{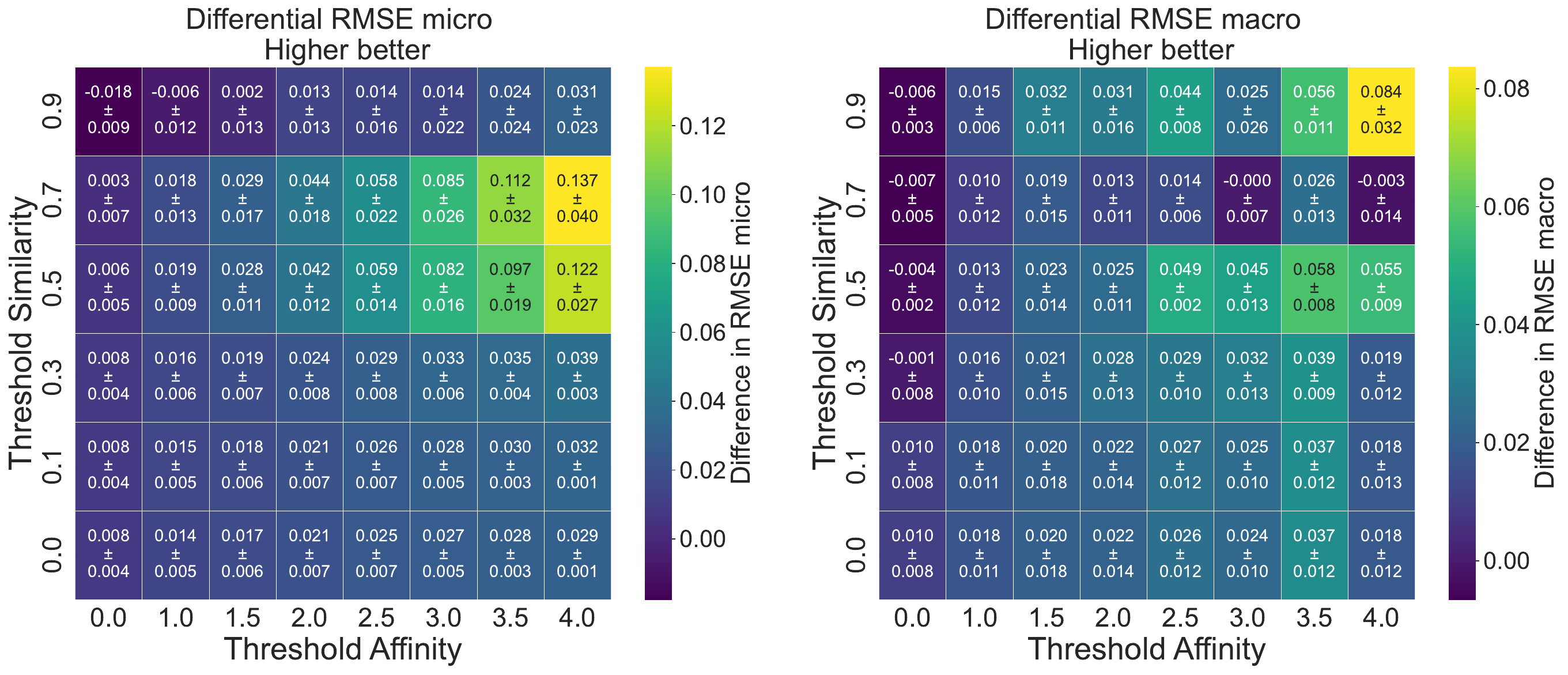}%
\caption{The differential heatmap of the $RMSE_{micro}$ (left) and $RMSE_{macro}$ (right) of the best DTI model (transfer learning involving only the drug encoder, \textbf{warm start}) for groups of compounds split by similarity and affinity thresholds for BindingDB (compound-based split).}%
\label{fig:DTI_BDB_cb_tl_ws_hm_diff}%
\end{figure}

\begin{figure}[H]
\centering
\includegraphics[scale=0.3]{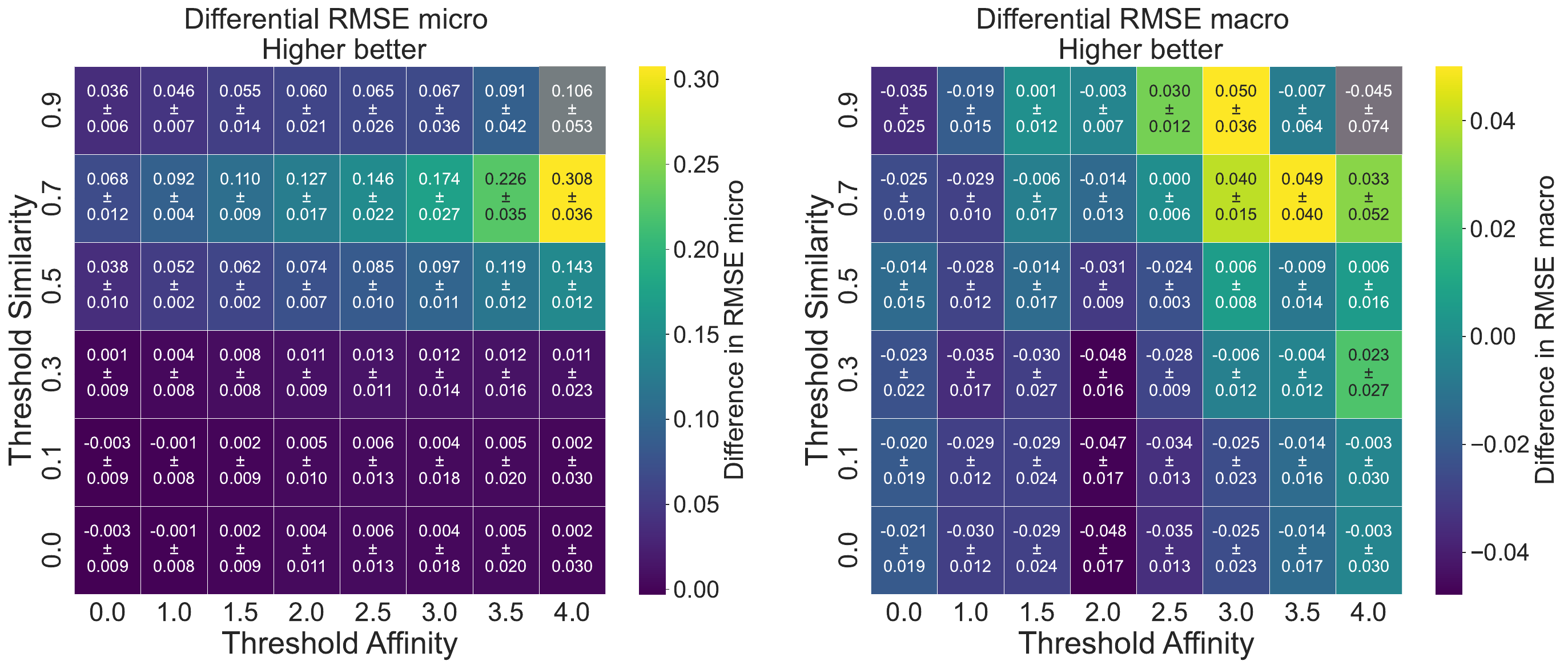}%
\caption{The differential heatmap of the $RMSE_{micro}$ (left) and $RMSE_{macro}$ (right) of the best DTI model (transfer learning involving only the drug encoder, \textbf{warm start}) for groups of compounds split by similarity and affinity thresholds for BindingDB (random split).}%
\label{fig:DTI_BDB_rs_tl_ws_hm_diff}%
\end{figure}

\subsection{Differential heatmaps when transferring both drug and target encoders}

\begin{figure}[H]
\centering
\includegraphics[scale=0.3]{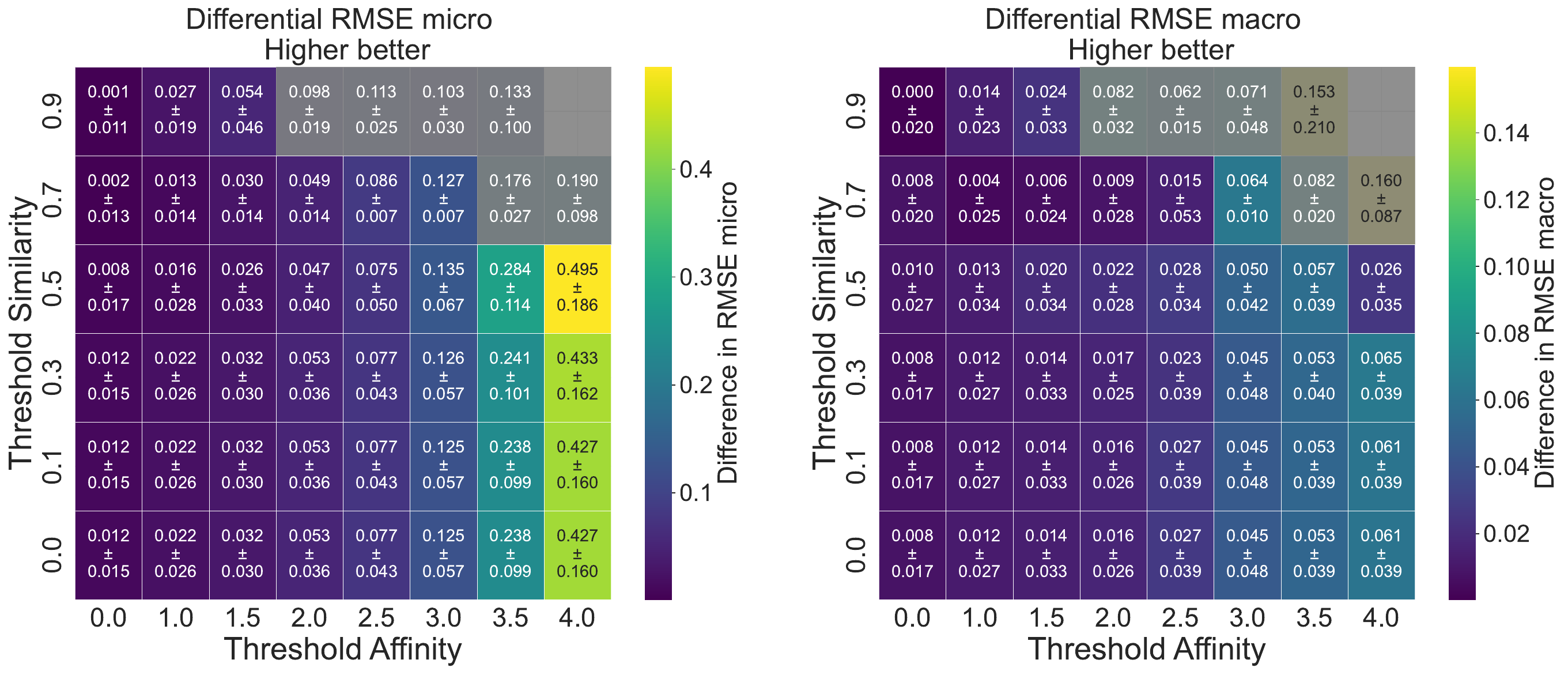}%
\caption{The differential heatmap of the $RMSE_{micro}$ (left) and $RMSE_{macro}$ (right) of the best DTI model (transfer learning involving both drug and target encoders, \textbf{warm start}) for groups of compounds split by similarity and affinity thresholds for KIBA (random split).}%
\label{fig:DTI_KIBA_rs_tl_t_enc_ws_hm_diff}%
\end{figure}

\begin{figure}[H]
\centering
\includegraphics[scale=0.3]{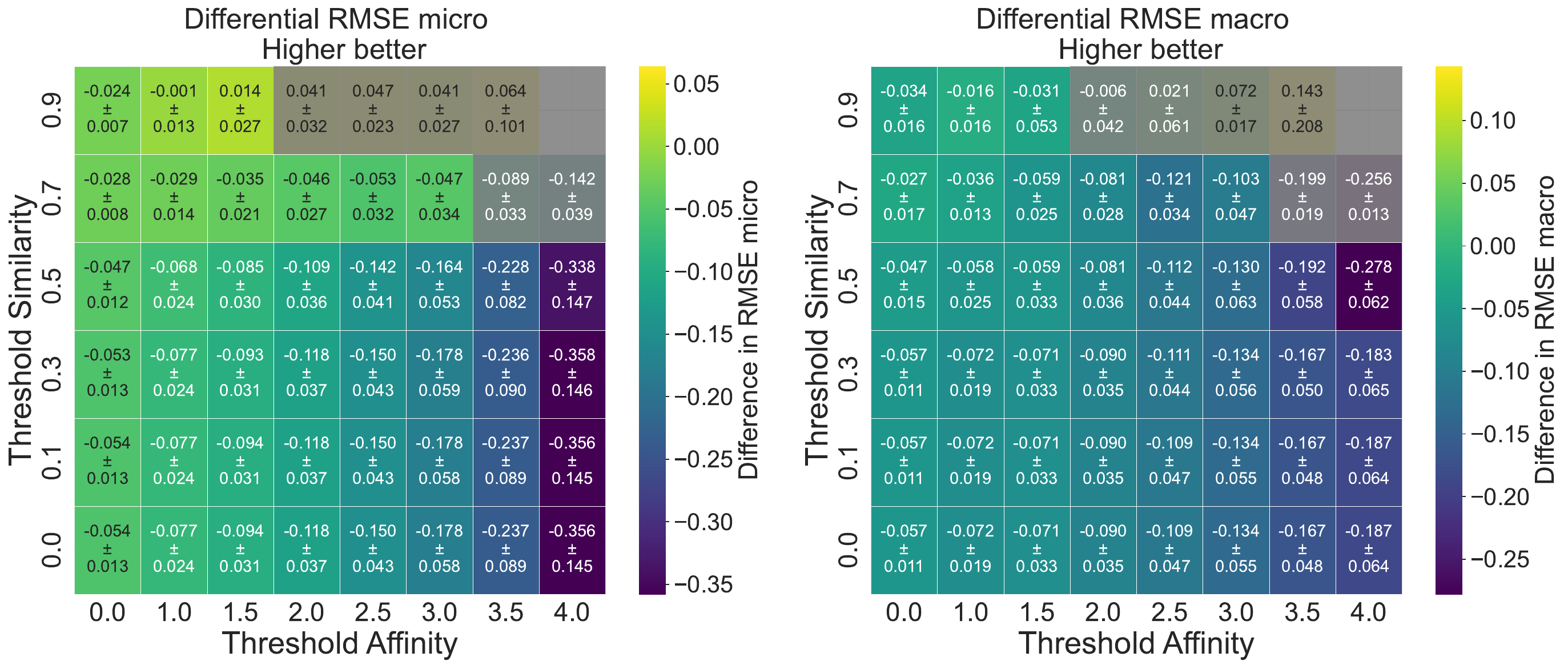}%
\caption{The differential heatmap of the $RMSE_{micro}$ (left) and $RMSE_{macro}$ (right) of the best DTI model (transfer learning involving both drug and target encoders, \textbf{with freezing weights}) for groups of compounds split by similarity and affinity thresholds for KIBA (random split).}%
\label{fig:DTI_KIBA_rs_tl_t_enc_f_hm_diff}%
\end{figure}

\begin{figure}[H]
\centering
\includegraphics[scale=0.3]{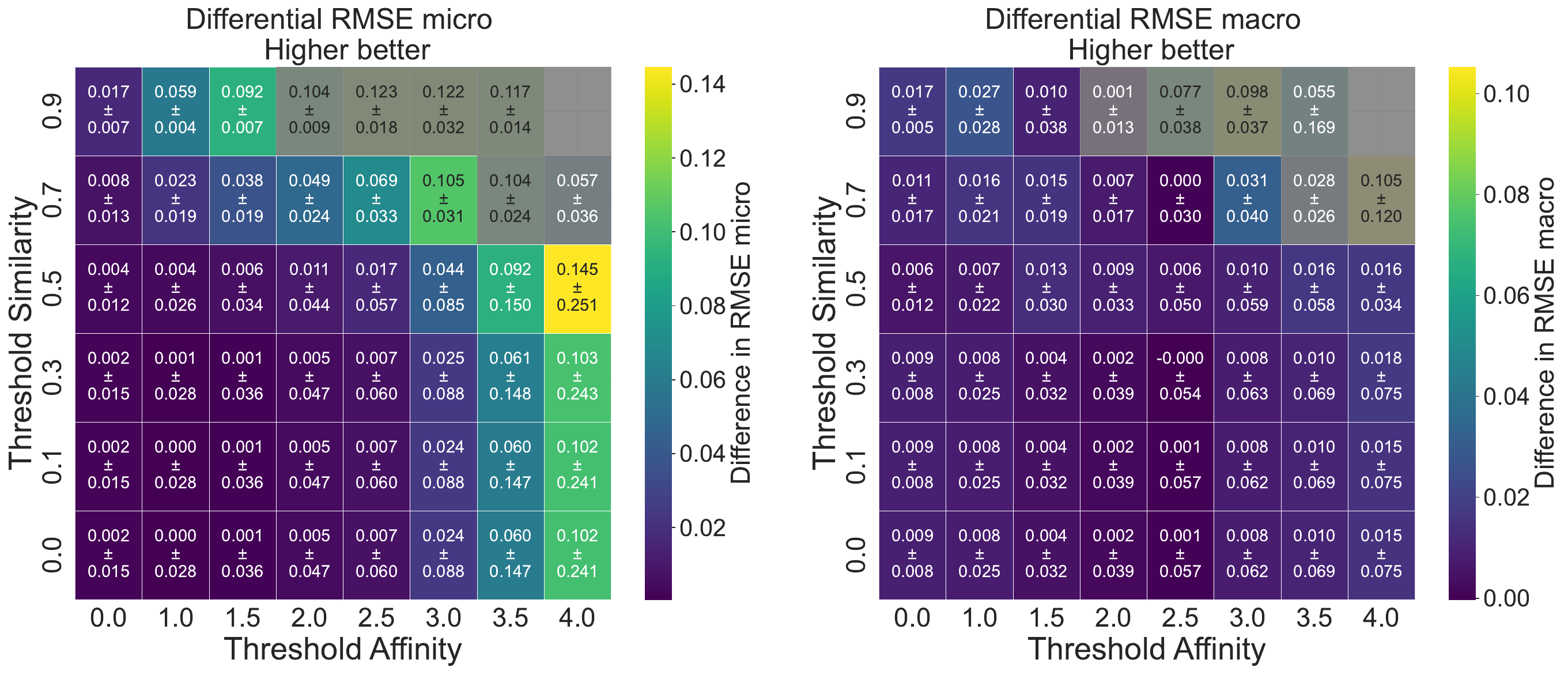}%
\caption{The differential heatmap of the $RMSE_{micro}$ (left) and $RMSE_{macro}$ (right) of the best DTI model (transfer learning involving both drug and target encoders, \textbf{with freezing weights and adding an extra layer}) for groups of compounds split by similarity and affinity thresholds for KIBA (random split).}%
\label{fig:DTI_KIBA_rs_tl_t_enc_f_el_hm_diff}%
\end{figure}

\begin{figure}[H]
\centering
\includegraphics[scale=0.3]{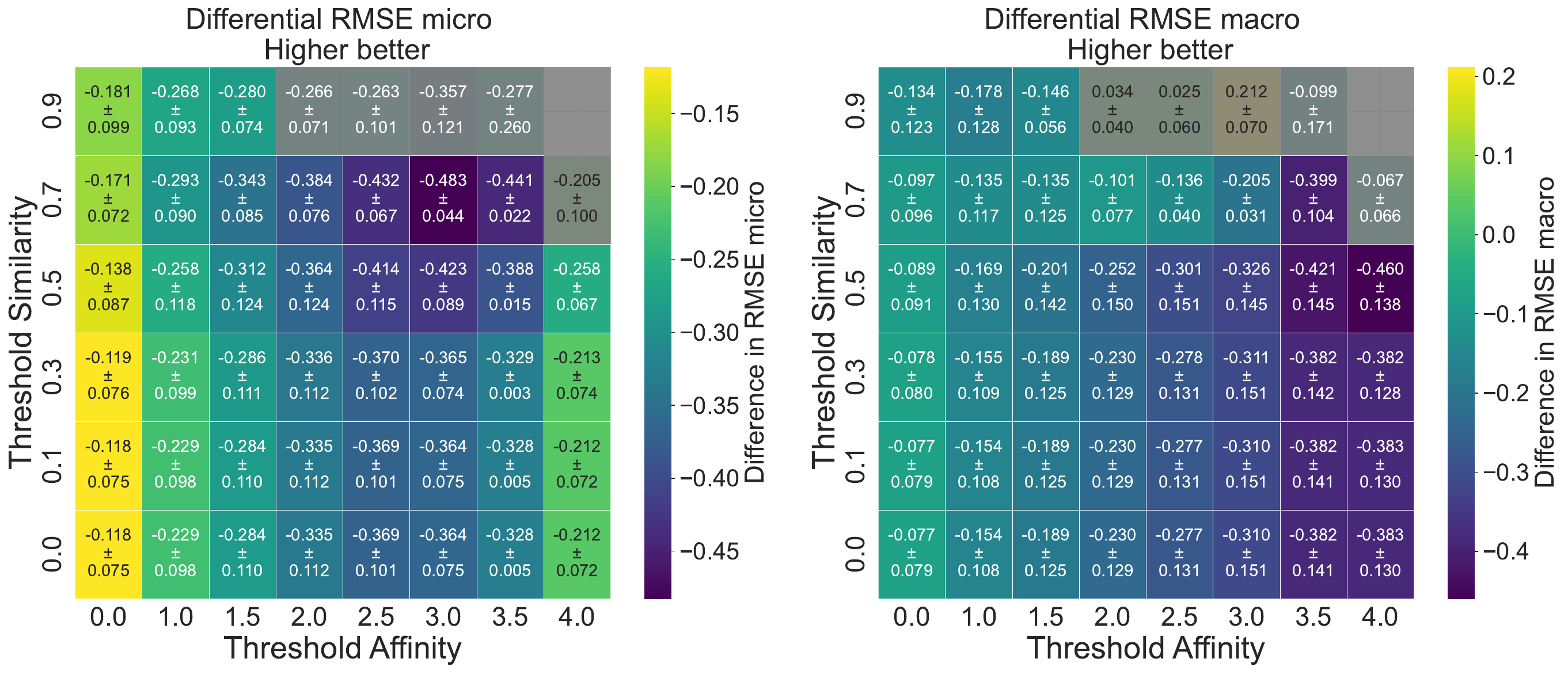}%
\caption{The differential heatmap of the $RMSE_{micro}$ (left) and $RMSE_{macro}$ (right) of the best DTI model (transfer learning involving both drug and target encoders, \textbf{with freezing weights}) for groups of compounds split by similarity and affinity thresholds for KIBA (compound-based split).}%
\label{fig:DTI_KIBA_cb_tl_t_enc_f_hm_diff}%
\end{figure}

\begin{figure}[H]
\centering
\includegraphics[scale=0.3]{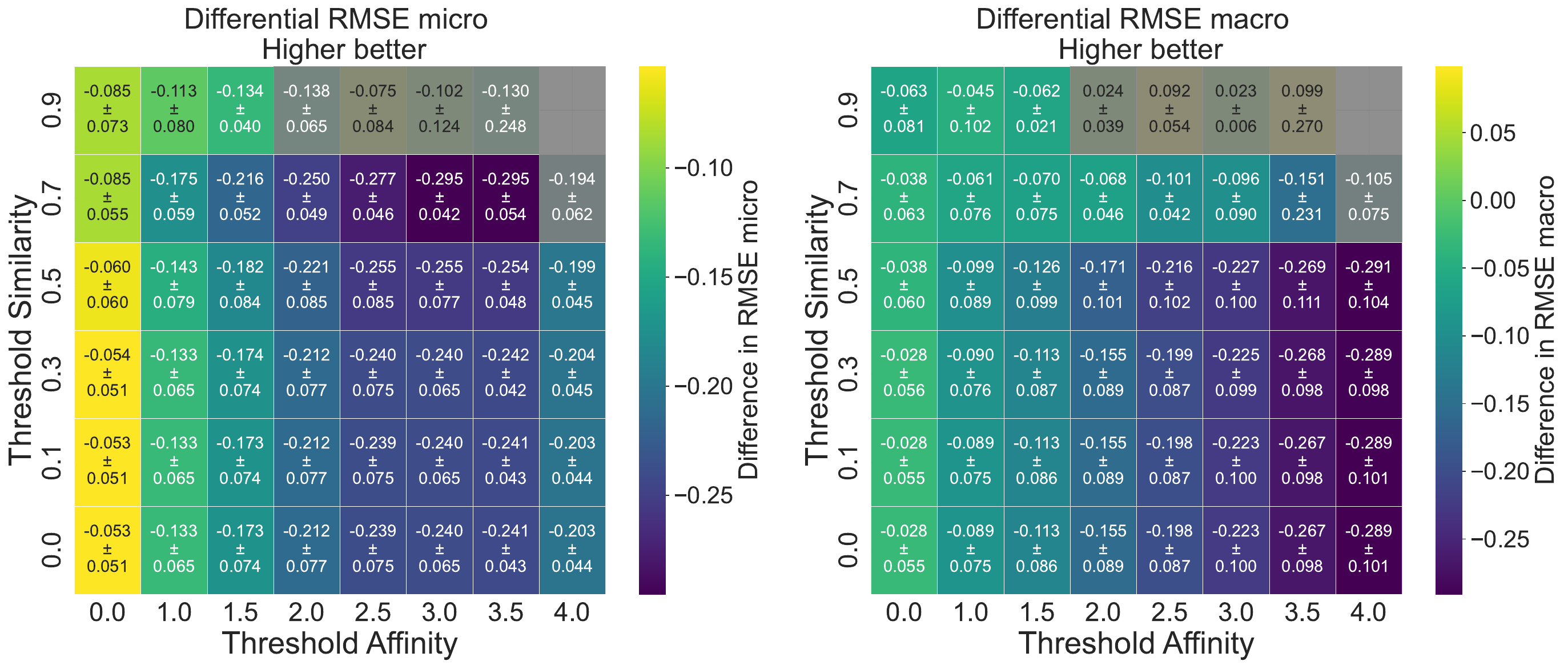}%
\caption{The differential heatmap of the $RMSE_{micro}$ (left) and $RMSE_{macro}$ (right) of the best DTI model (transfer learning involving both drug and target encoders, \textbf{with freezing weights and adding an extra layer}) for groups of compounds split by similarity and affinity thresholds for KIBA (compound-based split).}%
\label{fig:DTI_KIBA_cb_tl_t_enc_f_el_hm_diff}%
\end{figure}

\begin{figure}[H]
\centering
\includegraphics[scale=0.3]{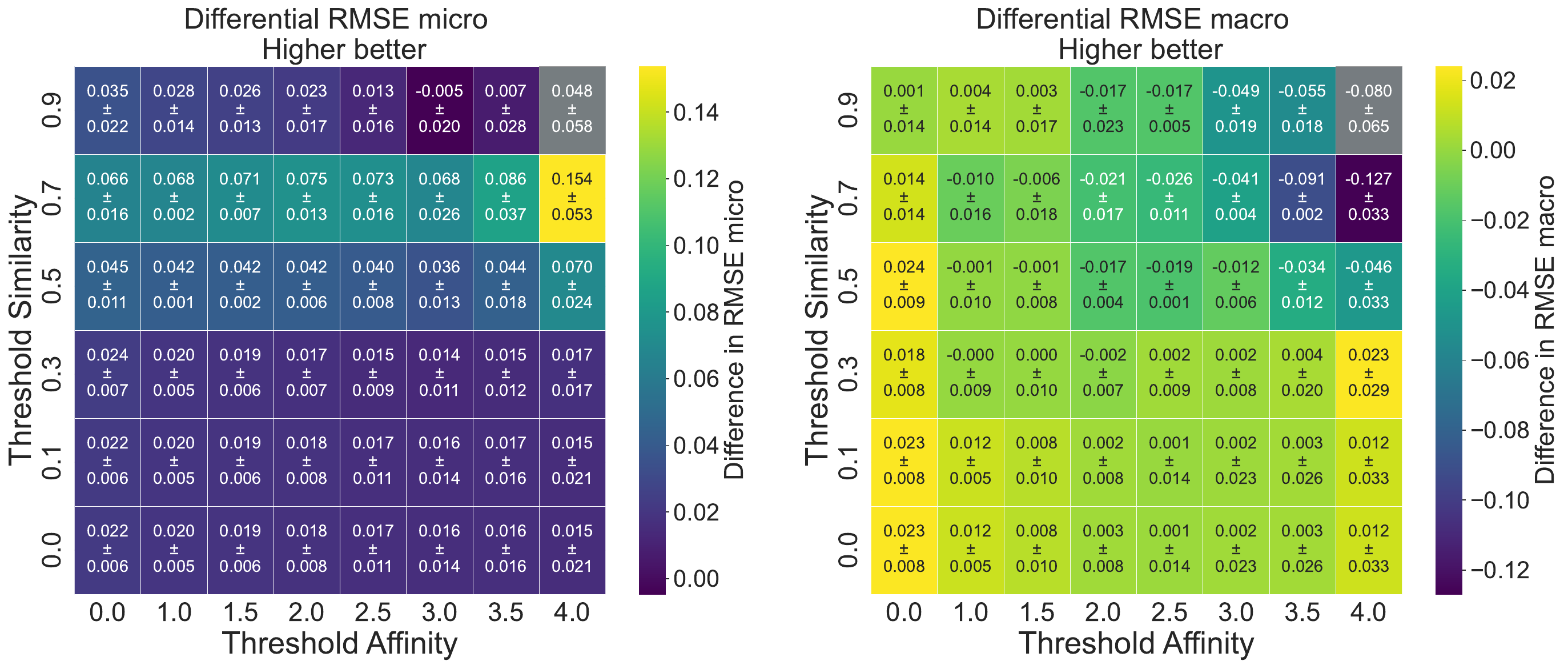}%
\caption{The differential heatmap of the $RMSE_{micro}$ (left) and $RMSE_{macro}$ (right) of the best DTI model (transfer learning involving both drug and target encoders, \textbf{warm start}) for groups of compounds split by similarity and affinity thresholds for BindingDB (random split).}%
\label{fig:DTI_BDB_rs_tl_t_enc_ws_hm_diff}%
\end{figure} 

\begin{figure}[H]
\centering
\includegraphics[scale=0.3]{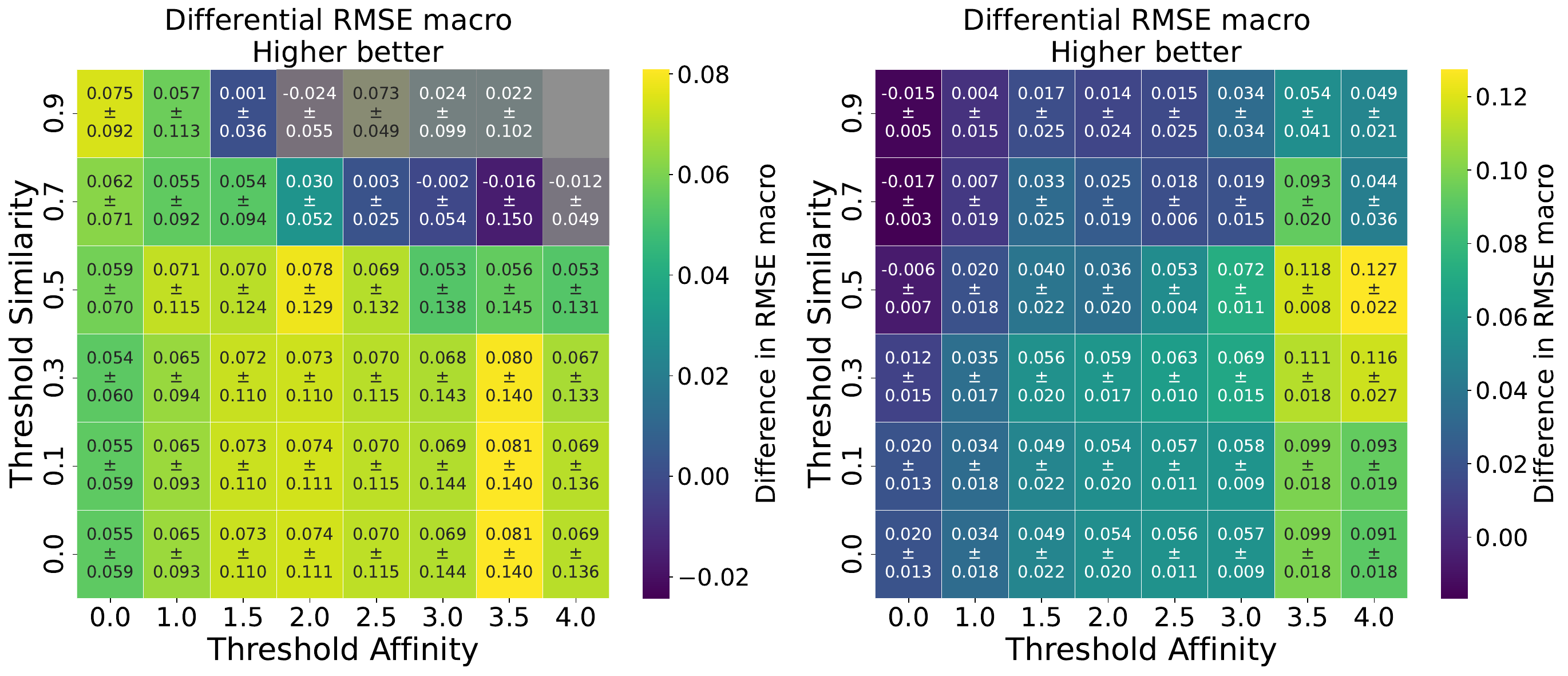}%
\caption{The differential heatmap of the $RMSE_{micro}$ for the best DTI model(transfer learning involving both drug and target encoders, \textbf{warm start}) for the KIBA (left) and BindingDB (right) datasets in the case of a compound-based splits, showing groups of compounds split by similarity and affinity thresholds.}%
\label{fig:DTI_KIBA_BDB_cb_tl_ws_t_enc_macro_hm_diff}%
\end{figure}

\end{document}